\documentclass[10pt,aps,prx,reprint,showpacs,superscriptaddress,nofootinbib]{revtex4-1}
\usepackage{amsthm,amsmath,amsfonts,amssymb,verbatim,dsfont,color}
\usepackage{graphicx,mathrsfs,stackrel}
\usepackage{bm}
\usepackage{epsfig,slashed}
\usepackage{physics}
\usepackage{tikz}
\usepackage{xparse}
\usepackage{ifthen}
\usepackage{lipsum}
\usepackage[colorlinks=true,citecolor=blue,
linkcolor=blue,urlcolor=blue]{hyperref}
\usepackage[caption=false]{subfig}
\theoremstyle{plain}

\theoremstyle{remark}

\begin{document}

\providecommand{\remarkname}{Remark}
\providecommand{\theoremname}{Theorem}

\newcommand{\bsigma}{\boldsymbol{\sigma}}
\newcommand{\re}{\mathop{\mathrm{Re}}}
\newcommand{\im}{\mathop{\mathrm{Im}}}
\renewcommand{\b}[1]{{\boldsymbol{#1}}}
\newcommand{\diag}{\mathrm{diag}}
\newcommand{\sign}{\mathrm{sign}}
\newcommand{\sgn}{\mathop{\mathrm{sgn}}}
\renewcommand{\c}[1]{\mathcal{#1}}

\newcommand{\mb}{\bm}
\newcommand{\ua}{\uparrow}
\newcommand{\da}{\downarrow}
\newcommand{\ra}{\rightarrow}
\newcommand{\la}{\leftarrow}
\newcommand{\mc}{\mathcal}
\newcommand{\bs}{\boldsymbol}
\newcommand{\lra}{\leftrightarrow}
\newcommand{\nn}{\nonumber}
\newcommand{\half}{{\textstyle{\frac{1}{2}}}}
\newcommand{\mf}{\mathfrak}
\newcommand{\MF}{\text{MF}}
\newcommand{\IR}{\text{IR}}
\newcommand{\UV}{\text{UV}}
\newcommand{\so}{\mathfrak{so}}

\title{Continuous transition between Ising magnetic order and a chiral spin liquid}
\author{G. Shankar}
\email{sankaran@ualberta.ca}
\affiliation{Department of Physics, University of Alberta, Edmonton, Alberta T6G 2E1, Canada}
\author{Chien-Hung Lin}
\affiliation{Department of Physics, University of Alberta, Edmonton, Alberta T6G 2E1, Canada}
\author{Joseph Maciejko}
\email{maciejko@ualberta.ca}
\affiliation{Department of Physics, University of Alberta, Edmonton, Alberta T6G 2E1, Canada}
\affiliation{Theoretical Physics Institute, University of Alberta, Edmonton, Alberta T6G 2E1, Canada}

\date\today

\begin{abstract}
The competition between fractionalized spin-liquid states and magnetically ordered phases is an important paradigm in frustrated magnetism. Spin-orbit coupled Mott insulators with Ising-like magnetic anisotropies, such as Kitaev materials, are a particularly rich playground to explore this competition. In this work, we use effective field theory methods to show that a direct quantum phase transition can occur in two-dimensional (2D) Ising spin systems between a topologically ordered chiral spin liquid and a phase with magnetic long-range order. Such a transition can be protected by lattice symmetries and is described by a theory of massless Majorana fields coupled to non-Abelian $SO(N)$ gauge fields with a Chern-Simons term. We further show that Euclidean Majorana zero modes bound to $\mathbb{Z}_2$ monopole-instantons in the emergent non-Abelian gauge field are key to understanding spontaneous symmetry breaking in the ordered phase.
\end{abstract}

\maketitle

\section{Introduction}

Quantum phase transitions out of fractionalized spin-liquid states in two dimensions (2D) are an active area of research in the study of quantum matter~\cite{savary2017,zhou2017,vojta2018}. From a theoretical standpoint, the universal properties of spin liquids are captured by slave-particle gauge theories with bosonic or fermionic spinons~\cite{WenBook}. In describing a transition from a spin liquid to a conventional ordered phase, two effects must be accounted for: spontaneous symmetry breaking, and confinement of excitations with nonzero gauge charge. In theories with bosonic spinons, such as Schwinger boson theories of $\mathbb{Z}_2$ spin liquids~\cite{read1991}, confinement concomitant with symmetry breaking results from the condensation of bosonic visons and/or spinons which carry nontrivial quantum numbers under global symmetries~\cite{xu2009}. In descriptions of spin liquids with fermionic gauge theories, such as the $U(1)$ gauge theory of the Dirac spin liquid~\cite{hermele2005}, condensation of a fermion bilinear leads to symmetry breaking and opening of a fermion mass gap~\cite{ghaemi2006,lu2017,song2019,dupuis2019,zerf2019,zerf2020,janssen2020,kruger2021}. The opening of this gap is followed by the proliferation of monopole-instantons, which induces confinement~\cite{polyakov1975,polyakov1977,polyakov1987}. In addition to those of fermion bilinears, the symmetry quantum numbers of monopole operators are important to determine the precise patterns of symmetry breaking~\cite{song2019,alicea2008,hermele2008,song2020,dupuis2021}.

An emerging platform for the observation of spin liquids and their competition with various ordered phases is spin-orbit coupled Mott insulators; such systems have been a focus of quantum materials research in recent years~\cite{witczak-krempa2014,rau2016}. In those systems, strong correlations promote the formation of local magnetic moments, while spin-orbit coupling entangles spin and orbital degrees of freedom and introduces anisotropy in the magnetic exchange interactions. The paradigmatic class of materials in this context is Kitaev materials~\cite{jackeli2009,takagi2019}, described at low energies by effective spin-1/2 moments on the honeycomb lattice and governed by a Kitaev-like Hamiltonian~\cite{kitaev2006} in which $SU(2)$ spin rotation symmetry is broken to a discrete subgroup. Recently, the Kitaev material $\alpha$-RuCl$_3$ has attracted much attention due to the observation of a magnetic-field-induced transition from a zigzag-ordered state at low fields to a paramagnetic state at higher fields~\cite{banerjee2016,banerjee2017,jansa2018,banerjee2018,balz2021,kasahara2018,yokoi2021}. Remarkably, the latter state appears to exhibit a quantized thermal Hall conductance $\kappa_{xy}/T=1/2$ in units of $\pi k_B^2/6\hbar$, suggestive of a gapped chiral spin liquid phase with intrinsic topological order~\cite{kasahara2018,yokoi2021}. In Ref.~\cite{zou2020}, a theory of the transition between the zigzag-ordered state and the chiral spin liquid was developed, based on dualities of (2+1)D gauge theories with $U(N)$ gauge groups.

Motivated by these recent developments, we ask the general question whether, from an effective field theory point of view, the phase diagram of spin systems with Ising spin-flip symmetry can exhibit a continuous transition from Ising magnetic order to a gapped chiral spin liquid. Given that the chiral spin liquid is a topological phase without a local order parameter, while the Ising-ordered phase exhibits conventional symmetry breaking, such a transition is necessarily an exotic non-Landau transition involving the fractionalized degrees of freedom of the spin liquid, and possibly monopole-instanton configurations in the associated emergent gauge field.

In this paper, we use effective field theory methods to show that such an exotic transition is in general possible. Our approach is based on a parton decomposition of the Ising spin operator, involving fractionalized Majorana fermion degrees of freedom coupled to an emergent non-Abelian $SO(N)$ gauge field. Our study can be viewed as a generalization of Ref.~\cite{Barkeshli1}---which studies transitions between Mott insulating, fractional quantum Hall, and superfluid states of bosons with continuous $U(1)$ symmetry---to systems of Ising spins with a discrete $\mathbb{Z}_2$ symmetry. Chern-number changing transitions between different topologically superconducting states of the Majorana partons correspond to different phases of the Ising spin system. While various spin-liquid states can be accessed in this way, including spin liquids with non-Abelian topological order, we focus on Abelian chiral spin liquids with the topological order of the bosonic fractional quantum Hall (Laughlin) state. In our construction, such a chiral spin liquid is naturally proximate to a trivial paramagnet and to a magnetically ordered phase with broken $\mathbb{Z}_2$ symmetry. Using recently conjectured dualities of $SO(N)$ gauge theories in (2+1)D, we find that the critical theory for the ordering transition from the trivial paramagnet is, as expected, dual to the standard 3D Ising Wilson-Fisher theory, while transitions involving the chiral spin liquid are described by theories of massless Majorana fields coupled to an $SO(N)$ gauge field with a Chern-Simons term. In particular, we show that a direct transition from the chiral spin liquid to the Ising-ordered phase is possible and can be protected by inversion symmetry on the honeycomb lattice.

Finally, in analogy with our previous work on bosons with $U(1)$ symmetry~\cite{shankar2021}, we show that the breaking of $\mathbb{Z}_2$ symmetry in the confined, ordered phase can be understood as a nontrivial consequence of Euclidean Majorana zero modes (ZMs) bound to monopole-instantons. By contrast with monopole-instantons in $U(1)$ theories, the latter carry here a $\mathbb{Z}_2$ topological charge under the $\mathbb{Z}_2^{\cal{M}}$ magnetic symmetry of $SO(N)$ gauge theory in (2+1)D. Under the assumption that the infrared effects of such instantons is adequately captured by a semiclassical instanton-gas treatment, the Euclidean ZMs lead to an effective interaction among Majorana fermions that is analogous to the 't~Hooft vertex in quantum chromodynamics~\cite{thooft1976,thooft1976a,thooft1986}. This interaction intertwines the Ising $\mathbb{Z}_2$ symmetry with the $\mathbb{Z}_2^{\cal{M}}$ magnetic symmetry. As a consequence of this intertwinement, the spontaneous breakdown of $\mathbb{Z}_2^{\cal{M}}$ magnetic symmetry expected in a confined phase~\cite{komargodski2018,benini2018} automatically results in long-range Ising order for the underlying spin system.

The rest of the paper is structured as follows. In Sec.~\ref{sec:U1}, we review the parton description of bosons with $U(1)$ symmetry~\cite{Barkeshli1,shankar2021}, as a means to introduce the basic ideas and methods that we will generalize to Ising spins with $\mathbb{Z}_2$ spin-flip symmetry. In Sec.~\ref{sec:ising}, we introduce our parton decomposition of Ising spins and discuss the various phases that can be accessed within the parton mean-field framework: chiral spin liquids, a trivial paramagnet, and an ordered phase with broken $\mathbb{Z}_2$ symmetry. Using $SO(N)$ dualities in (2+1)D, we discuss transitions between these phases. In Sec.~\ref{sec:Z2instantons}, we turn our focus to the broken phase. Although conventional from the microscopic standpoint, its description within the parton framework necessitates accounting for nonperturbative confinement effects. We discuss $\mathbb{Z}_2$ monopole-instantons in $SO(N)$ gauge theory, show that Euclidean fermion ZMs are bound to them, and resum the instanton gas to exhibit the 't~Hooft vertex that properly accounts for the broken $\mathbb{Z}_2$ symmetry. We conclude in Sec.~\ref{sec:conclusion} with a summary of our main results and suggestions for future research.

\section{Warm-up: bosons with $U(1)$ symmetry}
\label{sec:U1}

We begin by briefly reviewing the problem of continuous quantum phase transitions in systems of hardcore bosons with the global $U(1)$ symmetry associated with particle-number conservation. To aid the passage from $U(1)$ bosons to $\mathbb{Z}_2$ spins, we re-interpret the results of Ref.~\cite{Barkeshli1} in the context of dualities of (2+1)D quantum field theories with unitary gauge groups~\cite{ChenFisherWu,son2015,Metlitskiduality,Wangduality,SEIBERGduality,DualityKarchTong,murugan2016,
mross2016,kachru2017,karch2017}. We also point out the key role of monopole-instantons and the Euclidean fermion ZMs bound to them in accounting for the physics of broken-symmetry phases~\cite{shankar2021}.

\subsection{Parton construction}\label{sec:partonU1}

We begin by considering a system of charge-1 hardcore bosons on a 2D lattice described by operators $b(\b{r})$ ($b^\dag(\b{r})$) that annihilate (create) a boson on lattice site $\b{r}$. We then write the boson operator as
\begin{align}\label{partonU1}
b(\b{r})=f_1(\b{r})f_2(\b{r}),
\end{align}
where $f_1(\b{r})$ and $f_2(\b{r})$ are fermionic annihilation operators. This parton decomposition~\cite{Wenparton1,Wenparton2} introduces a local gauge redundancy. We consider parton mean-field ans\"atze such that $f_1$ forms a Chern insulator with Chern number 1 and $f_2$ forms a Chern insulator with Chern number $C$. In general, such ans\"atze have a $U(1)$ gauge structure with emergent gauge field $a_\mu$; we assume $f_1$ ($f_2$) carries gauge charge $-1$ ($+1$), and $f_2$ carries the unit global $U(1)$ charge of the boson system.

Upon integrating out the massive partons $f_1$ and $f_2$, we obtain the low-energy effective Lagrangian:
\begin{align}\label{LeffU1}
\mathcal{L}=\frac{1}{4\pi}ada+\frac{C}{4\pi}(a+A)d(a+A),
\end{align}
where $adb\equiv\epsilon^{\mu\nu\lambda}a_\mu\partial_\nu b_\lambda$ for any two gauge fields $a_\mu$ and $b_\mu$, and we have added a background gauge field $A_\mu$ which couples to the global $U(1)$ symmetry. Performing the shift $a_\mu\rightarrow a_\mu-\frac{C}{C+1}A_\mu$ to eliminate the cross terms, we obtain:
\begin{align}
\mathcal{L}=\frac{C+1}{4\pi}ada+\frac{1}{4\pi}\frac{C}{C+1}AdA.
\end{align}
For values of $C$ other than $C=-2,-1,0$, this describes an Abelian fractional quantum Hall state with ground-state degeneracy $|C+1|^g$ on a genus-$g$ surface and quantized Hall conductance $\sigma_{xy}=C/(C+1)$. For $C=0$, the ground state is unique and the Hall conductance vanishes: this is the Bose Mott insulator. For $C=-2$, the ground state is again unique, but the Hall conductance is nonzero, $\sigma_{xy}=2$: this is a bosonic integer quantum Hall state~\cite{lu12,senthil2013,chen13,barkeshli2013}. For $C=-1$, the Chern-Simons term for $a_\mu$ cancels and we must keep a Maxwell term. Integrating out $a_\mu$, we obtain in the low-energy limit,
\begin{align}\label{SeffU1SF}
S_\text{eff}[A_\mu]=\frac{1}{8\pi^2}\int\frac{d^3q}{(2\pi)^3}A_\mu(-q)\left(\eta^{\mu\nu}-\frac{q^\mu q^\nu}{q^2}\right)A_\nu(q),
\end{align}
where $\eta^{\mu\nu}$ is the (2+1)D Minkowski metric. Equation~(\ref{SeffU1SF}) describes the (transverse) Meissner response of a charged superfluid, thus the $C=-1$ phase is a superfluid of the $b$ bosons.

\subsection{Phase transitions and $U(1)$ dualities}\label{sec:U1critical}

At the parton mean-field level, transitions between the different bosonic phases mentioned above are Chern-number-changing (topological) transitions in the $f_2$ parton band structure. For simplicity, we focus on transitions between the Mott insulator ($C=0$), superfluid ($C=-1$), and $\nu=1/2$ bosonic fractional quantum Hall state ($C=1$)~\cite{kalmeyer1987}. To derive a critical theory for the transition, we can integrate out $f_1$, which remains gapped across the transition. This generates a $U(1)_1$ Chern-Simons term in the effective theory. By contrast, $f_2$ becomes gapless at the transition and must be kept in the critical theory. In the low-energy limit and near the transition, the $f_2$ band structure will generically consist of two Dirac points $\b{K}_+$ and $\b{K}_-$, such that $f_2(\b{r})$ can be expanded near the Dirac points: $f_2(\b{r})\approx\sum_{k=\pm}e^{i\b{K}_k\cdot\b{r}}\psi_{2k}(\b{r})$, where $\psi_{2+},\psi_{2-}$ are slow two-component Dirac fields with mass $m_+,m_-$ respectively. The low-energy effective theory interpolating between all three phases is thus:
\begin{align}\label{EFTpartonU1}
\mathcal{L}=\frac{1}{4\pi}ada+\sum_{k=\pm}\bar{\psi}_{2k}(i\slashed{D}-m_k)\psi_{2k},
\end{align}
where $\slashed{D}=\gamma^\mu D_\mu$ with $D_\mu=\partial_\mu-i(a_\mu+A_\mu)$ the gauge-covariant derivative, and $\gamma^\mu$ are (2+1)D Dirac matrices.

\begin{figure}[t]
\includegraphics[width=0.6\columnwidth]{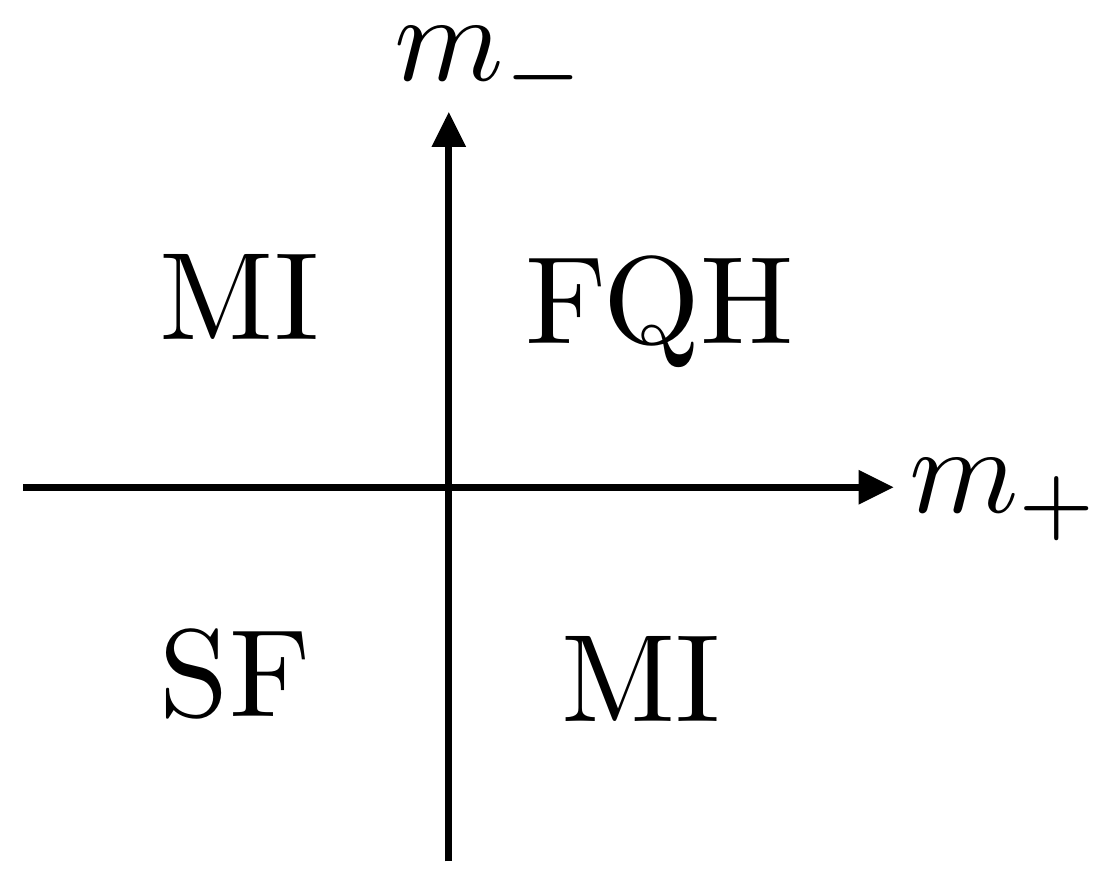}
\caption{Phase diagram for bosons with $U(1)$ symmetry as a function of the two tuning parameters $m_+$, $m_-$. MI: trivial Mott insulator; SF: superfluid; FQH: $\nu=1/2$ bosonic Laughlin state.}
\label{fig:phase1}
\end{figure}

The phases described in Sec.~\ref{sec:partonU1} are recovered when both Dirac fermions are massive and can be integrated out (Fig.~\ref{fig:phase1}). Since a single massive two-component Dirac fermion with mass $m$ carries a partial Chern number of $\frac{1}{2}\sgn m$~\cite{niemi1983,redlich1984,*redlich1984b}, when both $m_\pm>0$, we recover Eq.~(\ref{LeffU1}) with $C=1$, i.e., the bosonic $\nu=1/2$ Laughlin state.  When both $m_\pm<0$, we find the superfluid with $C=-1$. When the masses are of opposite sign, the partial Chern numbers from the two Dirac fermions cancel out and we obtain the trivial Mott insulator with $C=0$.

Transitions between the Mott insulator and superfluid and between the Mott insulator and the bosonic Laughlin state can be accessed by tuning $m_-$ through zero at $m_+<0$ and $m_+>0$, respectively. When $m_+<0$, integrating out $\psi_{2+}$ and setting $m_-=0$ gives
\begin{align}\label{U1:fermion1}
\mathcal{L}=\frac{1}{8\pi}ada+\bar{\psi}_{2-}i\slashed{D}\psi_{2-}-\frac{1}{4\pi}Ada-\frac{1}{8\pi}AdA,
\end{align}
a single two-component Dirac fermion coupled to a $U(1)$ Chern-Simons gauge field at level 1/2, which is conjectured~\cite{ChenFisherWu,SEIBERGduality,DualityKarchTong,murugan2016,kachru2017,karch2017} to be dual in the infrared to the (2+1)D Wilson-Fisher fixed point of a single complex scalar $\phi$,
\begin{align}\label{U1:boson1}
\mathcal{L}_\text{dual}=|(\partial_\mu-iA_\mu)\phi|^2-\lambda|\phi|^4.
\end{align}
We thus recover the known fact that the boson superfluid-Mott insulator transition is in the 3D XY universality class (in the presence of particle-hole symmetry, which is assumed here due to the relativistic Dirac dispersions). Furthermore, a fermion mass term for $\psi_{2-}$ maps to a mass term for the scalar of the same sign\footnote{The reason the fermion and scalar masses are of the same sign is that we are in fact using a time-reversed version of the duality in Ref.~\cite{SEIBERGduality}.}, such that $m_->0$ corresponds to the disordered (Mott insulating) phase of the scalar and $m_-<0$ to its broken symmetry (superfluid) phase. As is clear from Eq.~(\ref{U1:boson1}), the dual scalar field $\phi$ carries charge 1 under the background gauge field $A_\mu$ and can thus be directly interpreted as the continuum limit of the boson operator $b$ near the superfluid-insulator transition.

When $m_+>0$, integrating out $\psi_{2+}$ and setting $m_-=0$ yields
\begin{align}\label{U1:fermion2}
\mathcal{L}=\frac{3}{8\pi}ada+\bar{\psi}_{2-}i\slashed{D}\psi_{2-}+\frac{1}{4\pi}Ada+\frac{1}{8\pi}AdA,
\end{align}
a Dirac fermion coupled to $U(1)_{3/2}$ Chern-Simons theory, which is dual to a single complex scalar coupled to $U(1)_{-2}$ Chern-Simons theory,
\begin{align}\label{U1:boson2}
\mathcal{L}_\text{dual}=|(\partial_\mu-i\tilde{a}_\mu)\phi|^2-\lambda|\phi|^4-\frac{2}{4\pi}\tilde{a}d\tilde{a}+\frac{1}{2\pi}Ad\tilde{a}.
\end{align}
The derivation of this duality is reviewed in Appendix~\ref{app:U1dualities}. If we add a positive mass term for the scalar (corresponding to $m_->0$ for the fermion), the scalar can be integrated out and upon shifting $\tilde{a}\rightarrow\tilde{a}+\frac{1}{2}A$ we obtain $U(1)$ Chern-Simons terms of level $-2$ and $1/2$ for the $\tilde{a}$ and $A$ gauge fields, respectively, in accordance with the $\nu=1/2$ bosonic Laughlin state expected for $m_->0$. Setting $A=0$, approximate critical exponents for the Mott insulator-bosonic Laughlin state transition can be obtained by studying Eq.~(\ref{U1:fermion2}) in the $1/N_f$ expansion where $N_f$ denotes the number of Dirac fermion flavors~\cite{ChenFisherWu,ye1998}, or by studying the dual theory (\ref{U1:boson2}) in a bosonic $1/N_b$ expansion~\cite{WenWu}.

Finally, exact microscopic symmetries may force $m_+=m_-$ and protect a topological transition in which $C$ changes by 2 (e.g., inversion symmetry in the Haldane model~\cite{haldane1988}). In this case a direct transition from the superfluid to the bosonic Laughlin state is generically allowed, and the critical theory is (\ref{EFTpartonU1}) with $m_+=m_-=0$. Critical exponents can be estimated using the large-$N_f$ expansion mentioned above~\cite{ChenFisherWu,ye1998}.

\subsection{Instantons, fermion zero modes, and superfluidity}
\label{sec:U1instantons}

So far, we have ignored the compactness of the emergent $U(1)$ gauge field $a_\mu$. In the superfluid phase, the low-energy effective action does not feature a Chern-Simons term, thus confinement effects due to the proliferation of monopole-instantons~\cite{polyakov1975,polyakov1977,polyakov1987} are expected to play a key role. In the presence of fermions, such instantons can additionally lead to symmetry-breaking effects~\cite{affleck1982}. In an instanton background $a_\mu^{(g)}$ of topological charge $g$, a (2+1)D Dirac fermion $\psi$ of charge $e$ and mass $m$ possesses Euclidean ZMs given by~\cite{shankar2021}:
\begin{align}
\psi_0^+(r,\theta,\phi)&=\frac{\sqrt{2m}}{r}e^{-mr}\mathcal{Y}^{1/2}_{1/2,0,0}(\theta,\phi),\\
\psi_0^-(r,\theta,\phi)&=\frac{\sqrt{-2m}}{r}e^{mr}\mathcal{Y}^{1/2}_{-1/2,0,0}(\theta,\phi),
\end{align}
for $eg=+1/2$ and $eg=-1/2$, respectively, according to the Dirac quantization condition. Here $\mathcal{Y}^{j\pm 1/2}_{eg,j,m_j}(\theta,\phi)$ are monopole spinor harmonics with total angular momentum $j$ and its projection $m_j$~\cite{wu1976,kazama1977}, and $(r,\theta,\phi)$ are spherical coordinates in 3D Euclidean spacetime. In the semiclassical instanton gas approximation, those fermion ZMs induce an effective four-fermion interaction, known as the 't Hooft vertex~\cite{thooft1976,thooft1976a,thooft1986}:
\begin{align}\label{tHooftU1}
e^{-2\pi i\sigma}e^{-i\vartheta}\psi_{1+}^\intercal\gamma\psi_{2+}\psi_{1-}^\intercal\gamma\psi_{2-}+\mathrm{H.c.},
\end{align}
where $\psi_{1\pm}$ are the slow fields in the low-energy expansion $f_1(\b{r})\approx\sum_{k=\pm}e^{i\b{K}_k\cdot\b{r}}\psi_{1k}(\b{r})$, $\sigma$ is the dual photon, $\vartheta$ is a topological angle analogous to the theta angle of 4D Yang-Mills theory, and $\gamma$ is a certain $2\times 2$ matrix in spinor space~\cite{shankar2021}. The operator $\mathcal{M}(x)=e^{2\pi i\sigma(x)}$ is a monopole operator that inserts $2\pi$ flux at a given point $x$ in spacetime.

The importance of this term is that it properly accounts for the unique $U(1)$ global symmetry of the microscopic boson system. Absent instanton effects, the Lagrangian (\ref{EFTpartonU1}) has a spurious $U(1)\times U(1)_\text{top}$ symmetry, where the first factor corresponds to the conservation of $\psi_{2+}$ particle number, and the second factor represents the conservation of the topological current $j^\mu_\text{top}=\frac{1}{2\pi}\epsilon^{\mu\nu\lambda}\partial_\nu a_\lambda=\frac{1}{(2\pi)^2}\partial^\mu\sigma$. The 't~Hooft vertex (\ref{tHooftU1}) intertwines these two symmetries, reducing them to the diagonal $U(1)$ subgroup under which a phase rotation of $\psi_{2\pm}$ is compensated by a shift of $\sigma$. Assuming instanton proliferation in the confined phase, $\sigma$ acquires an expectation value, which breaks this $U(1)$ symmetry spontaneously and results in a superfluid phase.

\section{Ising spins}
\label{sec:ising}

We now turn to our main focus, a system of quantum Ising spins $\tau^z(\b{r})=\pm 1$ living on the sites $\b{r}$ of a 2D lattice. We assume the Hamiltonian of the system is invariant under the global $\mathbb{Z}_2$ symmetry $\tau^z(\b{r})\rightarrow-\tau^z(\b{r})$, for all $\b{r}$. We wish to describe ordered and disordered phases of this quantum spin system, as well as transitions between them, using a parton construction.

\subsection{Parton construction}\label{sec:partonZ2}

In analogy with (\ref{partonU1}), we introduce the parton decomposition
\begin{align}\label{partonZ2}
\tau^z(\b{r})=i^{N/2}\chi^1(\b{r})\chi^2(\b{r})\cdots\chi^N(\b{r}),
\end{align}
where $N$ is even and $\chi^\alpha(\b{r})$, $\alpha=1,\ldots,N$ are Hermitian (Majorana) fermion operators obeying the $SO(N)$ Clifford algebra $\{\chi^\alpha(\b{r}),\chi^\beta(\b{r}')\}=2\delta^{\alpha\beta}\delta_{\b{r}\b{r}'}$. One easily checks that (\ref{partonZ2}) implies the expected properties $\tau^z(\b{r})=\tau^z(\b{r})^\dag$, $\tau^z(\b{r})^2=1$, and that the $\tau^z(\b{r})$ commute on different sites. This parton decomposition introduces a local $SO(N)$ gauge redundancy $\chi(\b{r})\rightarrow R(\b{r})\chi(\b{r})$ under which $\tau^z(\b{r})$ remains invariant, where $R(\b{r})\in SO(N)$ and we group the Majorana operators into an $SO(N)$ vector $\chi=(\chi^1,\ldots,\chi^{N})^\intercal$. The charge under the global Ising symmetry can be assigned to any odd number of the $N$ Majorana modes; we choose to assign the global $\mathbb{Z}_2$ charge to $\chi^1$, i.e., $\chi$ transforms as $\chi(\b{r})\rightarrow W\chi(\b{r})$ where $W=\diag(-1,1,\ldots,1)$. This action of the global symmetry does not commute with $SO(N)$ gauge transformations in the parton Hilbert space. Therefore, unlike for the $U(1)$ boson problem (Sec.~\ref{sec:U1}), we cannot couple the parton system to a background $\mathbb{Z}_2$ gauge field while maintaining $SO(N)$ invariance at the Lagrangian level. However, the global symmetry action is well defined on gauge-invariant operators and gauge-invariant states. To the difference of other Majorana-based parton decompositions of spin operators~\cite{fu2018}, explicit expressions for the operators $\tau^x(\b{r})$ and $\tau^y(\b{r})$ in terms of the Majorana fermions $\chi^\alpha(\b{r})$ are expected to be nonlocal and cannot be easily written down. Nonetheless, in Appendix~\ref{app:SONLGT}, we show that the strong-coupling limit of an $SO(N)$ lattice gauge theory with Majorana matter naturally describes a quantum Ising spin system, thus lending support to (\ref{partonZ2}) as a valid parton representation.

We consider an $SO(N)$-invariant parton mean-field ansatz in which all $N$ partons of the multiplet form a class-D topological superconductor~\cite{read2000} with Chern number $C$. Considering fluctuations above the mean-field ground state, the partons couple to an emergent $SO(N)$ gauge field denoted by $a_\mu$. Integrating out the partons, we obtain the effective Lagrangian
\begin{align}\label{EFTpartonZ2}
\mathcal{L}=\text{CS}_{SO(N)_C}[a]+\ldots,
\end{align}
where $\text{CS}_{SO(N)_k}[a]$ denotes a level-$k$ non-Abelian $SO(N)$ Chern-Simons term for the gauge field $a$~\cite{dijkgraaf1990,Maxduality,Aharony2017},
\begin{align}
\text{CS}_{SO(N)_k}[a]=\frac{k}{2\cdot 4\pi}\tr\left(a\wedge da+\frac{2}{3}a\wedge a\wedge a\right),
\end{align}
with the trace in the vector representation of $SO(N)$. The dots in (\ref{EFTpartonZ2}) denote non-topological gauge invariant terms such as the Yang-Mills action $\propto\tr(f\wedge *f)$ where $f=da+a\wedge a$ is the non-Abelian field-strength 2-form.

We now investigate the different phases of the original spin system that can be reached by varying $C$. When $C=0$, the Chern-Simons term is absent and the Yang-Mills term dominates the action. At least when regularized on a lattice, as is the case here, pure $SO(N)$ gauge theory in $2+1$ dimensions and without a Chern-Simons term is believed to be confining at zero temperature~\cite{bursa2013,lau2017}. The confining theory is massive, thus $C=0$ corresponds to a gapped phase of the original Ising spin system. When $C=1$, the effective Lagrangian contains an $SO(N)_1$ Chern-Simons term,
\begin{align}
\mathcal{L}=\text{CS}_{SO(N)_1}[a]+\ldots
\end{align}
This theory is also massive, but the Chern-Simons term leads to deconfinement at zero temperature. Below the mass gap and in the long-wavelength limit, the system is described by a pure topological $SO(N)_1$ theory. Similarly to $U(1)_1$ Chern-Simons theory, this is an invertible topological quantum field theory with a unique ground state on all closed manifolds~\cite{seiberg2016}. Note that $SO(N)_k$ Chern-Simons theory is a consistent theory of microscopic bosons, even if $k$ is odd, provided that only fermionic matter fields couple to the $SO(N)$ gauge field~\cite{Maxduality}; both conditions are satisfied here. Thus for $C=1$ the original Ising spin system forms a gapped paramagnet without topological order. For $|C|>1$, one obtains a deconfined phase described by $SO(N)_C$ Chern-Simons theory. Such theories are not invertible, and thus the Ising spin system is in a phase with intrinsic topological order, i.e., a gapped spin liquid.

Another way to see that the $C=1$ phase corresponds to a gapped paramagnet without topological order is by looking at the edge degrees of freedom. At the mean-field level, the $C=1$ phase features $N$ free chiral Majorana modes on the boundary, which is a noninteracting conformal field theory (CFT) with chiral $\mathfrak{so}(N)_1$ current algebra and chiral central charge $c_-=N/2$~\cite{witten1984}. However, when projecting to the physical Hilbert space the $SO(N)$ symmetry is gauged, which gives a trivial coset CFT on the edge with vanishing chiral central charge.

More generally, for $C>1$ one obtains $NC$ free chiral Majorana modes on the boundary at the parton mean-field level, corresponding to a chiral $\mathfrak{so}(NC)_1$ current algebra (we assume $C$ is positive without loss of generality, as a sign reversal of $C$ simply corresponds to a reversal of chirality). In Appendix~\ref{app:SONCFT}, we show that this current algebra obeys the following conformal embedding:
\begin{align}
\mathfrak{so}(N)_{C}\otimes\mathfrak{so}(C)_{N}\subseteq\mathfrak{so}(NC)_1.
\end{align}
Gauging the $SO(N)$ symmetry leaves a chiral $\mathfrak{so}(C)_{N}$ current algebra~\cite{antoniadis1986}, with chiral central charge
\begin{align}
c_-=\frac{NC(C-1)}{2(N+C-2)}.
\end{align}
Thus the $C>1$ phases are chiral topological phases with protected edge modes described by the chiral $\mathfrak{so}(C)_{N}$ Wess-Zumino-Witten (WZW) CFT. Since the microscopic system consists of interacting Ising spins, these are chiral spin-liquid phases with broken time-reversal symmetry but unbroken $\mathbb{Z}_2$ spin-flip symmetry. For $C=2$, the edge theory is a $\mathfrak{so}(2)_{N}=u(1)_{N}$ CFT with $c_-=1$. This is consistent with the fact that the bulk $SO(N)_2$ Chern-Simons theory is equivalent to $SO(2)_{N}=U(1)_{N}$ by level-rank duality~\cite{naculich1990}. Thus the $C=2$ phase is an Abelian topological phase, with the topological order of a $\nu=1/N$ bosonic fractional quantum Hall state~\cite{kalmeyer1987,wen1989,wen1989b}. Such a chiral spin liquid has anyonic spinon excitations with statistical angle $\theta=\pi/N$. Likewise, for $C>2$ but $N=2$, the edge theory is a $\mathfrak{so}(C)_2$ CFT which is equivalent to $\mathfrak{su}(C)_1$~\cite{Aharony2017}, with chiral central charge $c_-=C-1$. This can be understood intuitively since $N=2$ corresponds to two Majorana fermions, which is equivalent to a single Dirac fermion. The mean-field state is a Chern insulator of this Dirac fermion with Chern number $C$, and gauging the internal $SO(2)$ symmetry corresponds to gauging the $U(1)$ symmetry of the Dirac fermion. For $N>2$ and $C>2$, the bulk $SO(N)_C$ Chern-Simons theory is level-rank dual to $SO(C)_{N}$, consistent with the $\mathfrak{so}(C)_{N}$ edge CFT. In the following, we will be interested exclusively in the case $N>2$, for reasons to be clarified shortly.

\subsection{Phase transitions and $SO(N)$ dualities}
\label{sec:SONQPT}

\begin{figure}[t]
\includegraphics[width=0.6\columnwidth]{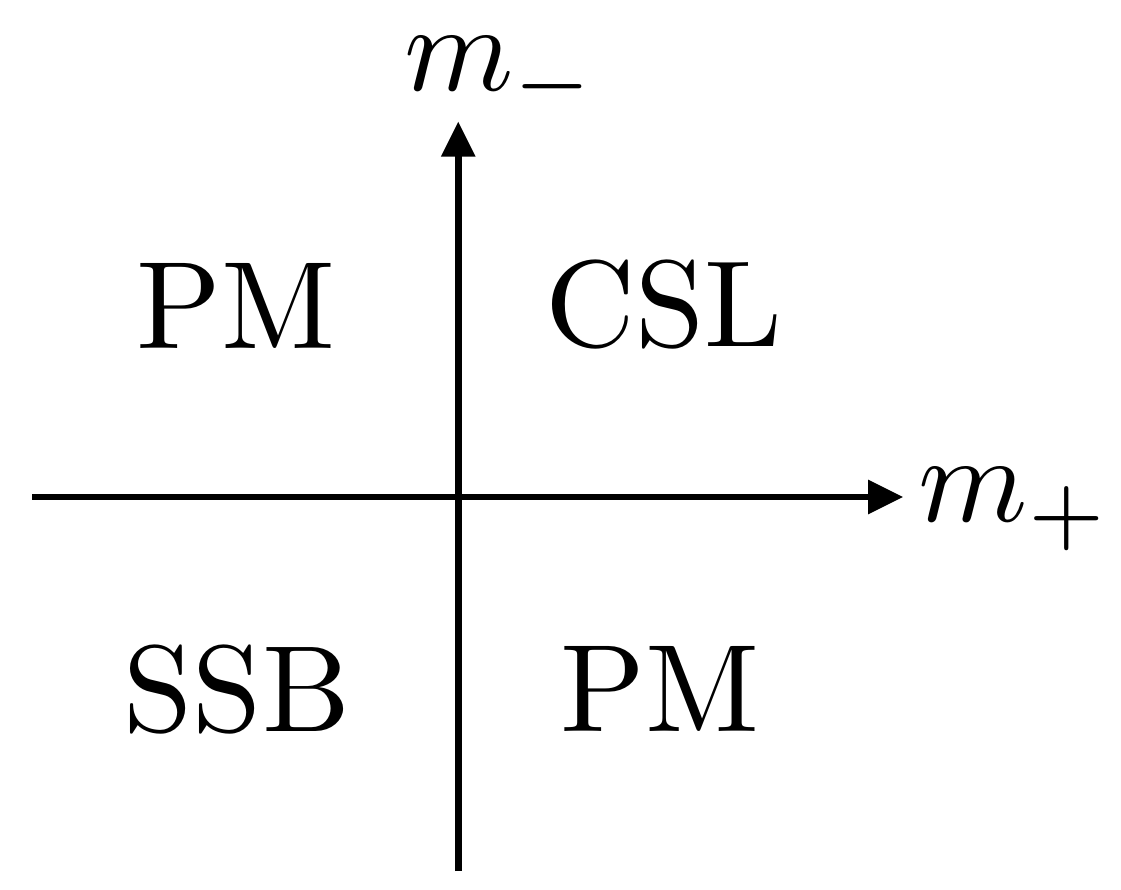}
\caption{Phase diagram for Ising spins as a function of the two tuning parameters $m_+$, $m_-$. PM: trivial paramagnet; SSB: ordered phase with $\mathbb{Z}_2$ spontaneous symmetry breaking; CSL: chiral spin liquid.}
\label{fig:phase2}
\end{figure}

For simplicity, and to make an analogy with Sec.~\ref{sec:U1critical}, we focus on transitions between the three phases with $C=0,1,2$. At the mean-field level, those are topological transitions that proceed by linear (Majorana) crossings of the Bogoliubov-de Gennes bands of the topological superconductor. We consider a parton bandstructure such that a low-lying band with Chern number one remains filled across the transition, and there is a linear crossing at zero energy of two other bands (for examples of multiband Majorana models with topological transitions, see Refs.~\cite{yang2019,mirmojarabian2020,farjami2020}). Since the total Chern number must be integer, and a single massive two-component Majorana fermion carries a partial Chern number of $\half\sgn m$, the low-energy bandstructure in the vicinity of the transitions can be described by two slow, two-component Majorana fields $\psi_+,\psi_-$ in the vector representation of $SO(N)$ with masses $m_+,m_-$ respectively. In Appendix~\ref{sec:pgt}, we give an example of Majorana hopping model that produces such a low-energy bandstructure with tunable masses $m_\pm$. Considering gauge fluctuations, the theory interpolating between all three phases is
\begin{align}\label{LcritSON}
\mathcal{L}=\text{CS}_{SO(N)_1}[a]+\frac{1}{4}\sum_{k=\pm}\psi_k^{\intercal}\c{C}(i\slashed{D}-m_k)\psi_k,
\end{align}
where $\slashed{D}=\gamma^\mu(\partial_\mu-ia_\mu)$ is a gauge-covariant derivative involving the internal $SO(N)$ gauge field $a_\mu$, $\c{C}$ is a charge-conjugation matrix, and the level-1 Chern-Simons term comes from the response of the low-lying band. The phases described in Sec.~\ref{sec:partonZ2} for $C=0,1,2$ are obtained when $m_+$ and $m_-$ are nonzero. When $m_\pm>0$, the two Majorana fermions $\psi_\pm$ can be integrated out, yielding an $SO(N)_2$ Chern-Simons term. As seen above, this is a chiral spin liquid with the topological order of the $\nu=1/N$ bosonic fractional quantum Hall state. When both $m_\pm<0$, the Chern-Simons level vanishes and one obtains a pure Yang-Mills theory which confines. As we discuss below, this is a phase with magnetic long-range order. Finally, when the masses are of opposite sign, one has an $SO(N)_1$ Chern-Simons term which corresponds to a topologically trivial, gapped paramagnet. A schematic phase diagram is given in Fig.~\ref{fig:phase2}.

The critical theories for the transitions in Fig.~\ref{fig:phase2} can all be written in the following general form:
\begin{align}\label{Lcrit}
\c{L}_{N_f,\nu}=\text{CS}_{SO(N)_\nu}[a]+\frac{1}{4}\sum_{j=1}^{N_f}\psi_j^{\intercal}\c{C}i\slashed{D}\psi_j,
\end{align}
where $\nu$ is the Chern-Simons level and $N_f$ is the number of Majorana fields that become massless at the transition. We first consider transitions involving the chiral spin liquid. The transition between the chiral spin liquid and the paramagnet is tuned by $m_+$ ($m_-$) crossing zero at constant $m_->0$ ($m_+>0$); integrating out the massive fermion, we find Eq.~(\ref{Lcrit}) with $N_f=1$ and $\nu=3/2$. A direct transition from the chiral spin liquid to the ordered phase is obtained by tuning the mass of both fermions through zero, and is described by Eq.~(\ref{Lcrit}) with $N_f=2$ and $\nu=1$. Such a transition can be protected by microscopic symmetries enforcing $m_+\!=\!m_-$; for example, this is achieved by requiring inversion symmetry in the honeycomb lattice model presented in Appendix~\ref{sec:pgt}. Assuming both theories flow to a {\it bona fide} critical fixed point in the infrared, they correspond to novel universality classes.

We next argue that the confining phase at $m_\pm<0$ (i.e., $C=0$) has a spontaneously broken $\mathbb{Z}_2$ symmetry: that is, it possesses Ising-type magnetic long-range order. The transition between the $C=1$ (paramagnetic) and $C=0$ phases is obtained by tuning $m_+$ ($m_-$) through zero at constant $m_-<0$ ($m_+<0$). It is thus described by $\c{L}_{1,1/2}$ in Eq.~(\ref{Lcrit}), i.e., a single flavor of Majorana fermions in the vector representation coupled to an $SO(N)_{1/2}$ theory. It was conjectured in Ref.~\cite{Aharony2017} that an $SO(k)_{-M+\frac{N_f}{2}}$ theory coupled to $N_f$ flavors of vector Majorana fermions is dual to an $SO(M)_k$ theory coupled to $N_f$ flavors of real scalars $\boldsymbol{\phi}$ in the vector representation with $(\boldsymbol{\phi}^2)^2$ interactions, i.e., an $SO(M)_k$ theory coupled to the bosonic $O(M)$ vector model at its Wilson-Fisher fixed point. For $N_f=M=1$, this stipulates that an $SO(k)_{-1/2}$ theory coupled to a single Majorana fermion is dual to a single real scalar $\phi$ at its Wilson-Fisher fixed point, since $SO(1)$ on the scalar side is trivial:
\begin{align}\label{SOduality}
\text{Majorana}+SO(k)_{-1/2}\longleftrightarrow\text{real scalar},
\end{align}
which can be viewed as a fermionization of the 3D Ising transition. Note that this duality is conjectured to hold only for $k>2$, which corresponds in our case to $N>2$, i.e., a minimum of four partons in the decomposition (\ref{partonZ2}). Equation~(\ref{SOduality}) can also be viewed as the Majorana counterpart of the $U(1)$ boson-fermion duality~\cite{ChenFisherWu,SEIBERGduality,DualityKarchTong,murugan2016,kachru2017,karch2017}
\begin{align}
\text{Dirac}+U(1)_{-1/2}\longleftrightarrow\text{complex scalar},
\end{align}
whose time-reversed version we have used in Eqs.~(\ref{U1:fermion1}-\ref{U1:boson1}), or as the ``inverse'' of the Majorana bosonization duality~\cite{Aharony2017,Maxduality}
\begin{align}
\text{Majorana}\longleftrightarrow\text{$O(M)$ scalar}+SO(M)_1,
\end{align}
obtained by considering $N_f=k=1$ (and $M\geq 3$). Setting $k=N$ and performing time reversal to reverse the sign of the Chern-Simons level, Eq.~(\ref{SOduality}) is precisely the critical theory $\c{L}_{1,1/2}$. Thus one obtains the dual critical theory
\begin{align}\label{Z2dual}
\mathcal{L}_\text{dual}=\frac{1}{2}(\partial_\mu\phi)^2-\frac{\lambda}{4!}\phi^4.
\end{align}
The dual Lagrangian (\ref{Z2dual}) has a global $\mathbb{Z}_2$ symmetry $\phi\rightarrow-\phi$ and two massive phases, the unbroken phase $\langle\phi\rangle=0$ and the phase $\langle\phi\rangle\neq 0$ with spontaneously broken $\mathbb{Z}_2$ symmetry. This transition can be tuned by adding a scalar mass term $\propto\phi^2$. Since $m_->0$ in the original theory corresponds to a trivial paramagnet with no broken symmetries, this must correspond to the unbroken phase in the dual theory, i.e., a positive scalar mass term. Therefore, $m_-<0$ must correspond to a negative scalar mass term in the dual theory, i.e., the phase with spontaneously broken $\mathbb{Z}_2$ symmetry.

It remains to be shown that the $\mathbb{Z}_2$ symmetry under $\phi\rightarrow-\phi$ of the dual theory (\ref{Z2dual}) is nothing but the original global $\mathbb{Z}_2$ symmetry of the spin system. As in Sec.~\ref{sec:U1instantons}, this is properly achieved by the consideration of nonperturbative instanton effects, to which we now turn.

\section{Instantons, Majorana zero modes, and Ising symmetry}
\label{sec:Z2instantons}

In this section, we wish to understand how the low-energy $SO(N)$ gauge theory accounts for the broken Ising symmetry in the phase with parton Chern number $C\!=\!0$. In our description of the $C\!=\!1\!\rightarrow\! C\!=\!0$ transition in Sec.~\ref{sec:SONQPT}, the $C\!=\!0$ Chern number of the occupied bands resulted from the cancellation between a spectator $C\!=\!1$ band and a nearly-critical $C\!=\!-1$ band (fermions $\psi_\pm$ with masses $m_\pm<0$). To understand the physics deep in the $C=0$ phase, it is simpler to work with a topologically equivalent theory, that is, a single $C\!=\!0$ band producing two continuum Majorana fields $\Psi_+$ and $\Psi_-$ with opposite masses $\pm m$ (see again Appendix~\ref{sec:pgt} for a lattice representative), specified by the Euclidean Lagrangian
\begin{align}\label{eq:Lnaive}
\mathcal{L}&=\frac{1}{4}\Psi_+^{\intercal}\c{C}(\slashed{\partial}-i\slashed{a}+m)\Psi_+
+\frac{1}{4}\Psi_-^{\intercal}\c{C}(\slashed{\partial}-i\slashed{a}-m)\Psi_-\nn\\
&\phantom{=}+\frac{1}{2g^2}\tr f^2,
\end{align}
where we have explicitly included a Yang-Mills term, in the absence of a net Chern-Simons level. As we discuss below, the breaking of Ising symmetry ultimately results from Euclidean ZMs supported by those massive Majorana fields in the presence of instantons in the $SO(N)$ gauge field, which have heretofore been ignored.

The rest of this section is structured as follows. We first discuss the monopole operator of $SO(N)$ gauge theory in (2+1)D, which is charged under a topological (magnetic) $\mathbb{Z}_2^{\c{M}}$ symmetry and becomes a $\mathbb{Z}_2$ instanton in Euclidean spacetime (Sec.~\ref{subsec:GNO}). Together with the global $\mathbb{Z}_2$ symmetry action $W$ on the Majorana partons $\chi$ (recall Sec.~\ref{sec:partonZ2}), the Lagrangian (\ref{eq:Lnaive}) has a spurious global $\mathbb{Z}_2\times\mathbb{Z}_2^{\c{M}}$ symmetry absent instanton effects. We show that in the presence of massive fermions coupled to the $SO(N)$ gauge field, these instantons are dressed by Euclidean Majorana ZMs bound to the instanton (Sec.~\ref{subsec:MZMs}). We then show that semiclassical resummation of the $\mathbb{Z}_2$ instanton gas produces an interaction term among the fermionic partons, the 't~Hooft vertex, that explicitly breaks this spurious $\mathbb{Z}_2\times\mathbb{Z}_2^{\c{M}}$ symmetry down to its diagonal $\mathbb{Z}_2$ subgroup (Sec.~\ref{subsec:tHooftZ2}). This intertwinement ensures that if the $\mathbb{Z}_2^{\c{M}}$ magnetic symmetry is spontaneously broken, as is typical in a confined phase~\cite{komargodski2018,benini2018}, the microscopic Ising symmetry $\tau^z\rightarrow-\tau^z$ is broken also. The $C=0$ confined phase is thus naturally identified as a broken-symmetry phase, in agreement with the duality arguments of Sec.~\ref{sec:SONQPT}.

\subsection{$\mathbb{Z}_{2}$ instantons in $SO(N)$ gauge theory}
\label{subsec:GNO}

In the absence of instantons, $SO(N)$ gauge theory in (2+1)D with $N\!>\!2$ possesses a magnetic $\mathbb{Z}_{2}^{\mathcal{M}}$ symmetry~\cite{aharony2013,Aharony2017,benini2017,cordova2018,komargodski2018,benini2018}, analogous to the topological $U(1)_{\mathrm{top}}$ symmetry of $U(1)$ gauge theory in (2+1)D. Unlike the latter, $\mathbb{Z}_{2}^{\mathcal{M}}$ lacks a conserved current, being a discrete symmetry. The similarity between the two is that both result in the existence of disorder operators
that create topological excitations, in this case monopole-instantons.

The operator charged under the $\mathbb{Z}_{2}^{\mathcal{M}}$ symmetry
is a local monopole operator $\mathcal{M}(x)$, whose charge is defined
by a nontrivial second Stiefel-Whitney class $w_{2}\in H^2(\Sigma,\mathbb{Z}_2)$
on a closed surface $\Sigma$ surrounding the operator insertion~\cite{[{For an introduction to Stiefel-Whitney classes in a condensed matter context, see, e.g., }]senthil2019,*ahn2018,*ahn2019}:
\begin{equation}
\int_\Sigma w_2\in\mathbb{Z}_2.
\end{equation}
A nontrivial Stiefel-Whitney class is an obstruction to lifting an $SO(N)$ bundle to its double cover, a $\mathrm{Spin}(N)$ bundle. For $\Sigma$ a sphere, the $\mathbb{Z}_2$ monopole charge corresponds to the nontrivial homotopy group $\pi_1(SO(N))\!\cong\!\mathbb{Z}_2$, $N\!>\!2$. It measures the winding number of an $SO(N)$ gauge transformation that relates the gauge fields $a_\mu^\text{I}$ and $a_\mu^\text{II}$ on the overlap of two coordinate charts I \& II on $\Sigma$. As in the $U(1)$ theory~\cite{polyakov1975,polyakov1977,polyakov1987}, the $\mathbb{Z}_2$ monopoles are regarded here as instantons in 3D Euclidean spacetime.

An explicit semiclassical representative~\cite{goddard1977} is obtained by placing a Dirac monopole in a specific $SO(2)$ subgroup of $SO(N)$, so that one may use the Wu-Yang connection 1-form~\citep{wu1976}:
\begin{equation}
\c{A}_{n}=\frac{n}{2}(1-\cos\theta)t_{c}\dd{\phi},\quad n\!\in\!\mathbb{Z},\label{eq:wuyang}
\end{equation}
where $n$ is the Dirac monopole charge, $\phi$ is the azimuthal coordinate on a sphere $\Sigma$ surrounding the
monopole, and $t_{c}\!\in\!\mathfrak{so}(N)$ generates a subgroup $SO(2)\!\subset\!SO(N)$.
By means of gauge rotations, any $t_{c}\!\in\!\mathfrak{so}(N)$ can
be rotated to a Cartan generator, which shall be taken as $t_{(12)}$,
the generator of rotations in the $(\chi^{1},\chi^{2})$ plane of the Majorana
vector $(\chi^{1},...,\chi^{N})$. Such a semiclassical characterization
explicitly breaks the $SO(N)$ gauge symmetry down to an $[O(2)\!\times\!O(N\!-\!2)]/\mathbb{Z}_{2}$
subgroup, where quotienting by $\mathbb{Z}_{2}$ restricts the determinant
of the overall transformation to be positive. Note that a suitable $SO(N)$ gauge transformation can invert the $SO(2)$ monopole charge $n$; thus an $n\!=\!2\!=\!1\!+\!1$ monopole is topologically equivalent to an $n\!=\!0\!=\!1\!+\!(-1)$ monopole, and the unique $\mathbb{Z}_2$-nontrivial monopole is given by $n\!=\!1$ (or $n\!=\!-1$). Although it does not preserve full $SO(N)$ gauge invariance, this semiclassical description will allow us to perform an explicit instanton-gas calculation analogous to that in Ref.~\citep{shankar2021} for $U(1)$ gauge theory. 

Incarnating the $SO(N)$ monopole as a Dirac monopole in a specific $SO(2)$
subgroup can be regarded as a partial gauge choice. Formally notating
this gauge condition as $G(a)\!=\!0$, the Euclidean path integral for the theory
can be expressed using the Faddeev-Popov (FP) method as 
\begin{equation}
\label{eq:gfZ}
Z=\int_{SO(N)}\mathscr{D}R\int\mathscr{D}a\mathscr{D}\Psi_{\pm}\,\Delta_{G}[a]\delta[G(a^{(R)})]e^{-S[a,\Psi_{\pm}]},
\end{equation}
where $a^{(R)}$ is related to $a$ by a gauge transformation $R(x)$, and we denote $\mathscr{D}\Psi_\pm\equiv\mathscr{D}\Psi_+\mathscr{D}\Psi_-$ for simplicity. The FP determinant $\Delta_{G}[a]$ and the delta functional $\delta[G(a^{(R)})]$ can be written respectively as ghost and gauge-fixing
terms in the Lagrangian. The essential idea expressed by Eq.~\eqref{eq:gfZ} is that one
can perform a path integral calculation in a fixed gauge (gauge slice), and then
integrate the result over its gauge orbit $(\int\mathscr{D}R)$ to recover
gauge invariance. This will allow us to use a $U(1)$ monopole operator that creates $2\pi$ flux in the  $SO(2)$ subgroup, for which an explicit expression is known. In the following, we will omit explicit integration over the gauge orbit but invoke heuristic arguments to (partially) restore $SO(N)$ gauge invariance at the end of the calculation, focusing on its physical consequences.

As stated earlier, incarnating the $SO(N)$ monopole as a Dirac monopole in an $SO(2)$ subgroup only partially fixes the gauge. The group $S[O(2)\!\times\!O(N\!-\!2)]\!\equiv\![O(2)\!\times\!O(N\!-\!2)]/\mathbb{Z}_2$ of global gauge rotations is a stabilizer for such a monopole configuration. Na{\"i}vely, the existence of a nontrivial stabilizer leads to ghost ZMs, which are ZMs in the FP determinant $\Delta_G[a]\!=\!\abs{\det\delta G/\delta\omega}$, where $\omega\!\in\!\mf{so}(N)$ generates the rotation $R\!=\!\exp(-\omega)$, and $G(a)$ is the aforementioned gauge function partially determining the gauge. By employing the background-field gauge method~\cite{thooft1976a}, we show in Appendix~\ref{app:instantons} that such ZMs can be removed from the FP determinant at the cost of introducing an overall factor of $\mathrm{vol}(SO(N)/S[O(2)\!\times\!O(N\!-\!2)])^\c{N}$ in the $\c{N}$-instanton contribution to the partition function. This is interpreted as the volume of the moduli space of ``gauge collective coordinates''---global gauge rotations that act to move the Dirac monopole to distinct $SO(2)$ subgroups of $SO(N)$, thus yielding other viable instanton solutions \cite{tong2005}.  Besides this, the ghost and gauge-fixing terms will simply spectate in the instanton gas calculation to follow, and will thus henceforth be suppressed to reduce clutter.

The coupling of fermions to finite-action fluctuations (``gluons'') around the instanton solution $\mathcal{A}_n$ is ignored in the semiclassical approximation~\cite{thooft1976,thooft1976a,thooft1986}. In our choice of gauge, the contribution to the path integral from $\mathbb{Z}_2$ instantons can be separated and written as~\citep{shankar2021}
\begin{align}
Z= & \int\mathscr{D}a\,e^{-\frac{1}{2g^{2}}\int d^3x\,\tr f^{2}}\sum_{\c{N}=0}^{\infty}\frac{\lambda^\c{N}}{\c{N}!}\nonumber \\
 & \times\prod_{k\!=\!1}^{\c{N}}\int\dd^{3}{z_{k}}\sum_{n_{k}\!=\!\pm1}\mathcal{M}_{n_{k}}^{(12)}(z_{k})\int\mathscr{D}\Psi_{\pm}e^{-S_{F}[\c{A}_{n_{k}},\Psi_\pm]},\label{eq:fullZ}
\end{align}
where $\mathcal{M}_{n}^{(12)}$ is a monopole operator that creates
a Dirac monopole of charge $n$ in the $SO(2)$ subgroup generated by $t_{(12)}\!\in\!\mf{so}(N)$,
and $\lambda$ is the fugacity of an $n=\pm 1$ instanton. In this fixed gauge, the monopole operator has an explicit representation $\exp(in\gamma_{(12)})$ in terms of the dual photon $\gamma_{(12)}$~\cite{polyakov1975,polyakov1977,polyakov1987,borokhov2002}. Unlike in $U(1)$ gauge theory, $\gamma_{(12)}$ is no longer
gauge invariant, as evident from the presence of the gauge-dependent
susbcript that selects an $SO(2)$ subgroup in $SO(N)$. The fermion action
in the instanton background is
\begin{multline}
S_{F}[\c{A}_{n}]=\frac{1}{4}\int\dd^{3}{x}[\Psi_{+}^{\intercal}\c{C}(\slashed{\partial}-i\slashed{\c{A}}_{n}+m)\Psi_{+}\\
+\Psi_{-}^{\intercal}\c{C}(\slashed{\partial}-i\slashed{\c{A}}_{n}-m)\Psi_{-}].\label{eq:SF}
\end{multline}

The inclusion of charge $n\!=\!\pm1$ monopoles in Eq.~\eqref{eq:fullZ} deserves further explanation in light of the $\mathbb{Z}_2$ nature of the topological charge of $SO(N)$ monopoles. A simple explanation is that in our fixed choice of gauge, these two charges are distinct configurations and must both be accounted for. Alternatively, one can resort to a stability argument. In one higher spacetime dimension (4D), $\mathbb{Z}_2$ monopoles feature as solitons in the gauge theory. A monopole of topological charge $0$ or $1$ can be ``dynamically'' represented as Dirac monopoles of charges $\{0,\pm 2,\pm 4,...\}$ or $\{\pm 1, \pm 3,...\}$ respectively, in some $SO(2)\!\subset\!SO(N)$. A stability analysis \cite{brandt1979, coleman1983} indicates that the uniquely stable dynamical configurations in the two topological classes are the charge $0$ and $\pm 1$ Dirac monopoles. This result can be used to determine the stable dynamical configuration of multimonopole solutions.  At distances large compared to their separation, two monopoles with Dirac charges $+1$ look like a single monopole of Dirac charge $2$, which is unstable to the charge $0$ configuration. This implies the instability of the $1+1$ to the $1-1$ configuration, which proceeds by the emission of gluon radiation. While such a stability analysis has been applied to monopoles as soliton excitations in 4D, we expect that a similar result holds for monopole-instantons, with the charge $0$ and $\pm 1$ configurations being the most probable instanton events. Since the instanton gas calculation is performed with the Dirac charge instead of the topological $\mathbb{Z}_2$ charge, one must account for both $\pm 1$ charges in the instanton sum \eqref{eq:fullZ}, as both are expected to be equally probable. Finally, we find that inclusion of both $\pm 1$ charges is required to maintain reflection positivity of the instanton-induced 't~Hooft vertex, as discussed in Sec.~\ref{subsec:tHooftZ2}.

\subsection{Euclidean Majorana zero modes}
\label{subsec:MZMs}

A natural question to ask now is if there are (Euclidean) Majorana
ZMs, associated with zero-eigenvalue modes of the Euclidean
Dirac operators
\begin{equation}
\mathcal{D}_{\pm}\equiv\slashed{\partial}-i\slashed{\c{A}}_{n}\pm m,
\end{equation}
appearing in the fermion action $S_{F}$, for Dirac instantons of charges
$n\!=\!\pm1$. In the absence of a Callias index theorem for Dirac instantons in Abelian $SO(2)\cong U(1)$ gauge theory~\cite{marston1990,unsal2008}, we resort to an explicit solution of the Dirac equation.

As stated previously, we assume a gauge in which the instanton
incarnates as a Dirac monopole in the $SO(2)$ subgroup generated
by
\begin{equation}\label{t12}
t_{(12)}=\left(
\begin{array}{cc|c}
0 & -i & \bm{0}\\
i & 0 & \bm{0}\\
\hline
\bm{0} & \bm{0} & \bm{0}
\end{array}
\right)\in\mathfrak{so}(N),
\end{equation}
where the upper-left block corresponds to the $(12)$ subspace, and $\bm{0}$ denotes a zero matrix of the appropriate size, involving the remaining $N\!-\!2$ directions in color space. Writing $\c{A}_n=a_n t_{(12)}$, and working in the Cartan (diagonal) basis of $\mathfrak{so}(N)$, the Dirac operators can be written as
\begin{equation}
\mathcal{D}_{\pm}=U
\left(
\begin{array}{cc|c}
\slashed{\partial}\!-\!i\slashed{a}_{n}\!\pm\!m & 0 & \bm{0}\\
0 & \slashed{\partial}\!+\!i\slashed{a}_{n}\!\pm\!m & \bm{0}\\
\hline
\bm{0} & \bm{0} & (\slashed{\partial}\!\pm\!m)\bm{1}
\end{array}
\right)
U^{\dagger},\label{eq:Dp}
\end{equation}
where $U$ is the unitary matrix that diagonalizes $t_{(12)}$. Borrowing
results from Ref.~\citep{shankar2021}, in an $n\!=\!1$
instanton background, $(\slashed{\partial}\!-\!i\slashed{a}_{+}\!+\!m)$
and $(\slashed{\partial}\!+\!i\slashed{a}_{+}\!-\!m)$ have the respective
normalizable ZMs:
\begin{align}
\psi_{0}^{+} & =\frac{\sqrt{2m}}{r}e^{-mr}\mathcal{Y}_{1/2,0,0}^{0}(\theta,\phi),\nonumber \\
\psi_{0}^{-} & =\frac{\sqrt{2m}}{r}e^{-mr}\mathcal{Y}_{-1/2,0,0}^{0}(\theta,\phi),\label{eq:u1zms}
\end{align}
where $\mathcal{Y}_{n/2,j,m_{j}}^{j\pm1/2}(\theta,\phi)$ are monopole
spinor harmonics. The rest of the operators on the diagonal of Eq.~(\ref{eq:Dp})
do not have any normalizable ZMs. Therefore, the normalizable ZMs
of $\mathcal{D}_{\pm}$ in an $n\!=\!1$ instanton background are,
respectively,
\begin{align}
u_{0} & =U(\psi_{0}^{+},0,...,0)^{\intercal}=\frac{\psi_{0}^{+}}{\sqrt{2}}(-i,1,0,...,0)^{\intercal},\nonumber \\
v_{0} & =e^{i\alpha}U(0,\psi_{0}^{-},0,...,0)^{\intercal}=\frac{e^{i\alpha}\psi_{0}^{-}}{\sqrt{2}}(i,1,0,...,0)^{\intercal}.\label{eq:pzms}
\end{align}
Any phase multiplying a ZM still produces a normalized ZM, and this
apparent freedom has been encoded in an arbitrary relative phase $e^{i\alpha}$. 

Similarly, in an $n\!=\!-1$ instanton background, the operators $\mathcal{D}_{\pm}$
have the ZMs 
\begin{align}
\tilde{u}_{0} & =U(0,\psi_{0}^{+},0,...,0)^{\intercal}=\frac{\psi_{0}^{+}}{\sqrt{2}}(i,1,0,...,0)^{\intercal},\nonumber \\
\tilde{v}_{0} & =e^{-i\beta}U(\psi_{0}^{-},0,...,0)^{\intercal}=\frac{e^{-i\beta}\psi_{0}^{-}}{\sqrt{2}}(-i,1,0,...,0)^{\intercal}.\label{eq:mzms}
\end{align}
The $\mathbb{Z}_2$ topological equivalence of the $n=\pm 1$ field configurations under the full $SO(N)$ gauge structure will be discussed later, as well as constraints on the relative phases $\alpha$ and $\beta$.

\subsection{The 't Hooft vertex and Ising symmetry}
\label{subsec:tHooftZ2}

In this subsection, we show that the Euclidean Majorana ZMs found in
the previous subsection induce symmetry-breaking interactions in the $SO(N)$ gauge theory. (As mentioned previously, Appendix~\ref{app:instantons} shows that FP ghosts do not give rise to physical ZMs bound to instantons.) Specifically, these
ZMs imply that instanton events are correlated with creation (or annihilation)
of Majorana fermions. Resumming the instanton gas results in a new
fermion interaction, called the 't~Hooft vertex, which reduces
the symmetry of the initial Lagrangian (\ref{eq:Lnaive}).

We now sketch a derivation of this 't~Hooft vertex; more details regarding
the structure of such a calculation can be found in Ref.~\citep{shankar2021}.
In the background of an $n\!=\!1$ instanton fixed at location $z_{+}$,
the measure of the fermion part of the path integral (\ref{eq:fullZ})
can be defined by means of the mode expansions 
\begin{align}
\Psi_{+}(x) & =u_{0}(x-z_{+})\eta_{0}+\sideset{}{'}\sum_{i}u_{i}(x-z_{+})\eta_{i},\nonumber \\
\Psi_{-}(x) & =v_{0}(x-z_{+})\chi_{0}+\sideset{}{'}\sum_{i}v_{i}(x-z_{+})\chi_{i},\label{eq:pmodexp}
\end{align}
where $\eta_{i},\chi_{i}$ are single-component Grassmann variables,
$u_{0}$ and $v_{0}$ are the respective ZMs of $\mathcal{D}_{+}$
and $\mathcal{D}_{-}$ in an $n\!=\!1$ instanton background, and
the primed sums denote non-ZM contributions. The functions that form
the non-ZM contributions can be taken to be eigenfunctions of a self-adjoint
extension of the Hermitian operator $\mathcal{D}_{\pm}^{\dagger}\mathcal{D}_{\pm}$,
whose non-ZM eigenfunctions occur in pairs that share the same eigenvalue~\citep{affleck1982,stone2020}.

Defining the fermion functional measure as 
\begin{equation}
\mathscr{D}\Psi_{\pm}=\mathscr{D}\Psi_{+}\mathscr{D}\Psi_{-}=\dd{\eta_{0}}\dd{\chi_{0}}\sideset{}{'}\prod_{i}\dd{\eta_{i}}\dd{\chi_{i}},
\end{equation}
we observe that the mode expansions (\ref{eq:pmodexp}) diagonalize
the fermion action $S_{F}$, but the ZMs do not appear in the diagonalized
action, by virtue of being annihilated by the Dirac operators $\mathcal{D}_{\pm}$.
This causes the integral over the ZMs $(\eta_{0},\chi_{0})$ to vanish,
killing the path integral. As in Ref.~\citep{shankar2021}, instantons
do not contribute to the partition function itself, but to correlation
functions that can ``soak up'' the ZMs, such as $\ev{\Psi_{+}^\alpha\Psi_{-}^\beta}$.
Such correlation functions generically violate the apparent $\mathbb{Z}_{2}\!\times\!\mathbb{Z}_{2}^{\c{M}}$
symmetry of the naive continuum Lagrangian (\ref{eq:Lnaive}).
To find the true effective theory, we add a weak symmetry-breaking
source to the action and re-evaluate the fermion part of the path
integral to linear order in the source $J$. Explicitly, using the
mode expansions (\ref{eq:pmodexp}),
\begin{align}
Z_{F}[A_{+},J]= & \int\mathscr{D}\Psi_{\pm}e^{-S_{F}[A_{+}]-\!\int\!\dd^{3}{(x,y)}\Psi_{+}^{\intercal}(x)J(x,y)\Psi_{-}(y)},\nonumber \\
= & \int\dd^{3}{(x,y)}u_{0}^{\intercal}(x\!-\!z_{+})J(x,y)v_{0}(y\!-\!z_{+})K,\label{eq:ZFJA}
\end{align}
where $K$ denotes the path integral over non-ZMs, and $\dd^{3}{(x,y)}\!=\!\dd^{3}{x}\dd^{3}{y}$.
Strictly, nonlocal expressions like the source term require an insertion
of Wilson lines to maintain gauge invariance, but we do not write
these explicitly, as the final form of the 't~Hooft vertex will turn
out to be local. This is also consistent with our neglect of fermion-gluon
interactions at this stage.

Demanding an effective theory that reproduces this path integral
amounts to ``integrating out'' the instantons in the full partition
function (\ref{eq:fullZ}). As an ansatz for the resulting
partition function, consider 
\begin{multline}
I_{+}[J]=\int\mathscr{D}\Psi_{\pm}e^{-S_{F}-\!\int\!\dd^{3}{(x,y)}\Psi_{+}^{\intercal}(x)J(x,y)\Psi_{-}(y)}\\
\times\int\dd^{3}{(x_{1},x_{2})}\rho\Psi_{-}^{\intercal}(x_{2})\omega_{2}\omega_{1}^{\intercal}\Psi_{+}(x_{1}),
\end{multline}
 where $\rho$ and $\omega_{1,2}$ are fixed by requiring equality with
$Z_{F}[A_{+},J]$ in Eq.~(\ref{eq:ZFJA}). Note that the action
$S_{F}$ written without source arguments is the free Majorana action.
This leads to 
\begin{align}
\rho & =K,\nonumber \\
\omega_{1} & =\frac{1}{2}\c{C}(\slashed{\partial}+m)u_{0},\nonumber \\
\omega_{2} & =\frac{1}{2}\c{C}(\slashed{\partial}-m)v_{0}.
\end{align}

The above calculations can be repeated for an $n\!=\!-1$ instanton
background using the mode expansions 
\begin{align}
\Psi_{+}(x) & =\tilde{u}_{0}(x-z_{-})\eta_{0}+\sideset{}{'}\sum_{i}\tilde{u}_{i}(x-z_{-})\eta_{i},\nonumber \\
\Psi_{-}(x) & =\tilde{v}_{0}(x-z_{-})\chi_{0}+\sideset{}{'}\sum_{i}\tilde{v}_{i}(x-z_{-})\chi_{i},\label{eq:mmodexp}
\end{align}
where $\tilde{u}_{0}$ and $\tilde{v}_{0}$ are the respective ZMs
of the Dirac operators $\mathcal{D}_{\pm}$ in an $n\!=\!-1$ background,
discussed in Sec.~\ref{subsec:MZMs}. The fermion path integral $Z_{F}[A_{-},J]$
can be shown to be equal to 
\begin{multline}
I_{-}[J]=\int\mathscr{D}\Psi_{\pm}e^{-S_{F}-\!\int\!\dd^{3}{(x,y)}\Psi_{+}^{\intercal}(x)J(x,y)\Psi_{-}(y)}\\
\times\int\dd^{3}{(x_{1},x_{2})}K\Psi_{-}^{\intercal}(x_{2})\tilde{\omega}_{2}\tilde{\omega}_{1}^{\intercal}\Psi_{+}(x_{1}),
\end{multline}
provided 
\begin{align}
\tilde{\omega}_{1} & =\frac{1}{2}\c{C}(\slashed{\partial}+m)\tilde{u}_{0},\nonumber \\
\tilde{\omega}_{2} & =\frac{1}{2}\c{C}(\slashed{\partial}-m)\tilde{v}_{0}.
\end{align}
Substituting $I_{\pm}[J]$ instead of $Z_{F}[A_{\mu}^{\pm},J]$
in the full partition function (\ref{eq:fullZ}) and resumming
the instanton gas leads to an instanton-induced action of the form
\begin{align}
S_{\mathrm{inst}}\!=\! & -\!\lambda K\!\int\!\dd^{3}{(x,y,z)}\nn\\
&\qquad\times\left\{ \Psi_{-}^{\intercal}(x)\left[e^{i\gamma_{(12)}(z)}\omega_{2}(x\!-\!z)\omega_{1}^{\intercal}(y\!-\!z)\right.\right.\nn\\
 & \qquad\left.\left.+e^{-i\gamma_{(12)}(z)}\tilde{\omega}_{2}(x\!-\!z)\tilde{\omega}_{1}^{\intercal}(y\!-\!z)\right]\Psi_{+}(y)\right\} .
\end{align}
As $\omega_{1,2},\tilde{\omega}_{1,2}$ are proportional to the radial
part ($e^{-mr}/r$) of the ZMs, the contribution to the $x$
and $y$ integrals are mainly from small neighborhoods of $x\!=\!z$
and $y\!=\!z$. A change of integration variables $x\!\to\!x\!+\!z$
and $y\!\to\!y\!+\!z$, and subsequent Taylor expansions of the fermion
fields $\Psi_{-}(x\!+\!z)$ and $\Psi_{+}(y\!+\!z)$ to leading (zeroth)
order in $x$ and $y$, yield a local action. Substituting the explicit
forms of $\omega_{1,2}$ and $\tilde{\omega}_{1,2}$, this local action
is 
\begin{equation}
S_{\mathrm{inst}}=\frac{\lambda K}{m}\int\dd^{3}{z}\,\Psi_{-}^{\intercal}(z)\Delta(z)\Psi_{+}(z),
\end{equation}
where the instanton-induced 't~Hooft vertex is defined as
\begin{align}\label{tHooft1}
\Delta(z)=\frac{1}{2}(-\sigma_{z}\!+\!i\sigma_{y})\left[
e^{i\gamma_{(12)}}e^{i\alpha}
\left(
\begin{array}{cc|c}
1 & -i & \bm{0}\\
i & 1 & \bm{0}\\ \hline
\bm{0} & \bm{0} & \bm{0}
\end{array}
\right)\nn\right.\\
\left.+e^{-i\gamma_{(12)}}e^{-i\beta}
\left(
\begin{array}{cc|c}
1 & i & \bm{0}\\
-i & 1 & \bm{0}\\ \hline
\bm{0} & \bm{0} & \bm{0}
\end{array}
\right)\right],
\end{align}
where the $z$ dependence comes through the dual photon $\gamma_{(12)}=\gamma_{(12)}(z)$.

We next address the issue of $SO(N)$ gauge invariance, which the derived 't~Hooft vertex currently
lacks. Indeed, its matrix structure is invariant only under the $[O(2)\!\times\!O(N\!-\!2)]/\mathbb{Z}_{2}$ subgroup, and involves gauge-dependent variables $\alpha$, $\beta$, and $\gamma_{(12)}$.
The key physical feature that full $SO(N)$ invariance brings, for
$N\!>\!2$, is the gauge equivalence between instantons and anti-instantons
in any $SO(2)$ subgroup, given the $\mathbb{Z}_2$ topological charge discussed in Sec.~\ref{subsec:GNO}. It is expected this feature will be restored
upon performing the $SO(N)$ Haar integral in Eq.~(\ref{eq:gfZ}),
but this is analytically intractable. A more physically transparent
way is to impose by hand the gauge equivalence between the $\pm$ monopole
operators 
\begin{equation}
e^{i\gamma_{(12)}}\sim e^{-i\gamma_{(12)}},\label{eq:pmeqal}
\end{equation}
where the $\sim$ implies the two operators can be made equal by an
$SO(N)$ gauge rotation. Writing the $SO(N)$-invariant
monopole operator as
\begin{align}
\c{M}= e^{i\gamma},
\end{align}
the constraint (\ref{eq:pmeqal}) requires that $\gamma\!\in\!\{0,\pi\}\mod 2\pi$. Accordingly, the continuous $U(1)$ shift symmetry of the dual photon reduces to a discrete $\mathbb{Z}_{2}^{\mathcal{M}}$ magnetic symmetry, under which the monopole operator is charged:
\begin{align}\label{Z2magnetic}
\mathbb{Z}_{2}^{\mathcal{M}}  :\gamma&\mapsto\gamma+\pi\nonumber \\
  \c{M}&\mapsto-\c{M}.
\end{align}
This is the behavior expected of monopole operators in 3D $SO(N\!>\!2)$ Yang-Mills theories, which are charged under $\mathbb{Z}_{2}^{\mathcal{M}}$~\cite{aharony2013,Aharony2017,benini2017,cordova2018,komargodski2018,benini2018}.

The 't~Hooft vertex can be further simplified by imposing reflection positivity of the Euclidean action as well as an anti-unitary time-reversal symmetry. First, reflection positivity\footnote{In Euclidean signature, reality of the Minkowski action requires reflection-positivity
of the Euclidean one. However, as the instanton-induced term is free
of time derivatives, it is also a term in the effective Hamiltonian,
which is required to be Hermitian, so it suffices to check Hermiticity.} sets $\alpha\!=\!\beta$ in Eq.~(\ref{tHooft1}), partially constraining the phases of the
ZM functions. This simplifies the vertex to
\begin{equation}
\Delta(z)=\c{M}(z)(-\sigma_{z}\!+\!i\sigma_{y})
\left(
\begin{array}{cc|c}
\cos\alpha & \sin\alpha & \bm{0}\\
-\sin\alpha & \cos\alpha & \bm{0}\\ \hline
\bm{0} & \bm{0} & \bm{0}
\end{array}\right).\label{eq:hooftmat}
\end{equation}
From this form of the vertex, it is clear that $\alpha$ is analogous
to the theta angle in compact $U(1)$ gauge theory in 3D~\citep{shankar2021},
bar complications arising here from the lack of $SO(N)$ gauge invariance. To fix the value of $\alpha$, we demand that the 't~Hooft vertex (\ref{eq:hooftmat}) satisfies the same discrete spacetime
symmetries as the rest of the action obtained from the Lagrangian
(\ref{eq:Lnaive}). As can be checked explicitly, the corresponding Hamiltonian possesses an anti-unitary time-reversal symmetry $\mathcal{T}$, which is defined by
\begin{align}
\mathcal{T}\Psi_{\pm}\mathcal{T}^{-1} & =i\sigma_{y}\Psi_{\mp},\nonumber \\
\mathcal{T}a_i\mathcal{T}^{-1} & =a_i.\label{eq:Tdef}
\end{align}
The nonstandard transformation of the vector potential $a_i$ comes from the fact that the generators of $\mathfrak{so}(N)$ are pure imaginary antisymmetric matrices [e.g., Eq.~(\ref{t12})] which pick up an additional minus sign under complex conjugation. To determine the action of $\c{T}$ on $\c{M}$, we use a physical argument. $\c{T}$ can at most reverse the direction of $SO(2)$ flux created by the monopole operator; but monopoles and anti-monopoles are gauge equivalent. Thus we conclude that $\c{M}$ transforms trivially under $\c{T}$.

To study the effect of $\mathcal{T}$ on the 't~Hooft vertex, we first
rewrite it using the Majorana condition (\ref{eq:majcon1}) as
\begin{align}
\mathcal{L}_{\mathrm{inst}} & =\frac{\lambda K}{2m}(\Psi_{-}^{\dagger}\sigma_{x}\Delta\Psi_{+}+\Psi_{+}^{\dagger}\Delta^{\dagger}\sigma_{x}\Psi_{-})\nonumber \\
 & =\frac{\lambda K}{2m}(\Psi_{-}^{\dagger}\Delta\Psi_{+}+\Psi_{+}^{\dagger}\Delta^{\dagger}\Psi_{-}).
\end{align}
It is then readily observed that 
\begin{align}
\mathcal{T}\mathcal{L}_{\mathrm{inst}}\mathcal{T}^{-1} & =-\frac{\lambda K}{2m}(\Psi_{+}^{\dagger}\sigma_{y}\Delta^{*}\sigma_{y}\Psi_{-}+\Psi_{-}^{\dagger}\sigma_{y}\Delta^{\intercal}\sigma_{y}\Psi_{+}).
\end{align}
Demanding $\mathcal{T}$ invariance then yields the condition 
\begin{equation}
\sigma_{y}\Delta^{\intercal}\sigma_{y}=\Delta,
\end{equation}
which requires that the $SO(N)$ matrix in (\ref{eq:hooftmat}) be antisymmetric. Thus we obtain
\begin{equation}
\alpha\in\left\{\frac{\pi}{2},\frac{3\pi}{2}\right\}\mod 2\pi,
\end{equation}
which can be interpreted as a $\mathbb{Z}_2$ theta angle. The two resulting 't~Hooft vertices only differ by an overall sign that can be absorbed in the coupling constant. Choosing $\alpha\!=\!\pi/2$, the effective
Lagrangian that accounts for instanton effects is 
\begin{align}
\mathcal{L}_{\mathrm{eff}}&=\frac{1}{4}\Psi_{+}^{\intercal}\c{C}(\slashed{\partial}-i\slashed{a}+m)\Psi_{+}+\frac{1}{4}\Psi_{-}^{\intercal}\c{C}(\slashed{\partial}-i\slashed{a}-m)\Psi_{-}\nn\\
&\phantom{=}+\frac{i\lambda K}{m}\c{M}\Psi_{-}^{\intercal}(-\sigma_{z}\!+\!i\sigma_{y})t_{(12)}\Psi_{+}+\frac{1}{2g^{2}}\tr f^{2},\label{eq:Leff}
\end{align}
where the coupling of fermions to a gluon field $a$ has been restored.

Clearly, $\mathcal{L}_{\mathrm{eff}}$ is still only gauge invariant
under $[O(2)\!\times\!O(N\!-\!2)]/\mathbb{Z}_{2}$, the presence
of $t_{(12)}$ indicating memory of the specific $SO(2)$ subgroup
the instanton was placed in. Yet, $\mathcal{L}_{\mathrm{eff}}$ encapsulates
all the correct physical symmetries expected of the gauge-invariant
Lagrangian. Without instanton corrections, the parton theory has the spurious global symmetry $\mathbb{Z}_{2}\!\times\!\mathbb{Z}_{2}^{\mathcal{M}}$,
where $\mathbb{Z}_{2}$ is the microscopic parton representation of
the Ising symmetry, under which
\begin{equation}
\Psi_{\pm}\to W\Psi_{\pm},\qquad W\!=\!\mathrm{diag}(-1,1,\ldots,1)_{N\!\times\!N},\label{eq:Z2xZ2}
\end{equation}
as per our choice of global charge assignment in Sec.~\ref{sec:partonZ2}, and $\mathbb{Z}_2^{\c{M}}$ is the magnetic symmetry (\ref{Z2magnetic}). This enlarged symmetry is absent in the physical spin model. The low-energy effective theory (\ref{eq:Leff}) indicates that instantons
have the effect of explicitly breaking this spurious $\mathbb{Z}_{2}\!\times\!\mathbb{Z}_{2}^{\mathcal{M}}$
symmetry to the diagonal subgroup, under which 
\begin{equation}
\Psi_{\pm}\to W\Psi_{\pm},\qquad\c{M}\to-\c{M}.
\end{equation}
Indeed, since $W^{\intercal}t_{(12)}W\!=\!-t_{(12)}$, the fermion bilinear in the 't~Hooft vertex acquires a minus sign under the action of the first $\mathbb{Z}_2$ factor, which can be compensated by another minus sign coming from the $\mathbb{Z}_2^{\c{M}}$ symmetry action on the monopole operator $\c{M}$. This diagonal symmetry is finally understood as the correct incarnation, in the low-energy parton theory, of the microscopic Ising symmetry $\tau^z\rightarrow-\tau^z$. Although it is not presently clear whether nor how this may be derived analytically, we speculate that full averaging over the $SO(N)$ gauge orbit ($\int\mathscr{D}R$) in the partition function (\ref{eq:gfZ},\ref{eq:fullZ}) produces a fully $SO(N)$-invariant 't~Hooft vertex of the form
\begin{align}
\c{L}_\text{eff}\stackrel{?}{\sim}\c{M}\epsilon_{\alpha_{1}\cdots\alpha_{N}}&\Psi_{-}^{\alpha_{1}}(-\sigma_{z}\!+\!i\sigma_{y})\Psi_{+}^{\alpha_{2}}\cdots\nn\\
&\phantom{\sim}\times\cdots\Psi_{-}^{\alpha_{N-1}}(-\sigma_{z}\!+\!i\sigma_{y})\Psi_{+}^{\alpha_{N}}.
\end{align}
Under the $\mathbb{Z}_2$ symmetry $W$, the fermionic ``baryon'' operator is multiplied by a factor $\det W=-1$ which is compensated by the transformation of the monopole operator $\c{M}$ under the $\mathbb{Z}_2^{\c{M}}$ magnetic symmetry. Note that those transformation properties are now properly independent of the choice of global charge assignment to the fermionic partons, since $\det W$ is invariant under gauge-equivalent redefinitions $W\rightarrow RWR^\intercal$ with $R\in SO(N)$.

In either its $[O(2)\!\times\!O(N\!-\!2)]/\mathbb{Z}_{2}$ or $SO(N)$ invariant incarnations, the 't~Hooft vertex implies that a breakdown of magnetic symmetry, which is typically associated with confinement~\cite{komargodski2018,benini2018}, is concomitant with a breakdown of the Ising symmetry implemented by $W$ in the parton theory. We thus conclude that the $C=0$ phase, which is described by a confining pure Yang-Mills theory at low energies, is indeed a phase in which the microscopic Ising symmetry is spontaneously broken.

\section{Conclusion}
\label{sec:conclusion}

In summary, we have employed slave-particle methods to discuss universal aspects of quantum phase transitions between magnetically ordered, trivially paramagnetic, and gapped topological phases of Ising spin systems. Our theory can be viewed as a generalization of the work of Ref.~\cite{Barkeshli1} from hardcore bosons with $U(1)$ symmetry to Ising spins with $\mathbb{Z}_2$ symmetry. Using a slave-particle decomposition of Ising spins in terms of fermionic Majorana partons with $SO(N)$ gauge structure, we argued that placing the partons in topologically superconducting mean-field states with Chern number $C=0,1,2$ corresponds respectively to magnetically ordered, trivially paramagnetic, and chiral spin liquid phases of the constituent spins. Accounting for gauge fluctuations beyond mean-field, the corresponding Chern-number changing transitions were described by theories of Majorana fields coupled to $SO(N)$ gauge fields with a Chern-Simons term. Using recently conjectured $SO(N)$ dualities with Majorana fermions, the critical theory for the ordering transition from the trivial paramagnet was found to be dual to the usual Wilson-Fisher theory with a single scalar field, as expected for a standard Ising transition. We found that a direct ordering transition from the chiral spin liquid was also possible, and could be protected by lattice symmetries such as inversion symmetry on the honeycomb lattice.

Finally, we turned our attention to the ordered phase itself, in order to identify the symmetry-breaking mechanism from the point of view of the parton gauge theory. The latter was characterized by a spurious apparent $\mathbb{Z}_2\times\mathbb{Z}_2^{\c{M}}$ symmetry, with the first $\mathbb{Z}_2$ factor a global symmetry action on the Majorana partons, and $\mathbb{Z}_2^{\c{M}}$ the magnetic symmetry associated with $SO(N)$ monopole operators. We then showed that the resolution of this problem is to account for nonperturbative instanton effects. First, the massive Majorana fields of the $C=0$ phase support Euclidean ZMs bound to instantons. Second, resumming the instanton gas using semiclassical methods produces an interaction vertex ('t~Hooft vertex) involving Majorana fields and monopole operators, that is only invariant under the diagonal $\mathbb{Z}_2$ subgroup of $\mathbb{Z}_2\times\mathbb{Z}_2^{\c{M}}$. Under the plausible assumption of spontaneously broken $\mathbb{Z}_2^{\c{M}}$ magnetic symmetry in the (confined) $C=0$ phase, the 't~Hooft vertex naturally led to simultaneous breaking of the global Ising symmetry in the parton sector. Thus, as in our earlier work on $U(1)$ bosons~\cite{shankar2021}, we found that nonperturbative instanton effects are instrumental in accounting for spontaneous symmetry breaking in the relevant parton gauge theory. The precise pattern of symmetry breaking (e.g., ferromagnetism vs antiferromagnetism) in the physical spin system depends on the microscopic interpretation of the continuum Majorana spinors $\Psi_\pm$ in a specific lattice model.

We finally outline a few avenues for future research. First, it would be interesting to perform tests of the fermionization duality (\ref{SOduality}) using large-$N$ methods, as done in Ref.~\cite{ChenFisherWu} for the fermionization (\ref{U1:fermion1}) of the 3D XY transition, or in Refs.~\cite{hui2018,hui2019} for non-Abelian dualities with unitary gauge groups. In particular, the duality predicts that the scaling dimension of the Majorana mass operator $[\psi^\intercal\c{C}\psi]=3-\nu^{-1}$, which is dual to the $\phi^2$ operator on the scalar side and related to the correlation length exponent $\nu$, should be independent of the rank of the $SO(N)$ gauge group. It would be interesting to test this prediction by performing computations in the 't~Hooft limit with $N\rightarrow\infty$~\cite{hooft1974}. Second, while the transition between magnetic order and trivial paramagnet is ultimately a standard Ising transition, a direct transition between magnetic order and the $\nu=1/N$ chiral spin liquid is described by a theory of $N_f=2$ massless Majorana fermions coupled to an $SO(N)_1$ Chern-Simons term. This presumably defines a new universality class of Ising transitions in 2+1 dimensions, and it would be interesting to compute critical exponents using either large-$N_f$ or large-$N$ expansions. Third, to complement the semiclassical instanton gas calculation we have presented here, it would be interesting to study the scaling dimensions of $\mathbb{Z}_2$ monopole operators in critical $SO(N)$ gauge theories with Majorana matter, using the state-operator correspondence of conformal field theory~\cite{borokhov2002}. The latter has been successfully used in $U(1)$ gauge theories with massless Dirac matter~\cite{borokhov2002,pufu2014,dupuis2019,dupuis2021,dupuis2022}. Finally, from a more microscopic standpoint, it would be desirable to construct variational many-body wave functions based on the parton ans\"atze discussed here (i.e., $N$-flavor wave functions of Majorana fermions projected to the $SO(N)$ gauge-invariant sector) and use them to study frustrated lattice models of interacting spins with Ising symmetry. Such models could include antiferromagnetic quantum Ising models defined on geometrically frustrated lattices like the kagome lattice, or on non-frustrated lattices but with competing anisotropic interactions, as in the Kitaev model on the honeycomb lattice.

\acknowledgements

We thank M.~Barkeshli, F.~Benini, M.~Cheng, \L.~Fidkowski, Z.-C.~Gu, P.-S.~Hsin, A.~Kovner, M.~Levin, J.~McGreevy, M.~Mulligan, and P.~Ye for helpful discussions. This work was supported by NSERC Discovery Grants \#RGPIN-2020-06999 and \#RGPAS-2020-00064; the Canada Research Chair (CRC) Program; the Government of Alberta's Major Innovation Fund (MIF); the University of Alberta; the Tri-Agency New Frontiers in Research Fund (NFRF, Exploration Stream); and the Pacific Institute for the Mathematical Sciences (PIMS) Collaborative Research Group program.

\appendix

\section{Dualities for bosons with $U(1)$ symmetry}
\label{app:U1dualities}

For the reader's convenience, we provide here a derivation of the dualities between the fermionic critical theories (\ref{U1:fermion1}), (\ref{U1:fermion2}) and their respective bosonic duals (\ref{U1:boson1}), (\ref{U1:boson2}), respectively, based on Ref.~\cite{DualityKarchTong}.

The starting point is the duality of relativistic flux attachment, whereby coupling a level-1 Chern-Simons gauge field to a relativistic complex scalar attaches one flux quantum to the latter and turns it into a Dirac fermion~\cite{polyakov1988}. This can be expressed by the following equivalence between the partition functions
\begin{align}\label{fluxattachment}
Z_\psi[A]e^{\frac{i}{2}S_\text{CS}[A]}=\int\mathscr{D}a\,Z_\phi[a]e^{-iS_\text{CS}[a]+iS_\text{BF}[a,A]},
\end{align}
where we define the fermionic and bosonic partition functions
\begin{align}
Z_\psi[A]&=\int\mathscr{D}\bar{\psi}\mathscr{D}\psi\,e^{i\int d^3x\,\bar{\psi}i(\slashed{\partial}-i\slashed{A})\psi},\\
Z_\phi[A]&=\int\mathscr{D}\phi^*\mathscr{D}\phi\,e^{i\int d^3x\left(|(\partial_\mu-iA_\mu)\phi|^2-\lambda|\phi|^4\right)},\label{Zphi}
\end{align}
and the bosonic action is understood as being tuned to criticality. We define the Chern-Simons and $BF$ actions as
\begin{align}
S_\text{CS}[a]&=\frac{1}{4\pi}\int ada,\label{SCS}\\
S_\text{BF}[a,b]=S_\text{BF}[b,a]&=\frac{1}{2\pi}\int adb.
\end{align}
Equation~(\ref{fluxattachment}) for $A=0$ is simply the relativistic version of the statement that attaching a flux quantum to a boson turns it into a fermion. The $BF$ term for nonzero $A$ expresses the fact that the conserved $U(1)$ fermion current $\bar{\psi}\gamma^\mu\psi$ corresponds to $\frac{1}{2\pi}\epsilon^{\mu\nu\lambda}\partial_\nu a_\lambda$ in the bosonic theory~\cite{fradkin1994}. To understand the level $1/2$ Chern-Simons term for $A$, consider a massive deformation of the theory. If we add a mass term $-r|\phi|^2$ for the scalar with $r>0$, the scalar is gapped and can be integrated out. At low energies the factor $Z_\phi[a]$ in Eq.~(\ref{fluxattachment}) only contains irrelevant terms and reduces to a constant; integrating out $a$ then produces a Chern-Simons term of level $1$ for $A$. The fermionic theory should also be gapped. Assuming a fermionic mass term $\propto-r\bar{\psi}\psi$, if $r>0$ integrating out the fermion produces a Chern-Simons term at level 1/2 for $A$; an additional level-1/2 Chern-Simons term must be added to the fermionic action for the two sides to match. For this assignment to be consistent, the two sides should match also when $r<0$. In this case, on the fermionic side integrating out the fermion cancels out the Chern-Simons term and the Hall response vanishes. On the bosonic side, the scalar condenses and $a$ is Higgsed; the Chern-Simons term for $a$ becomes irrelevant and the Hall response also vanishes upon integration over $a$.

We now turn to deriving the duality between (\ref{U1:fermion1}) and (\ref{U1:boson1}). The partition function $Z[A]$ for the fermionic theory (\ref{U1:fermion1}) is given by
\begin{align}
Z[A]=\int\mathscr{D}a\,Z_\psi[a+A]e^{\frac{i}{2}S_\text{CS}[a]-\frac{i}{2}S_\text{BF}[a,A]-\frac{i}{2}S_\text{CS}[A]}.
\end{align}
Shifting $a\rightarrow a-A$ and using
\begin{align}
S_\text{CS}[a-A]&=S_\text{CS}[a]-S_\text{BF}[a,A]+S_\text{CS}[A],\\
S_\text{BF}[a-A,A]&=S_\text{BF}[a,A]-2S_\text{CS}[A],
\end{align}
we obtain
\begin{align}
Z[A]=\int\mathscr{D}a\,Z_\psi[a]e^{\frac{i}{2}S_\text{CS}[a]-iS_\text{BF}[a,A]+iS_\text{CS}[A]}.
\end{align}
Apart from an additional Chern-Simons term, this can be interpreted as applying the $S$ operation of Witten's $SL(2,\mathbb{Z})$ action on (2+1)D CFTs with a global $U(1)$ symmetry~\cite{witten2003} to the left-hand side of the flux-attachment duality (\ref{fluxattachment}). Using (\ref{fluxattachment}), $Z[A]$ becomes
\begin{align}
Z[A]&=\int\mathscr{D}a\mathscr{D}\tilde{a}\,Z_\phi[\tilde{a}]e^{-iS_\text{CS}[\tilde{a}]+iS_\text{BF}[\tilde{a},a]-iS_\text{BF}[a,A]+iS_\text{CS}[A]}\nn\\
&=\int\mathscr{D}\tilde{a}\,Z_\phi[\tilde{a}]e^{-iS_\text{CS}[\tilde{a}]+iS_\text{CS}[A]}\int\mathscr{D}a\,e^{\frac{i}{2\pi}\int ad(\tilde{a}-A)}.
\end{align}
Integrating over $a$ enforces $\tilde{a}=A+d\chi$ where $\chi$ is an arbitrary function. Exploiting the gauge invariance of the bosonic partition function (\ref{Zphi}) and the Chern-Simons action (\ref{SCS}), we find simply $Z[A]=Z_\phi[A]$, thus the gauged Wilson-Fisher theory (\ref{U1:boson1}) is dual to the fermionic theory (\ref{U1:fermion1}).

We now derive the duality between (\ref{U1:fermion2}) and (\ref{U1:boson2}). The partition function corresponding to (\ref{U1:fermion2}) is
\begin{align}
Z[A]=\int\mathscr{D}a\,Z_\psi[a+A]e^{\frac{3i}{2}S_\text{CS}[a]+\frac{i}{2}S_\text{BF}[a,A]+\frac{i}{2}S_\text{CS}[A]}.
\end{align}
Performing the shift $a\rightarrow a-A$ as before, we obtain
\begin{align}
Z[A]=\int\mathscr{D}a\,Z_\psi[a]e^{\frac{3i}{2}S_\text{CS}[a]-iS_\text{BF}[a,A]+iS_\text{CS}[A]}.
\end{align}
This can be interpreted as applying the combined $ST$ operation of Witten's $SL(2,\mathbb{Z})$ action to (\ref{fluxattachment}), whereby one first shifts the Chern-Simons level of the background gauge field by one before making it dynamical~\cite{witten2003}. Using (\ref{fluxattachment}) once again, we have
\begin{align}
Z[A]&=\int\mathscr{D}a\mathscr{D}\tilde{a}\,Z_\phi[\tilde{a}]e^{-iS_\text{CS}[\tilde{a}]+iS_\text{CS}[a]+iS_\text{BF}[a,\tilde{a}-A]+iS_\text{CS}[A]}\nn\\
&=\int\mathscr{D}\tilde{a}\,Z_\phi[\tilde{a}]e^{-2iS_\text{CS}[\tilde{a}]+iS_\text{BF}[\tilde{a},A]},
\end{align}
performing the path integral over $a$. Thus a single Dirac fermion coupled to $U(1)_{3/2}$ Chern-Simons theory [Eq.~(\ref{U1:fermion2})] is dual to the gauged Wilson-Fisher fixed point coupled to $U(1)_{-2}$ Chern-Simons theory [Eq.~(\ref{U1:boson2})].

\section{Majorana $SO(N)$ lattice gauge theory in the strong-coupling limit}
\label{app:SONLGT}

In this Appendix, we show that in the limit of strong gauge coupling, a theory of $N$ colors of Majorana fermions ($N$ even) coupled to an $SO(N)$ lattice gauge field naturally reduces to a theory of Ising spins corresponding to the gauge-invariant Majorana baryons (\ref{partonZ2}).

\subsection{Euclidean vs Hamiltonian approach}

First, we relate the Euclidean and Hamiltonian descriptions of $SO(N)$ lattice gauge theory with Majorana fermions in the vector representation, following the approach of Refs.~\cite{kogut1975,kogut1979,creutz}. We begin with a Euclidean action in discrete 3D spacetime,
\begin{align}
S=S_\chi+S_U,
\end{align}
where
\begin{align}
S_U=-\frac{\beta}{2}\sum_\square\tr UUUU-\frac{\beta_\tau}{2}\sum_{\square_\tau}\tr UUUU+\mathrm{c.c.},
\end{align}
is the gauge-field action, and
\begin{align}
S_\chi=\frac{it}{4}\sum_{i,\mu}\chi^T_ih_{i,i+\hat{\mu}}U_{i,i+\hat{\mu}}\chi_{i+\hat{\mu}}+\frac{t_\tau}{4}\sum_i\chi_i^TU_{i,i+\hat{\tau}}\chi_{i+\hat{\tau}},
\end{align}
is the gauged Majorana action. Here $i,j$ denote spacetime lattice sites, $\hat{\mu}=(\hat{x},\hat{y})$ denotes lattice vectors in the two space directions, and $\hat{\tau}$ denotes the lattice vector in the imaginary-time direction. We write $\chi_i=(\chi_i^1,\ldots,\chi_i^N)$ for the $N$-component vector of Majorana fields on site $i$, $U_{ij}\in SO(N)$ for the link variable on nearest-neighbor spacetime link $ij$, with $U_{ji}=U_{ij}^{-1}$, and $\square$ and $\square_\tau$ for spacelike and timelike plaquettes, respectively. The real antisymmetric matrix $h$ describes Majorana hopping in the absence of gauge fields~\cite{kitaev2006}, but we have factored out the hopping strength $t$. We consider spacetime-anisotropic couplings in anticipation of taking the $\tau$-continuum limit to relate the discrete-time action formulation to the Hamiltonian formulation~\cite{fradkin1978}. For the same reason, we take the lattice constant in the spatial direction to be unity, and the lattice constant in the temporal direction to be $\epsilon\ll 1$. The action is invariant under local $SO(N)$ gauge transformations,
\begin{align}
\chi_i\rightarrow R_i\chi_i,\hspace{5mm}U_{ij}\rightarrow R_iU_{ij}R_j^{-1},\hspace{5mm}R_i\in SO(N).
\end{align}

First, we use this gauge freedom to work in the temporal gauge: $U_{i,i+\hat{\tau}}=1$ on all temporal links. The Majorana action becomes,
\begin{align}
S_\chi&=\sum_\tau\biggl(\frac{t}{4}\sum_{r,\mu}\chi_r^T(\tau)U_{r,r+\hat{\mu}}(\tau)\chi_{r+\hat{\mu}}(\tau)\nn\\
&\hspace{10mm}+\frac{t_\tau}{4}\sum_r\chi_r^T(\tau)\chi_r(\tau+\epsilon)+\mathrm{c.c.}\biggr)\nn\\
&\approx\epsilon\sum_\tau\biggl(\frac{t}{4\epsilon}\sum_{r,\mu}\chi_r^T(\tau)U_{r,r+\hat{\mu}}(\tau)\chi_{r+\hat{\mu}}(\tau)\nn\\
&\hspace{10mm}+\frac{t_\tau}{4}\sum_r\chi_r^T(\tau)\partial_\tau\chi_r(\tau)+\mathrm{c.c.}\biggr),
\end{align}
to leading order in $\epsilon$, ignoring additive constants. Here we use $i=(r,\tau)$ to denote the dependence on space $r$ and time $\tau$ coordinates separately.

The gauge-field action is more subtle. The contribution from spatial plaquettes is obvious; we now focus on temporal plaquettes. In the temporal gauge, we have:
\begin{align}
\tr UUUU\bigr|_{\square_\tau}&=\tr U_{r,r+\hat{\mu}}(\tau)U_{r+\hat{\mu},r}(\tau+\epsilon)\nn\\
&=\tr U_{r,r+\hat{\mu}}^{-1}(\tau+\epsilon)U_{r,r+\hat{\mu}}(\tau),
\end{align}
using the cyclic property of the trace. To work towards the Hamiltonian formulation, we seek an operator $\hat{O}$ such that
\begin{align}\label{trotterg}
\langle U_{rr'}(\tau+\epsilon)|e^{-\epsilon\hat{O}}|U_{rr'}(\tau)\rangle=e^{\beta_\tau\Re\tr g},
\end{align}
where $g=U_{rr'}^{-1}(\tau+\epsilon)U_{rr'}(\tau)\in SO(N)$, and the equality holds in the limit $\epsilon\ll 1$. We focus on a given spatial link $rr'$. The state $|U_{rr'}\rangle$ is an eigenstate of the matrix-valued link operator $\hat{U}_{rr'}$,
\begin{align}
\hat{U}_{rr'}|U_{rr'}\rangle=U_{rr'}|U_{rr'}\rangle.
\end{align}
We define an electric-field operator $\hat{E}_{rr'}^a$ that is (almost) a canonical conjugate to $\hat{U}_{rr'}$,
\begin{align}\label{canonical}
[\hat{E}_{rr'}^a,\hat{U}_{rr'}]=-T^a\hat{U}_{rr'},
\end{align}
where $a=1,\ldots,N(N-1)/2$ ranges over the generators $T^a$ of $SO(N)$. (Note that on the right-hand side of Eq.~(\ref{canonical}), there is matrix multiplication between the $c$-number matrix $T^a$ and the matrix-valued operator $\hat{U}_{rr'}$, while for a given $a$, the operator $\hat{E}^a_{rr'}$ is a scalar.) The electric-field operators satisfy the $\mathfrak{so}(N)$ Lie algebra,
\begin{align}
[\hat{E}^a_{rr'},\hat{E}^b_{rr'}]=if^{abc}\hat{E}^c_{rr'},
\end{align}
where $f^{abc}$ are the $\mathfrak{so}(N)$ structure constants. Now consider the operator
\begin{align}
\hat{R}_{rr'}(g)=e^{-i\omega^a\hat{E}^a_{rr'}},
\end{align}
where the $SO(N)$ matrix is parametrized as $g=e^{-i\omega^aT^a}$. We have the property
\begin{align}\label{rotU}
\hat{R}_{rr'}(g)|U_{rr'}\rangle=|gU_{rr'}\rangle.
\end{align}
Indeed, using Eq.~(\ref{canonical}), we can show that $\hat{R}_{rr'}(g)|U_{rr'}\rangle$ is an eigenstate of $\hat{U}_{rr'}$ with eigenvalue $gU_{rr'}$:
\begin{align}
\hat{U}_{rr'}\left(\hat{R}_{rr'}(g)|U_{rr'}\rangle\right)
&=\left(\hat{U}_{rr'}e^{-i\omega^a\hat{E}^a_{rr'}}\hat{U}_{rr'}^{-1}\right)\hat{U}_{rr'}|U_{rr'}\rangle\nn\\
&=e^{-i\omega^a\hat{U}_{rr'}\hat{E}^a_{rr'}\hat{U}_{rr'}^{-1}}U_{rr'}|U_{rr'}\rangle\nn\\
&=e^{-i\omega^a(\hat{E}^a_{rr'}+T^a)}U_{rr'}|U_{rr'}\rangle\nn\\
&=e^{-i\omega^a\hat{E}^a_{rr'}}e^{-i\omega^aT^a}U_{rr'}|U_{rr'}\rangle\nn\\
&=gU_{rr'}\left(\hat{R}_{rr'}(g)|U_{rr'}\rangle\right).
\end{align}
In the fourth line, we use the fact that $[\hat{E}_{rr'}^a,T^a]=0$ because $T^a$ is a $c$-number matrix while $\hat{E}^a_{rr'}$ is a scalar operator.

Using property (\ref{rotU}), we claim that Eq.~(\ref{trotterg}) is satisfied if
\begin{align}\label{trotter2}
e^{-\epsilon\hat{O}}=\int dg\,e^{\beta_\tau\Re\tr g}\hat{R}_{rr'}(g),
\end{align}
where $dg$ denotes the Haar measure on $SO(N)$. Indeed, we then have
\begin{align}
&\langle U_{rr'}(\tau+\epsilon)|e^{-\epsilon\hat{O}}|U_{rr'}(\tau)\rangle\nn\\
&\hspace{15mm}=\int dg\,e^{\beta_\tau\Re\tr g}\langle U_{rr'}(\tau+\epsilon)|\hat{R}_{rr'}(g)|U_{rr'}(\tau)\rangle\nn\\
&\hspace{15mm}=\int dg\,e^{\beta_\tau\Re\tr g}\langle U_{rr'}(\tau+\epsilon)|gU_{rr'}(\tau)\rangle\nn\\
&\hspace{15mm}=\int dg\,e^{\beta_\tau\Re\tr g}\delta_{U_{rr'}(\tau+\epsilon),gU_{rr'}(\tau)}\nn\\
&\hspace{15mm}=e^{\beta_\tau\Re\tr U_{rr'}(\tau+\epsilon)U^{-1}_{rr'}(\tau)}\nn\\
&\hspace{15mm}=e^{\beta_\tau\Re\tr U_{rr'}^{-1}(\tau+\epsilon)U_{rr'}(\tau)},
\end{align}
where in the last line, we have used the fact that $\Re\tr g=\Re\tr g^\dag$, and $g^{-1}=g^\dag$ for $g\in SO(N)$. Finally, we consider the $SO(N)$ Haar integral in (\ref{trotter2}). Since $g=e^{-i\omega^aT^a}$, we have $\Re\tr g=\half\tr(g+g^\dag)=\tr\cos\omega^aT^a$. The integral over $g$ can be converted to an integral over $\omega$:
\begin{align}
e^{-\epsilon\hat{O}}=\left(\prod_a\int d\omega_a\right)\c{J}(\omega)e^{\beta_\tau\tr\cos\omega^aT^a}e^{-i\omega^a\hat{E}^a_{rr'}},
\end{align}
where $\c{J}$ is the Jacobian of the transformation. We further write $\beta_\tau=1/(\epsilon J)$ with fixed $J$, and consider the limit $\epsilon\ll 1$. In that limit, we can use a saddle-point approximation: the integral is dominated by Gaussian fluctuations around the maximum of $\tr\cos\omega^aT^a$, which is at $\omega^a=0$. Using
\begin{align}
\tr\cos\omega^aT^a&=\tr\left(1-\frac{1}{2}\omega^a\omega^bT^aT^b+\ldots\right)\nn\\
&=N-\frac{1}{4}\omega^a\omega^a+\mathcal{O}(\omega^4),
\end{align}
assuming the $SO(N)$ generators are normalized as $\tr T^aT^b=\frac{1}{2}\delta^{ab}$. We thus obtain
\begin{align}
e^{-\epsilon\hat{O}}&\propto \c{J}(0)\int d\omega\,e^{-\frac{\beta_\tau}{4}\omega^a\omega^a}e^{-i\omega^a\hat{E}^a_{rr'}}\nn\\
&\propto e^{-\epsilon J\hat{E}_{rr'}^a\hat{E}_{rr'}^a}.
\end{align}
Taking the logarithm on both sides and ignoring an irrelevant additive constant, we thus conclude that the desired operator $\hat{O}$ is
\begin{align}
\hat{O}=J\sum_{r,\mu}\hat{E}_{r,r+\hat{\mu}}^a\hat{E}_{r,r+\hat{\mu}}^a,
\end{align}
where we have generalized Eq.~(\ref{trotter2}) to include a product over all spatial links, since all spatial links decouple in the sum over temporal plaquettes. Finally, writing $\beta=\epsilon K$ with fixed $K$, $t=\epsilon\kappa$ with fixed $\kappa$, and normalizing the action such that $t_\tau=1$, we obtain
\begin{align}
S&\approx\epsilon\sum_\tau\left(\frac{1}{4}\sum_r\chi_r^T\partial_\tau\chi_r+H\right)\nn\\
&\approx\int d\tau\left(\frac{1}{4}\sum_r\chi_r^T\partial_\tau\chi_r+H\right),
\end{align}
where the Hamiltonian is, now dropping hats on operators,
\begin{align}
H&=\frac{i\kappa}{4}\sum_{r,\mu}\chi_r^T h_{r,r+\hat{\mu}}U_{r,r+\hat{\mu}}\chi_{r+\hat{\mu}}+J\sum_{r,\mu}\tr E_{r,r+\hat{\mu}}^2\nn\\
&\phantom{=}+K\sum_\square\Re\tr UUUU,
\end{align}
where we have defined the matrix-valued electric-field operator $E_{r,r+\hat{\mu}}\equiv E_{r,r+\hat{\mu}}^aT^a$ to arrive at a basis-independent expression (and have absorbed a factor of $\frac{1}{2}$ into $J$). Note that the hopping matrix $h_{rr'}=-h_{r'r}=h_{rr'}^*$ has no dependence on color indices. To be more precise, we have $\chi_r^Th_{rr'}U_{rr'}\chi_{r'}\equiv\chi_r^\alpha h_{rr'}U_{rr'}^{\alpha\beta}\chi_{r'}^\beta$, where $\alpha,\beta=1,\ldots,N$ are the color indices. We can check that the constraints of Fermi statistics and Hermiticity of the Hamiltonian both separately imply that $U_{r'r}^{\beta\alpha}=U_{rr'}^{\alpha\beta}$, i.e., that $U_{r'r}^T=U_{rr'}$, which is satisfied for $SO(N)$ gauge fields since $U_{r'r}^T=U_{r'r}^{-1}=U_{rr'}$. Thus lattice Majorana fermions can be consistently coupled to lattice $SO(N)$ gauge fields.

\subsection{Strong-coupling limit}

In the $\tau$-continuum limit, we saw that the relationship between the couplings in the spacetime lattice action $\beta,\beta_\tau$ and those in the Hamiltonian $J,K$ is $\beta=\epsilon K$ and $\beta_\tau=1/(\epsilon J)$. We now consider the ``electric'' limit in the Hamiltonian problem: $J\rightarrow\infty$ and $K\rightarrow 0$. We see that in this limit, $\beta,\beta_\tau\rightarrow 0$. Going back to the Euclidean lattice action, the plaquette term $S_U$ disappears in this limit, and the physics is purely governed by the gauged Majorana action: $S(J\rightarrow\infty,K\rightarrow 0)\approx S_\chi$, where
\begin{align}
S_\chi= \frac{it}{4}\sum_{i,\mu}\chi^T_ih_{i,i+\hat{\mu}}U_{i,i+\hat{\mu}}\chi_{i+\hat{\mu}}+\frac{1}{4}\sum_i\chi_i^TU_{i,i+\hat{\tau}}\chi_{i+\hat{\tau}}.
\end{align}
In this limit, all links decouple, and the functional integral over the gauge field reduces to a product of one-link Haar integrals over $SO(N)$~\cite{rossi1984,wolff1985}:
\begin{align}\label{SC}
Z&=\int\mathscr{D}\chi\mathscr{D}U\,e^{-S}\nn\\
&=\int\mathscr{D}\chi\left(\prod_{i,\mu}\int dU\,e^{-\frac{it}{4}\chi_i^Th_{i,i+\hat{\mu}}U\chi_{i+\hat{\mu}}}\right)
\nn\\
&\hspace{15mm}\times\left(\prod_i\int dU\,e^{-\frac{1}{4}\chi_i^TU\chi_{i+\hat{\tau}}}\right).
\end{align}
Consider first the spatial-link term. We perform a formal expansion in the hopping parameter:
\begin{align}
\int dU\,e^{-\frac{it}{4}\chi_i^Th_{i,i+\hat{\mu}}U\chi_{i+\hat{\mu}}}&=\sum_{n=0}^\infty\frac{1}{n!}\left(-\frac{it}{4}h_{i,i+\hat{\mu}}\right)^n\nn\\
&\phantom{=}\times\chi_i^{\alpha_1}\chi_{i+\hat{\mu}}^{\beta_1}\cdots\chi_i^{\alpha_n}\chi_{i+\hat{\mu}}^{\beta_n}\nn\\
&\phantom{=}\times\int dU\,U^{\alpha_1\beta_1}\cdots U^{\alpha_n\beta_n}.
\end{align}
Polynomial integrals over compact Lie groups can in principle be computed exactly in the framework of Weingarten calculus~\cite{collins2009}. Here we will not attempt to do this, but only use general properties of those integrals to illustrate the physics~\cite{chen2018}. The necessary results are given in Ref.~\cite{collins2009} for the orthogonal group $O(N)$. To compute integrals over $SO(N)$, we insert the factor $(1+\det U)/2$ in the integrand:
\begin{align}\label{weingarten}
&\int_{SO(N)}dU\,U^{\alpha_1\beta_1}\cdots U^{\alpha_n\beta_n}\nn\\
&\hspace{10mm}=\int_{O(N)}dU\,\left(\frac{1+\det U}{2}\right)U^{\alpha_1\beta_1}\cdots U^{\alpha_n\beta_n}\nn\\
&\hspace{10mm}=\frac{1}{2}\int_{O(N)}dU\,U^{\alpha_1\beta_1}\cdots U^{\alpha_n\beta_n}\nn\\
&\hspace{15mm}+\frac{1}{2}\epsilon^{\gamma_1\cdots\gamma_N}\int_{O(N)}dU\,U^{1,\gamma_1}\cdots U^{N,\gamma_N}
\nn\\
&\hspace{42mm}\times U^{\alpha_1\beta_1}\cdots U^{\alpha_n\beta_n}.
\end{align}
Consider the first term. For it to be nonzero, $n$ must be even: $n=2k$, and the sets $\{\alpha_1,\ldots,\alpha_{2k}\}$ and $\{\beta_1,\ldots,\beta_{2k}\}$ must each contain $k$ pairs of identical entries~\cite{collins2009}. Consider such pairs $\alpha_i=\alpha_j=\alpha$ and $\beta_i=\beta_j=\beta$; the corresponding Majorana term is $(\chi_i^\alpha)^2(\chi_{i+\hat{\mu}}^\beta)^2=\text{const}.$ Thus the first term in Eq.~(\ref{weingarten}) can be ignored (the gauge-invariant Majorana ``mesons'' are trivial). Turning to the second term, the pairing rule first requires that the set $\{1,\ldots,N,\alpha_1,\ldots,\alpha_n\}$ can be grouped into pairs. The smallest $n$ for which this occurs is $n=N$, which implies that $\{\alpha_1,\ldots,\alpha_N\}=\{1,\ldots,N\}$ in some order. Likewise, the set $\{\gamma_1,\ldots,\gamma_N,\beta_1,\ldots,\beta_n\}$ must obey the same pair constraint, which also implies that $\{\beta_1,\ldots,\beta_N\}=\{1,\ldots,N\}$ in some order since $\{\gamma_1,\ldots,\gamma_N\}=\{1,\ldots,N\}$ by virtue of the epsilon tensor. But since both $\{\alpha_1,\ldots,\alpha_N\}$ and $\{\beta_1,\ldots,\beta_N\}$ must equal $\{1,\ldots,N\}$ in some order, then
\begin{align}
\chi_i^{\alpha_1}\chi_{i+\hat{\mu}}^{\beta_1}\cdots\chi_i^{\alpha_n}\chi_{i+\hat{\mu}}^{\beta_n}
&\propto(\chi_i^1\ldots\chi_i^N)(\chi_{i+\hat{\mu}}^1\cdots\chi_{i+\hat{\mu}}^N)\nn\\
&\phantom{\propto}\times\epsilon^{\alpha_1\cdots\alpha_N}\epsilon^{\beta_1\cdots\beta_N},
\end{align}
since Majorana fields anticommute. Absorbing into a constant $B$ the following integral,
\begin{align}
B&\propto\epsilon^{\gamma_1\cdots\gamma_N}\epsilon^{\alpha_1\cdots\alpha_N}
\epsilon^{\beta_1\cdots\beta_N}\nn\\
&\phantom{\propto}\times\int_{O(N)}dU\,U^{1,\gamma_1}\cdots U^{N,\gamma_N}U^{\alpha_1\beta_1}\cdots U^{\alpha_N\beta_N},
\end{align}
we obtain:
\begin{align}
\int dU\,e^{-\frac{it}{4}\chi_i^Th_{i,i+\hat{\mu}}U\chi_{i+\hat{\mu}}}&=1+\frac{B(-t)^Nh_{i,i+\hat{\mu}}^N}{4^NN!}\tau_i^z\tau_{i+\hat{\mu}}^z+\ldots\nn\\
&\approx e^{J_{i,i+\hat{\mu}}\tau_i^z\tau_{i+\hat{\mu}}^z+\ldots},
\end{align}
where we have introduced the Ising baryon
\begin{align}
\tau_i^z=i^{N/2}\chi_i^1\ldots\chi_i^N,
\end{align}
with $N$ even, and an effective nearest-neighbor exchange $J_{i,i+\hat{\mu}}=B(-t)^Nh_{i,i+\hat{\mu}}^N/(4^NN!)$. Likewise for the temporal link integral in Eq.~(\ref{SC}), the formal expansion gives:
\begin{align}
\int dU\,e^{-\frac{1}{4}\chi_i^TU\chi_{i+\hat{\tau}}}&=1+\frac{Bi^N}{4^NN!}\tau_i^z\tau_{i+\hat{\tau}}^z+\ldots\nn\\
&\approx e^{K\tau_i^z\tau_{i+\hat{\tau}}^z+\ldots},
\end{align}
where $K=B(-1)^{N/2}/(4^NN!)$ is the nearest-neighbor coupling in the temporal direction. One thus obtains an effective spacetime lattice Ising action,
\begin{align}
S_\text{eff}[\tau^z]=-\sum_{i,\mu}J_{i,i+\hat{\mu}}\tau_i^z\tau_{i+\hat{\mu}}^z-K\sum_i\tau_i^z\tau_{i+\hat{\tau}}^z+\ldots,
\end{align}
which corresponds to an effective quantum Ising Hamiltonian in the $\tau$-continuum limit~\cite{fradkin1978},
\begin{align}
H_\text{eff}[\hat{\tau}^z,\hat{\tau}^x]=-\sum_{r,\mu}J'_{r,r+\hat{\mu}}\hat{\tau}_r^z\hat{\tau}_{r+\hat{\mu}}^z
-K'\sum_r\hat{\tau}_r^x+\ldots,
\end{align}
with a suitably defined exchange coupling $J'_{r,r+\hat{\mu}}$ and transverse field $K'$, neglecting higher-order multi-spin interactions that correspond to neglected higher-order baryon processes in the strong-coupling (hopping) expansion. Thus it is clear that, at least from a strong-coupling perspective, the $SO(N)$ Majorana gauge theory that results from the parton decomposition (\ref{partonZ2}) is a theory of interacting Ising spins.

\section{Conformal embeddings in $\mathfrak{so}(n)$ Wess-Zumino-Witten models}
\label{app:SONCFT}

In this Appendix, we explain the meaning of the conformal embedding~\cite{naculich1990}:
\begin{align}\label{embedding}
\mathfrak{so}(N)_k\otimes\mathfrak{so}(k)_N\subseteq\mathfrak{so}(Nk)_1,
\end{align}
which is a generalization of the embedding $\mathfrak{so}(k)_k\otimes\mathfrak{so}(k)_k\subseteq\mathfrak{so}(k^2)_1$ used in Refs.~\cite{sahoo2016,cheng2018}.

\subsection{Free chiral Majorana fields}

The starting point is the 2D CFT of $Nk$ free chiral Majorana fermions $\chi_\alpha(z)$, $\alpha=1,\ldots,Nk$. The (holomorphic) energy-momentum tensor for this free theory is~\cite{CFT}:
\begin{align}\label{Tz}
T(z)=-\frac{1}{2}\sum_\alpha\chi_\alpha\partial\chi_\alpha,
\end{align}
where $\partial\equiv\partial_z$. The chiral central charge $c_-$ for this theory is $1/2$ per flavor of Majorana fermion, i.e., $c_-=Nk/2$. This theory is equivalent to the critical $\mathfrak{so}(n)$ WZW model at level 1, with $n=Nk$. To establish this, we define the $\mathfrak{so}(n)$ currents:
\begin{align}\label{soncurrents}
j^a(z)\equiv\frac{i}{2}\chi^T(z)T^a\chi(z),
\end{align}
where $a=1,\ldots,n(n-1)/2$ ranges over the real antisymmetric generators $T^a$ of the $\mathfrak{so}(n)$ Lie algebra. These currents satisfy a nontrivial algebra (current algebra) in the sense of the operator product expansion (OPE). To compute the OPE for free fields, we simply need to use Wick's theorem. For now we are only interested in the singular part of the OPE, which is given by the sum of all Wick contractions:
\begin{align}\label{wick}
j^a(z)j^b(w)&\sim-\frac{1}{4}\sum_\text{Wick}\chi_\alpha(z)T^a_{\alpha\beta}\chi_\beta(z)\chi_\gamma(w)T^b_{\gamma\delta}\chi_\delta(w)\nn\\
&\sim-\frac{1}{4}T^a_{\alpha\beta}T^b_{\gamma\delta}\Bigl[-\langle\chi_\alpha(z)\chi_\gamma(w)\rangle\chi_\beta(z)\chi_\delta(w)\nn\\
&\hspace{20mm}-\langle\chi_\beta(z)\chi_\delta(w)\rangle\chi_\alpha(z)\chi_\gamma(w)\nn\\
&\hspace{20mm}+\langle\chi_\beta(z)\chi_\gamma(w)\rangle\chi_\alpha(z)\chi_\delta(w)\nn\\
&\hspace{20mm}+\langle\chi_\alpha(z)\chi_\delta(w)\rangle\chi_\beta(z)\chi_\gamma(w)\nn\\
&\hspace{20mm}-\langle\chi_\alpha(z)\chi_\gamma(w)\rangle\langle\chi_\beta(z)\chi_\delta(w)\rangle\nn\\
&\hspace{20mm}+\langle\chi_\alpha(z)\chi_\delta(w)\rangle\langle\chi_\beta(z)\chi_\gamma(w)\rangle\Bigr].
\end{align}
Next, we use the free Majorana Green's function:
\begin{align}
\langle\chi_\alpha(z)\chi_\beta(w)\rangle=\frac{\delta_{\alpha\beta}}{z-w},
\end{align}
and, in the operator-valued terms, expand $\chi_\alpha(z)=\chi_\alpha(w)+(z-w)\partial\chi_\alpha(w)+\ldots$. Keeping only terms singular as $z\rightarrow w$, we obtain:
\begin{align}
j^a(z)j^b(w)\sim-\frac{1}{4}\biggl(&\frac{2}{z-w}\chi^T[T^a,T^b]\chi\nn\\
&+\frac{2}{(z-w)^2}\tr T^a T^b\biggr),
\end{align}
using $\chi^TT^aT^b\chi=-\chi^TT^bT^a\chi$, from Grassmann anticommutation and the antisymmetry of the $\mathfrak{so}(n)$ generators. We assume the (anti-Hermitian) generators obey the following properties:
\begin{align}\label{Lie}
[T^a,T^b]=f^{abc}T^c,\hspace{5mm}\tr T^a T^b=-2\delta^{ab},
\end{align}
where $f^{abc}$ are the structure constants of $\mathfrak{so}(n)$. We then obtain:
\begin{align}\label{jjOPE}
j^a(z)j^b(w)\sim\frac{\delta^{ab}}{(z-w)^2}+\frac{if^{abc}}{z-w}j^c(w),
\end{align}
which is the $\mathfrak{so}(n)_1$ current algebra (Kac-Moody algebra)~\cite{CFT}. The energy-momentum tensor can be expressed in terms of these currents using the Sugawara construction:
\begin{align}\label{Tson1}
T_{\mathfrak{so}(n)_1}(z)=\frac{1}{2(n-1)}\sum_a\colon j^a(z)j^a(z)\colon,
\end{align}
where the colons denote normal ordering, i.e., the product $j^a(z)j^b(w)$ in the limit $z\rightarrow w$ (that is, the OPE) but with all singular terms subtracted. To do this computation, we use the identity in Eq.~(15.204) of Ref.~\cite{CFT}:
\begin{align}\label{OPE}
\sum_{\alpha\beta}\left(\colon(\chi_\alpha\chi_\beta)(\chi_\alpha\chi_\beta)\colon-\colon(\chi_\alpha\chi_\beta)(\chi_\beta\chi_\alpha)\colon\right)&=4(n-1)\nn\\
&\times\sum_\alpha\chi_\alpha\partial\chi_\alpha.
\end{align}
We also choose a particular basis for the $\mathfrak{so}(n)$ generators~\cite{CFT},
\begin{align}\label{songen}
T^{(r,s)}_{\alpha\beta}=\delta_\alpha^r\delta_\beta^s-\delta_\beta^r\delta_\alpha^s,
\end{align}
which is properly normalized according to Eq.~(\ref{Lie}). Here the generators are labeled by the $n(n-1)/2$ pairs $(r,s)$ with $1\leq r<s\leq n$. Using the identity
\begin{align}
\sum_{(r,s)}T_{\alpha\beta}^{(r,s)}T_{\gamma\delta}^{(r,s)}=\delta_{\alpha\gamma}\delta_{\beta\delta}-\delta_{\beta\gamma}\delta_{\alpha\delta},
\end{align}
and Eq.~(\ref{OPE}), we easily find that the Sugawara energy-momentum tensor (\ref{Tson1}) reproduces Eq.~(\ref{Tz}). The chiral central charge can also be checked. The Sugawara energy-momentum tensor of the $\mathfrak{g}_k$ WZW CFT has the general form~\cite{CFT}
\begin{align}\label{sugawara}
T_{\mathfrak{g}_k}(z)=\frac{1}{2(k+g)}\sum_a\colon J^aJ^a\colon,
\end{align}
where $k$ is the Kac-Moody level and $g$ is the dual Coxeter number. The chiral central charge is then
\begin{align}\label{c-}
c_-[\mathfrak{g}_k]=\frac{k\dim\mathfrak{g}}{k+g},
\end{align}
where $\dim\mathfrak{g}=\delta_{aa}$, i.e., the number of generators of the Lie algebra $\mathfrak{g}$. Here we have $\dim\so(n)=n(n-1)/2$, and $k=1$. By comparing (\ref{sugawara}) and (\ref{Tson1}), we find $g=n-2$, and thus
\begin{align}
c_-[\mathfrak{so}(n)_1]=\frac{n(n-1)/2}{1+n-2}=\frac{n}{2},
\end{align}
as expected for $n$ free Majorana fermions.

\subsection{Conformal embedding}

The conformal embedding (\ref{embedding}) arises from a natural embedding of the Lie algebras $\so(N)$ and $\so(k)$ into $\so(Nk)$. We first represent the indices for $\so(Nk)$ matrices as a pair $\boldsymbol{\alpha}=(\alpha,\tilde{\alpha})$ where $\alpha=1,\ldots,N$ and $\tilde{\alpha}=1,\ldots,k$. We construct an embedding $\so(N)\rightarrow\so(Nk)$ as
\begin{align}
\Sigma^a_{\boldsymbol{\alpha}\boldsymbol{\beta}}=(T^a)_{\alpha\beta}\delta_{\tilde{\alpha}\tilde{\beta}},
\end{align}
i.e., $\Sigma^a=T^a\otimes\mathds{1}_{k}$, where $T^a$ are $\so(N)$ generators and $\mathds{1}_{k}$ the $k\times k$ identity matrix. Likewise, we construct an embedding $\so(k)\rightarrow\so(Nk)$ as
\begin{align}
\tilde{\Sigma}^a_{\boldsymbol{\alpha}\boldsymbol{\beta}}=\delta_{\alpha\beta}(\tilde{T}^a)_{\tilde{\alpha}\tilde{\beta}},
\end{align}
i.e., $\tilde{\Sigma}^a=\mathds{1}_{N}\otimes\tilde{T}^a$, where $\tilde{T}^a$ are $\so(k)$ generators and $\mathds{1}_{N}$ the $N\times N$ identity matrix. We then define $\so(N)$ and $\so(k)$ currents, respectively, as
\begin{align}
J^a(z)\equiv\frac{i}{2}\chi^T(z)\Sigma^a\chi(z),\hspace{5mm}
\tilde{J}^a(z)\equiv\frac{i}{2}\chi^T(z)\tilde{\Sigma}^a\chi(z),
\end{align}
analogously to Eq.~(\ref{soncurrents}). By computing the OPE, we now show these satisfy the $\so(N)_k$ and $\so(k)_N$ current algebras, respectively. Using the explicit forms
\begin{align}\label{Jexplicit}
J^a(z)=\frac{i}{2}\chi_{\alpha\tilde{\alpha}}T^a_{\alpha\beta}\chi_{\beta\tilde{\alpha}},\hspace{5mm}
\tilde{J}^a(z)=\frac{i}{2}\chi_{\alpha\tilde{\alpha}}\tilde{T}^a_{\tilde{\alpha}\tilde{\beta}}\chi_{\alpha\tilde{\beta}},
\end{align}
and following the same steps as in Eqs.~(\ref{wick}-\ref{jjOPE}), we find:
\begin{align}
J^a(z)J^b(w)&\sim\frac{k\delta^{ab}}{(z-w)^2}+\frac{if^{abc}}{z-w}J^c(w),\\
\tilde{J}^a(z)\tilde{J}^b(w)&\sim\frac{N\delta^{ab}}{(z-w)^2}+\frac{if^{abc}}{z-w}\tilde{J}^c(w),
\end{align}
which are indeed the $\so(N)_k$ and $\so(k)_N$ current algebras, respectively. By following similar steps and using the fact that $\tr T^a=\tr\tilde{T}^a=0$, we can show that the mixed $J^a\tilde{J}^b$ OPE has no singular terms. Thus the two current algebras decouple.

Finally, we show that the energy-momentum tensor (\ref{Tson1}) of the $\so(Nk)_1$ theory decomposes into the sum of the energy-momentum tensors of the $\so(N)_k$ and $\so(k)_N$ theories:
\begin{align}
T_{\mathfrak{so}(Nk)_1}(z)=T_{\mathfrak{so}(N)_k}(z)+T_{\mathfrak{so}(k)_N}(z).
\end{align}
To do this, we need the following formula~\cite{sahoo2016,cheng2018}:
\begin{align}
\colon(\chi_\alpha\chi_\beta)(\chi_\alpha\chi_\beta)\colon=\chi_\alpha\partial\chi_\alpha+\chi_\beta\partial\chi_\beta,\hspace{5mm}\alpha\neq\beta,
\end{align}
without summation over $\alpha,\beta$. We also use Eqs.~(\ref{songen}) for $T^a,\tilde{T}^a$ and (\ref{Jexplicit}) to write:
\begin{align}
J^{(r,s)}=i\sum_{\tilde{\alpha}=1}^k\chi_{r\tilde{\alpha}}\chi_{s\tilde{\alpha}},\hspace{5mm}
\tilde{J}^{(\tilde{r},\tilde{s})}=i\sum_{\alpha=1}^N\chi_{\alpha\tilde{r}}\chi_{\alpha\tilde{s}},
\end{align}
with $1\leq r<s\leq N$ and $1\leq\tilde{r}<\tilde{s}\leq k$. We have:
\begin{widetext}
\begin{align}\label{JJN}
\sum_{(r,s)}\colon J^{(r,s)}J^{(r,s)}\colon&=-\sum_{(r,s)}\sum_{\tilde{\alpha}\tilde{\beta}}\colon(\chi_{r\tilde{\alpha}}\chi_{s\tilde{\alpha}})(\chi_{r\tilde{\beta}}\chi_{s\tilde{\beta}})\colon\nn\\
&=-\sum_{r<s}\left(\sum_{\tilde{\alpha}}\colon(\chi_{r\tilde{\alpha}}\chi_{s\tilde{\alpha}})(\chi_{r\tilde{\alpha}}\chi_{s\tilde{\alpha}})\colon
+\sum_{\tilde{\alpha}\neq\tilde{\beta}}\colon\chi_{r\tilde{\alpha}}\chi_{s\tilde{\alpha}}\chi_{r\tilde{\beta}}\chi_{s\tilde{\beta}}\colon\right)\nn\\
&=-\sum_{r<s}\left[\sum_{\tilde{\alpha}}(\chi_{r\tilde{\alpha}}\partial\chi_{r\tilde{\alpha}}+\chi_{s\tilde{\alpha}}\partial\chi_{s\tilde{\alpha}})+2\sum_{\tilde{\alpha}<\tilde{\beta}}\chi_{r\tilde{\alpha}}\chi_{s\tilde{\alpha}}\chi_{r\tilde{\beta}}\chi_{s\tilde{\beta}}\right]\nn\\
&=-\frac{1}{2}\sum_{\tilde{\alpha}}\left[\sum_{rs}(\chi_{r\tilde{\alpha}}\partial\chi_{r\tilde{\alpha}}+\chi_{s\tilde{\alpha}}\partial\chi_{s\tilde{\alpha}})-2\sum_r\chi_{r\tilde{\alpha}}\partial\chi_{r\tilde{\alpha}}\right]-2\sum_{r<s}\sum_{\tilde{\alpha}<\tilde{\beta}}\chi_{r\tilde{\alpha}}\chi_{s\tilde{\alpha}}\chi_{r\tilde{\beta}}\chi_{s\tilde{\beta}}\nn\\
&=-(N-1)\sum_{r\tilde{\alpha}}\chi_{r\tilde{\alpha}}\partial\chi_{r\tilde{\alpha}}-2O_{\chi\chi\chi\chi},
\end{align}
\end{widetext}
where we define the four-fermion operator
\begin{align}
O_{\chi\chi\chi\chi}\equiv\sum_{r<s}\sum_{\tilde{\alpha}<\tilde{\beta}}\chi_{r\tilde{\alpha}}\chi_{s\tilde{\alpha}}\chi_{r\tilde{\beta}}\chi_{s\tilde{\beta}}.
\end{align}
Note that this operator does not need further normal ordering since all fields in the product are different. Similarly, we find:
\begin{align}
\sum_{(\tilde{r},\tilde{s})}\colon J^{(\tilde{r},\tilde{s})}J^{(\tilde{r},\tilde{s})}\colon&=
-(k-1)\sum_{\alpha\tilde{r}}\chi_{\alpha\tilde{r}}\partial\chi_{\alpha\tilde{r}}\nn\\
&\phantom{=}-2\sum_{\alpha<\beta}\sum_{\tilde{r}<\tilde{s}}\chi_{\alpha\tilde{r}}\chi_{\alpha\tilde{s}}\chi_{\beta\tilde{r}}\chi_{\beta\tilde{s}}.
\end{align}
By performing the changes of dummy summation variables $\alpha,\beta\rightarrow r,s$ and $\tilde{r},\tilde{s}\rightarrow\tilde{\alpha},\tilde{\beta}$, we find:
\begin{align}\label{JJk}
\sum_{(\tilde{r},\tilde{s})}\colon J^{(\tilde{r},\tilde{s})}J^{(\tilde{r},\tilde{s})}\colon&=
-(k-1)\sum_{r\tilde{\alpha}}\chi_{r\tilde{\alpha}}\partial\chi_{r\tilde{\alpha}}\nn\\
&\phantom{=}-2\sum_{r<s}\sum_{\tilde{\alpha}<\tilde{\beta}}\chi_{r\tilde{\alpha}}\chi_{r\tilde{\beta}}\chi_{s\tilde{\alpha}}\chi_{s\tilde{\beta}}\nn\\
&=-(k-1)\sum_{r\tilde{\alpha}}\chi_{r\tilde{\alpha}}\partial\chi_{r\tilde{\alpha}}\nn\\
&\phantom{=}+2\sum_{r<s}\sum_{\tilde{\alpha}<\tilde{\beta}}\chi_{r\tilde{\alpha}}\chi_{s\tilde{\alpha}}\chi_{r\tilde{\beta}}\chi_{s\tilde{\beta}}\nn\\
&=-(k-1)\sum_{r\tilde{\alpha}}\chi_{r\tilde{\alpha}}\partial\chi_{r\tilde{\alpha}}+2O_{\chi\chi\chi\chi}.
\end{align}
Based on Eq.~(\ref{sugawara}) with $g=n-2$ for $\mathfrak{g}=\so(n)$, we expect the following Sugawara forms:
\begin{align}
T_{\so(N)_k}(z)&=\frac{1}{2(k+N-2)}\sum_{(r,s)}\colon J^{(r,s)}J^{(r,s)}\colon,\\
T_{\so(k)_N}(z)&=\frac{1}{2(N+k-2)}\sum_{(\tilde{r},\tilde{s})}\colon J^{(\tilde{r},\tilde{s})}J^{(\tilde{r},\tilde{s})}\colon.
\end{align}
Using Eqs.~(\ref{JJN}) and (\ref{JJk}), we thus find:
\begin{align}
T_{\so(N)_k}(z)+T_{\so(k)_N}(z)&=-\frac{1}{2}\sum_{r\tilde{\alpha}}\chi_{r\tilde{\alpha}}\partial\chi_{r\tilde{\alpha}}\nn\\
&=T_{\mathfrak{so}(Nk)_1}(z),
\end{align}
where we see that the four-fermion contributions $\propto O_{\chi\chi\chi\chi}$ cancel. Note that the $\so(N)_k$ and $\so(k)_N$ theories are interacting theories, since their energy-momentum tensors contain four-fermion terms, but their sum is a free theory.

Using Eq.~(\ref{c-}), we can also check that the chiral central charges add:
\begin{align}
c_-[\so(N)_k]&=\frac{kN(N-1)}{2(k+N-2)},\nn\\
c_-[\so(k)_N]&=\frac{Nk(k-1)}{2(N+k-2)},\nn\\
c_-[\so(N)_k]+c_-[\so(k)_N]&=\frac{1}{2}Nk=c_-[\so(Nk)_1].
\end{align}

Finally, Ref.~\cite{antoniadis1986} shows that a theory of $N$ flavors of Majorana fermions $\psi_a^i$ with an internally gauged $SO(k)$ symmetry ($a=1,\ldots,N$, $i=1,\ldots,k$, thus $Nk$ Majorana fermions in total) is equivalent to the $\so(N)_k$ WZW model. This is consistent with projecting out the $\so(k)_N$ sector in the conformal embedding (\ref{embedding}). The $\mathfrak{su}(n)$ analog~\cite{affleck1986,naculich1990} of this embedding was used previously in a similar manner to understand the edge physics of fractional quantum Hall states obtained from a parton construction~\cite{Wenparton1}.


\section{Kitaev-Kekul\'e model}
\label{sec:pgt}

In this Appendix, we give an example of noninteracting Majorana hopping model whose low-energy bandstructure consists of two continuum Majorana fields $\Psi_+,\Psi_-$ with tunable masses $m_+,m_-$~\cite{yang2019,farjami2020,mirmojarabian2020}. We begin with nearest-neighbor Majorana hopping on the honeycomb lattice, which produces two massless Majorana fields at low energies~\cite{kitaev2006}. We then add two perturbations: a second-neighbor hopping term of strength $\kappa$, which gives a Haldane-type mass~\cite{haldane1988,kitaev2006} of the same sign for both Majorana fields, and a Kekul\'e distortion term~\cite{chamon2000,hou2007} of strength $\lambda$, which gives masses of opposite sign for the Majorana fields. By tuning both $\kappa$ and $\lambda$, the low-energy Majorana masses $m_\pm$ resulting from the combined effect of both perturbations can be tuned independently.

The hopping model consists of two terms:
\begin{align}
H=H_\text{Kit}+H_\text{Kek},
\end{align}
where
\begin{align}
\label{eq:kitcomb}
H_\text{Kit}&=\frac{i}{4}\sum_{j,k}A_{jk}c_jc_k,
\end{align}
is the model specified by Eq.~(48) of Ref.~\cite{kitaev2006}, with nearest-neighbor hopping amplitude $J$ and second-neighbor hopping amplitude $\kappa$ for Majorana fermions $c_j$ on the honeycomb lattice. This model gives a topological superconductor with Chern number equal to $\sgn\kappa$. The second term is a spatially non-uniform modulation of the nearest-neighbor hopping amplitude:
\begin{align}
H_\text{Kek}=\frac{i}{4}\sum_{j,k}t_{jk}c_jc_k,
\end{align}
where $t_{jk}$ specifies the Kekul\'e distortion pattern:
\begin{equation}
t_{jk}=\begin{cases}
-\frac{t}{\sqrt{3}}e^{i\b{K}_{+}\cdot\b{\delta}_{n}}e^{i\b{G}\cdot \b{r}_{j}}+\mathrm{c.c.}, & \mathrm{if}\quad \b{r}_{k}\!=\!\b{r}_{j}\!+\!\b{\delta}_{n},\\
0, & \mathrm{otherwise}.
\end{cases}
\end{equation}
Here, $\b{\delta}_{n}$ are the 3 nearest-neighbor vectors on the honeycomb
lattice, $\b{K}_{\pm}\!=\!(\pm4\pi/3,0)$
are the two gapless Dirac points obtained in the limit $\kappa\!=\!t\!=\!0$, and $\b{G}\!=\!\b{K}_+\!-\!\b{K}_-$ is the momentum connecting the two Dirac points.
The complex parameter $t$ is such that $\abs{t}$ controls the strength
of the distortion. This distortion triples the size of the unit cell of the honeycomb
lattice, and thus folds the Brillouin zone three times. The Dirac points are mapped to the $\Gamma$ point of the reduced Brillouin zone. Since there
are six inequivalent sites in the Kekul\'e-distorted lattice, there
are six bands in the bandstructure. When $\kappa$ and $|t|$ are small compared to $J$, the low-energy physics is dominated by two bands with avoided crossings near the $\Gamma$ point. The low-energy degrees of freedom are the spinors 
\begin{equation}
\label{eq:eta}
\eta_{\pm}(\b{k})\!\equiv\!(c_{\b{K}_{\pm}\!+\!\b{k}}^{A}\quad c_{\b{K}_{\pm}\!+\!\b{k}}^{B})^{\intercal},\qquad\eta_{\pm}^{\dagger}(\b{k})\!=\!\eta_{\mp}^{\intercal}(-\b{k}),
\end{equation}
where $A$ and $B$ superscripts indicate the two sublattices of the
honeycomb lattice, and the second equality above is a Majorana condition. Linearizing the low-energy bandstructure near the Dirac ($\Gamma$) point, we obtain~\citep{yang2019}:
\begin{align}
H  \approx\frac{J\sqrt{3}}{4}\int&\frac{\dd^{2}{k}}{(2\pi)^{2}}\biggl[\eta_{+}^{\dagger}(\b{k})\begin{pmatrix}6\kappa/J & -k_{y}\!-\!ik_{x}\\
-k_{y}\!+\!ik_{x} & -6\kappa/J
\end{pmatrix}\eta_{+}(\b{k})\nonumber \\
 & +\eta_{-}^{\dagger}(\b{k})\begin{pmatrix}6\kappa/J & -k_{y}\!-\!ik_{x}\\
-k_{y}\!+\!ik_{x} & -6\kappa/J
\end{pmatrix}\eta_{-}(\b{k})\nonumber \\
 & +\eta_{+}^{\dagger}(\b{k})\begin{pmatrix}0 & 2it/J\\
-2it/J & 0
\end{pmatrix}\eta_{-}(\b{k})\nonumber \\
 & +\eta_{-}^{\dagger}(\b{k})\begin{pmatrix}0 & 2it^{*}/J\\
-2it^{*}/J & 0
\end{pmatrix}\eta_{+}(\b{k})\biggr].\label{eq:latH}
\end{align}
The Kekul\'e distortion couples the gapless excitations from the $\b{K}_{\pm}$
valleys. To diagonalize the Hamiltonian, we define the new spinors
\begin{align}
\label{eq:psimaj}
\Psi_{+}(\b{k}) & \equiv\frac{1}{\sqrt{2}}\bigl(-i\sigma_{z}\eta_{+}(\b{k})+\sigma_{y}\eta_{-}(\b{k})\bigr),\nonumber \\
\Psi_{-}(\b{k}) & \equiv\frac{1}{\sqrt{2}}\bigl(\sigma_{z}\eta_{+}(\b{k})-i\sigma_{y}\eta_{-}(\b{k})\bigr),\nonumber \\
\Psi_{\pm}^{\dagger}(\b{k}) & =\Psi^{\intercal}_\pm(-\b{k})\sigma_{x}.
\end{align}
In this model, the $\pm$ indices in $\Psi_{\pm}$ are no longer valley indices. Indeed, it is obvious from their definition that
the $\Psi_{\pm}$ fermions mix the fermions $\eta_{\pm}$ from the
valleys $\b{K}_{\pm}$. Furthermore, we set the Kekul\'e coupling $t\!=\!i\lambda$,
where $\lambda\!\in\!\mathbb{R}$, thus removing the phase degree of freedom
in the distortion. This diagonalizes in flavor space the linearized Hamiltonian
(\ref{eq:latH}), which is now written as 
\begin{align}
H\!=\!\frac{J\sqrt{3}}{4}\!\sum_{i=\pm}\!\int\!\frac{\dd^{2}{k}}{(2\pi)^{2}}\Psi_{i}^{\dagger}(\b{k})(k_{y}\sigma_{x}-k_{x}\sigma_{y}+m_i\sigma_{z})\Psi_{i}(\b{k}),
\end{align}
where the low-energy Majorana masses
\begin{align}
\label{eq:majmass}
m_\pm=\frac{6\kappa\pm\lambda}{J},
\end{align}
can be tuned independently by the lattice couplings $\kappa$ and $\lambda$. When $m_+$ and $m_-$ are of the same sign (i.e., when $\kappa$ dominates), the system is a topological superconductor with Chern number $\pm 1$; when $m_+$ and $m_-$ are of opposite sign (i.e., when $\lambda$ dominates), the system is a trivial superconductor. Using the Euclidean gamma matrix representation $(\gamma_0,\gamma_1,\gamma_2)=(\sigma_z,\sigma_x,\sigma_y)$, and rescaling the couplings to set the Majorana velocity to unity, we obtain the Euclidean Lagrangian in position space,
\begin{align}
\label{eq:kitlagr}
\mathcal{L}=\frac{1}{4}\sum_{i=\pm}\Psi_i^{\intercal}\c{C}(\slashed{\partial}+m_i)\Psi_i,
\end{align}
whose gauged version appears in Eq.~(\ref{LcritSON}). Here, $\c{C}\!=\!-i\gamma_2$ is a charge-conjugation matrix, and the fermionic fields $\Psi_\pm$ obey the Majorana condition
\begin{equation}\label{eq:majcon1}
\bar{\Psi}_{\pm}\equiv\Psi_{\pm}^{\dagger}\gamma_0=\Psi_{\pm}^{\intercal}\c{C}.
\end{equation}

For general nonzero $\kappa$ and $\lambda$, the masses \eqref{eq:majmass} break the microscopic time-reversal ($\mathcal{T}$) and inversion ($\mathcal{I}$) symmetries. These can be represented on the Majorana fields $\eta_{\pm}$ in Eq.~\eqref{eq:eta} as
\begin{align}
\label{eq:etaTRinv}
\mc{T}\eta_\pm(\bm{k})\mc{T}^{-1} &= -i\sigma_z K \eta_\mp(-\bm{k}),\nonumber\\
\mc{I}\eta_\pm(\bm{k})\mc{I}^{-1} &= i\sigma_y \eta_\mp(-\bm{k}),
\end{align}
where we denote complex conjugation by $K$. These nonstandard transformations deserve further explanation. We recall that the Kitaev honeycomb model \eqref{eq:kitcomb} is in fact a $\mathbb{Z}_2$ gauge theory with static gauge fields $u_{jk}$ that modulate the nearest-neighbor hopping amplitude $J$. The model \eqref{eq:kitcomb} is obtained as the effective Hamiltonian in the ground-state (zero-flux) sector in standard gauge $u_{jk}\!=\!1$, for all $j\!\in\!A,\,k\!\in\!B$. The standard definitions of time-reversal ($c^{A/B}_j\!\to\!Kc^{A/B}_j$) and inversion ($c^{A/B}_{j}\!\to\!c^{B/A}_{-j}$) also flip the sign of $u_{jk}$, and thus do not preserve the standard gauge. However, the sign change of the latter can be compensated by a $\mathbb{Z}_2$ gauge transformation on either the $j$ or $k$ sites. The definitions \eqref{eq:etaTRinv} denote such composite transformations, and are thus projective representations of time reversal and inversion on the Majorana partons. These filter down to the modified spinors $\Psi_\pm(\bm{k})$ in Eq.~\eqref{eq:psimaj} as
\begin{align}
\label{eq:psiTRinv}
\mc{T}\Psi_\pm(\bm{k})\mc{T}^{-1} &= i\sigma_y K \Psi_\mp(-\bm{k}),\nonumber\\
\mc{I}\Psi_\pm(\bm{k})\mc{I}^{-1} &= i\sigma_z \Psi_\mp(-\bm{k}).
\end{align}
Using these transformations on the Lagrangian \eqref{eq:kitlagr}, one observes that the Kekul\'e distortion $\lambda$ provides a $\mc{T}$-invariant mass but breaks $\mc{I}$, whereas the Haldane mass $\lambda$ breaks $\mc{T}$, but preserves $\mc{I}$. Imposing $\mc{I}$, we obtain $m_+=m_-=6\kappa/J$, and tuning $\kappa$ through zero induces a direct continuous transition between the chiral spin liquid and the Ising-ordered phase in Fig.~\ref{fig:phase2}.

\section{Instanton calculus in the background field gauge}
\label{app:instantons}

To perform the instanton gas calculation in this paper, we use a representation of $\mathbb{Z}_2$ monopoles in $SO(N)$ gauge theory as Dirac monopoles in an $SO(2)$ subgroup. This representation breaks the $SO(N)$ invariance down to a $S[O(2)\!\times\!O(N\!-\!2)]\!\equiv\![O(2)\!\times\!O(N\!-\!2)]/\mathbb{Z}_2$ subgroup. This is interpreted as a partial choice of gauge, and na{\"i}vely leads to ZMs in the Faddeev-Popov (FP) determinant. In this Appendix, we employ the background field gauge to show that such ZMs can be removed \cite{thooft1976a,bernard1979,osborn1981,marino2015}, at the cost of introducing ``gauge collective coordinates'', which rotate the Dirac monopole between distinct $SO(2)$ subgroups of $SO(N)$.

We shall begin by formulating and gauge-fixing the $\c{N}$-instanton contribution to the partition function. Formally decomposing the gauge field $A$ into an instanton background $\bar{A}$ and
a fluctuation part $a$,
\begin{equation}
A=\bar{A}+a,
\end{equation}
the associated field strength decomposes to\footnote{Given a gauge group $G$ and a linear representation $\rho\!:\!G\!\to\!\mathrm{Aut}(V)$,
the exterior derivative with respect to a $\mathfrak{g}$-valued connection
$A$ is $d_{A}=d+d\rho(A)\wedge$, where $d\rho$ is the induced representation of $\mathfrak{g}$ on
$V$.} 
\begin{equation}
F=\bar{F}+d_{\bar{A}}a+a\wedge a.
\end{equation}
Defining a gauge-invariant inner product on the space of $\mf{so}(N)$-valued
forms as 
\begin{equation}
\ev{\alpha,\beta}=\frac{1}{2g^{2}}\int\tr(\alpha\wedge\ast\beta),\label{eq:inprod}
\end{equation}
the Yang-Mills action can be decomposed as $S_{\mathrm{YM}}=\bar{S}_{\mathrm{YM}}+S_{a}[\bar{A}]$ where $\bar{S}_{\mathrm{YM}}=\ev{\bar{F},\bar{F}}$ and
\begin{align}
S_{a}[\bar{A}]\equiv\ev{d_{\bar{A}}a,d_{\bar{A}}a}+2\ev{\bar{F},a\wedge a}+\text{\ensuremath{\mathcal{O}(a^{3})}}.
\end{align}
Terms linear in $a$ vanish as $\bar{F}$ satisfies the equations
of motion. The fermion action can be similarly decomposed, 
\begin{equation}
S_{F}=\frac{1}{4}\Psi^{\intercal}C\slashed{a}\Psi+\frac{1}{4}\Psi^{\intercal}C(\slashed{\partial}+\slashed{\bar{A}}+m)\Psi.
\end{equation}
A single fermion $\Psi$ is considered here, but the derivation is straightforwardly generalized to the case of multiple fermion flavors relevant for the main text.

The net action is invariant under the infinitesimal gauge transformation
\begin{align}
\bar{A}+a & \to\bar{A}+a+d_{\bar{A}}\omega+[a,\omega],\nonumber \\
\Psi & \to e^{-\omega}\Psi=\Psi-\omega\Psi.
\end{align}
The fermions will just spectate in the following discussion, and so
will not be discussed further. Since $\bar{A}$ is a classical background
field (not integrated over in the path integral), a true gauge transformation
must act only on the fluctuation $a$, so that 
\begin{align}
\label{eq:truegt}
\delta_\omega a & =d_{\bar{A}}\omega+[a,\omega],\nonumber\\
\delta_\omega \bar{A} & =0.
\end{align}
However, it is useful to define a ``pseudo'' gauge transformation
\begin{align}
\delta_{\mathrm{pseudo}}\bar{A} & =d_{\bar{A}}\omega,\nonumber \\
\delta_{\mathrm{pseudo}}a & =[a,\omega],\label{eq:pseudogt}
\end{align}
under which the action remains invariant. As far as the parent theory
with $A\!=\!\bar{A}\!+\!a$ is concerned, the pseudo and true gauge
transformations are identical. We shall shortly see that the background
field method is a clever choice of gauge that retains invariance under
the pseudo gauge transformations \eqref{eq:pseudogt} while gauge-fixing the fluctuation part of the path
integral 
\begin{equation}
Z=e^{-\bar{S}_{\mathrm{YM}}}\int\mathscr{D}\Psi\mathscr{D}a\,e^{-S_{a}[\bar{A}]-S_{F}[\bar{A},a]}.
\end{equation}
To gauge-fix this path integral, we select an $\mathfrak{so}(N)$-valued gauge function $G(a)$ and employ the FP method
by inserting into $Z$ the identity in the form $\Delta_{\mathrm{FP}}^{-1}\Delta_{\mathrm{FP}}$,
where the gauge-invariant FP determinant is defined
as the inverse of 
\begin{align}
\Delta_{\mathrm{FP}}^{-1} & =\int\mathscr{D}\omega\,\delta[G(a+\delta_{\omega}a)],\nonumber \\
 & =\int\mathscr{D}G\,\delta[G(a+\delta_{\omega}a)]\,\abs{\det\frac{\delta G}{\delta\omega}}^{-1},\nonumber \\
 & =\abs{\det\frac{\delta G}{\delta\omega}}_{G=0}^{-1}.\label{eq:fpinv}
\end{align}
We will choose the background field gauge,
\begin{equation}
G(a)=\bar{D}_{\mu}a_{\mu}=\partial_\mu a_\mu + [\bar{A}_\mu,a_\mu]=0,
\end{equation}
where $\bar{D}_\mu$ is the gauge covariant derivative with respect to the instanton field $\bar{A}$.
The reason for this choice is that the gauge function will eventually
feature in a gauge-fixing term $\tr G^{2}$ in the Lagrangian, and
it is easy to show that such a term is invariant under the pseudo
gauge transformation (\ref{eq:pseudogt}), but \emph{not }under a
true gauge transformation (\ref{eq:truegt}) of the fluctuation field
$a$. In this manner, the gauge invariance of the parent theory with
$A\!=\!\bar{A}\!+\!a$ is retained.

The FP determinant is easily evaluated to be
\begin{equation}
\det\frac{\delta G}{\delta\omega}=\mathrm{det}_{0}\bar{D}_{\mu}D_{\mu},\label{eq:FPdet}
\end{equation}
where the subscript $0$ indicates that the determinant is to be evaluated
in the space of $\mf{so}(N)$-valued 0-forms $\omega(x)$, and the (bar-less) covariant derivative $D_\mu$ is with respect to the total field $A\!=\!\bar{A}\!+\!a$, so that $D_\mu \omega\!=\!\partial_\mu\omega\!+\![\bar{A}_\mu\!+\!a_\mu,\omega]$. If $\bar{A}\!=\!0$,
then this reduces to the familiar result for Lorenz gauge. A good
gauge function must satisfy $G(a\!+\!\delta_{\omega}a)\!\neq\!G(a)$
for any $\omega\!\neq\!0$, so that the gauge slice $G(a)\!=\!0$ contains
only inequivalent configurations of $a$. If there exists an $\omega$
that violates this requirement, then this would result in a ZM
contribution to (\ref{eq:FPdet}); these can be interpreted as would-be FP ghost ZMs [see Eq.~(\ref{Sgh})]. To see this explicitly, note that
the FP operator is
\begin{align}
\bar{D}_{\mu}D_{\mu}\omega & =\bar{D}_{\mu}^{2}\omega+[\bar{D}_{\mu}a_{\mu},\omega]+[a_{\mu},\bar{D}_{\mu}\omega],\nonumber \\
 & =\bar{D}_{\mu}^{2}\omega+[G,\omega]+[a_{\mu},\bar{D}_{\mu}\omega].
\end{align}
If $\bar{D}_{\mu}\omega\!=\!0$, then $\omega$ is a ZM of
the FP operator evaluated on the gauge slice $G(a)\!=\!0$ {[}see Eq.\,(\ref{eq:fpinv}){]}.
Noting that $\bar{D}_{\mu}\omega\!=\!0$ infinitesimally means $e^{-\omega}(\bar{A}+d)e^{\omega}=\bar{A}$, we find that ZMs exist if there is a nontrivial stabilizer
(denoted $\mathcal{H}$) of $\bar{A}$ in the group of gauge transformations
$\mathcal{G}$.\footnote{If $M$ is spacetime and $G\!=\!SO(N)$ the gauge group, the group $\mathcal{G}\!:\!M\!\to\!G$ of gauge transformations acts as $\mathcal{G}\!:\!x\!\mapsto\!\exp(\omega(x))$.} For instance, if $\bar{A}\!=\!0$ then the stabilizer consists of
all global gauge transformations so that $\mathcal{H}\!=\!SO(N)$, the
gauge group. Here, $\bar{A}$ is the instanton embedded in an
$SO(2)$ subgroup, which has a stabilizer subgroup of global rotations
in $S[O(2)\!\times\!O(N\!-\!2)]$.

Due to the presence of ZMs in the FP determinant, one must split the
domain of the $\omega$ integral in Eq.\,(\ref{eq:fpinv}) into a
ZM space consisting of all $\omega\!\in\!\ker_{0}\bar{D}_{\mu}$,
and its orthogonal complement $\ker_{0}\bar{D}_{\mu}^{\perp}$, where
the subscript 0 indicates that the domain of $\bar{D}_{\mu}$ is restricted
to 0-forms. Such a grading can be achieved by means of the inner product
(\ref{eq:inprod}) defined on this space. Then any gauge transformation
can be decomposed as
\begin{equation}
\omega=\phi+\lambda,\quad\phi\!\in\!\ker_{0}\bar{D}_{\mu},\quad\lambda\!\in\!\ker_{0}\bar{D}_{\mu}^{\perp}.
\end{equation}
 This also means that 
\begin{align}
\Delta_{\mathrm{FP}}^{-1} & =\int\mathscr{D}\omega\,\delta[G(a+\delta_{\omega}a)],\nonumber \\
 & =\int_{\ker_{0}\bar{D}_{\mu}}\mathscr{D}\phi\int_{\ker_{0}\bar{D}_{\mu}^{\perp}}\mathscr{D}\lambda\delta[G(a+\delta_{\lambda}a)],\nonumber \\
 & =\mathrm{vol}(\mathcal{H})\abs{\det\nolimits'_{0}\bar{D}_{\mu}D_{\mu}}^{-1},
\end{align}
where the prime indicates that the determinant of $\bar{D}_{\mu}D_{\mu}$
is evaluated in the space $\ker_{0}\bar{D}_{\mu}^{\perp}$, with the
ZMs removed. The gauge-fixed path integral is then,
\begin{align}
Z & =\frac{e^{-\bar{S}_{\mathrm{YM}}}}{\mathrm{vol}(\mathcal{H})}\!\int\mathscr{D}a\,\mathscr{D}\Psi\int\mathscr{D}\omega\nonumber \\
 & \qquad\times\delta[G(a+\delta_{\omega}a)]\abs{\det\nolimits'_{0}\bar{D}_{\mu}D_{\mu}}e^{-S_{a}[\bar{A}]-S_{F}[\bar{A},a]},\nonumber \\
 & =\left(\int\mathscr{D}\omega\right)\frac{e^{-\bar{S}_{\mathrm{YM}}}}{\mathrm{vol}(\mathcal{H})}\!\int\mathscr{D}a\mathscr{D}\Psi\nonumber \\
 & \qquad\times\delta[G(a)]\abs{\det\nolimits'_{0}\bar{D}_{\mu}D_{\mu}}\,e^{-S_{a}[\bar{A}]-S_{F}[\bar{A},a]},
\end{align}
where the second line is obtained on a gauge transformation by $-\omega$, keeping all gauge invariant quantities fixed. The integral over
$\omega$ is an infinite constant that can be dropped by defining
a suitable normalization.

The fluctuation integral is subject to the gauge condition $G(a)\!=\!\bar{D}_{\mu}a_{\mu}\!=\!0$.
Once again, the inner product (\ref{eq:inprod}) can be used to split
the space of $\mf{so}(N)$-valued 1-forms into $\ker_{1}\bar{D}_{\mu}$
and its orthogonal complement $\ker_{1}\bar{D}_{\mu}^{\perp}$, where
the subscript 1 now indicates that the domain of $\bar{D}_{\mu}$
is the space of 1-forms. Using the inner product (\ref{eq:inprod}),
it is readily seen (using integration by parts) that any nontrivial
element of $\ker_{1}\bar{D}_{\mu}^{\perp}$ is of the form $\bar{D}_{\mu}\varphi$
for some 0-form $\varphi\!\in\!\ker_{0}\bar{D}_{\mu}^{\perp}$. Then,
\begin{equation}
a_{\mu}=\alpha_{\mu}+\bar{D}_{\mu}\varphi,\quad\alpha_{\mu}\!\in\!\ker_{1}\bar{D}_{\mu},\quad\varphi\!\in\!\ker_{0}\bar{D}_{\mu}^{\perp}.
\end{equation}
A change of variables from $a$ to $\alpha$ and $\varphi$ in the
path integral now has a nontrivial Jacobian, found by examining
the metric in this functional space, 
\onecolumngrid
\begin{align}
\norm{a_{\mu}}^{2}  =\ev{\alpha_{\mu}+\bar{D}_{\mu}\varphi,\alpha_{\mu}+\bar{D}_{\mu}\varphi}
  =\langle \begin{pmatrix}\alpha_{\mu} & \varphi\end{pmatrix}\begin{pmatrix}1 & 0\\
0 & -\bar{D}_{\mu}^{2}
\end{pmatrix}\begin{pmatrix}\alpha_{\mu}\\
\varphi
\end{pmatrix}\rangle.
\end{align}
The Jacobian is thus $[\det'_{0}(-\bar{D}_{\mu}^{2})]^{1/2}$, where
the operator acts on $\varphi\!\in\!\ker_{0}\bar{D}_{\mu}^{\perp}$,
so there are no ZMs in this determinant. Since $-\bar{D}_{\mu}^{2}$ is a positive-definite operator on $\ker_{0}\bar{D}_{\mu}^{\perp}$, its square root is well defined, and an absolute value sign is
redundant. The path integral then simplifies to
\begin{align}
Z & =\frac{e^{-\bar{S}_{\mathrm{YM}}}}{\mathrm{vol}(\mathcal{H})}\!\int_{\ker_{1}\bar{D}_{\mu}}\!\mathscr{D}\alpha\!\int_{\ker_{0}\bar{D}_{\mu}^{\perp}}\!\mathscr{D}\varphi\,\delta[\bar{D}_{\mu}^{2}\varphi]\sqrt{\mathrm{det}'_{0}(\!-\!\bar{D}_{\mu}^{2})}\abs{\mathrm{det}'_{0}\bar{D}_{\mu}D_{\mu}}\int\mathscr{D}\Psi e^{-S_{\alpha\!+\!\bar{D}_{\mu}\varphi}[\bar{A}]\!-\!S_{F}[\bar{A},\alpha\!+\!\bar{D}_{\mu}\varphi]},\nonumber \\
 & =\frac{e^{-\bar{S}_{\mathrm{YM}}}}{\mathrm{vol}(\mathcal{H})}\!\int_{\ker_{1}\bar{D}_{\mu}}\mathscr{D}\alpha\,[\mathrm{det}'_{0}(-\bar{D}_{\mu}^{2})]^{-1/2}\abs{\mathrm{det}'_{0}\bar{D}_{\mu}D_{\mu}}\int\mathscr{D}\Psi e^{-S_{\alpha}[\bar{A}]-S_{F}[\bar{A},\alpha]},\label{eq:Zgen}
\end{align}
where the second line is obtained by using $\delta[\bar{D}_{\mu}^{2}\varphi]\!=\![\det_{0}(-\bar{D}_{\mu}^{2})]^{-1}\delta[\varphi]$, and performing the $\varphi$ integral.

The path integral (\ref{eq:Zgen}) has actually been derived for a general
background $\bar{A}$. We will now specialize to the case of instantons
in $SO(N)$ gauge theory, and evaluate an $\c{N}$-instanton contribution
to the path integral, such as appears in Eq.~(\ref{eq:fullZ}). Each instanton incarnates as a Dirac monopole
in some $SO(2)$ subgroup, and the background $\bar{A}$ is a simple
sum of the single instanton 1-form (\ref{eq:wuyang}) in the dilute
gas approximation. Within such an approximation, the stabilizer
for such an instanton configuration on a spacetime $M$ is simply
$\mathcal{H}^\c{N}$, where the single instanton stabilizer is $\mathcal{H}\!:\!M\!\to\!(H\!=\!S[O(2)\!\times\!O(N\!-\!2)])$.
Writing a general element of $\mathcal{H}$ as $\exp[-\phi^{a}(x)h^{a}]$,
where $a\!\in\!\{1,...,\dim H\}$,
\begin{align}
\mathrm{vol}(\mathcal{H})  =\int_{\ker_{0}\bar{D}_{\mu}}\mathscr{D}\phi(x)
  =\int_{H}\prod_{a=1}^{\dim H}d\phi^{a}\sqrt{\frac{\mathrm{vol}M}{2g^{2}}}
  =\left(\frac{\mathrm{vol}M}{2g^{2}}\right)^{\dim H/2}\mathrm{vol}(H).
\end{align}

The determinant $[\mathrm{det}'_{0}(-\bar{D}_{\mu}^{2})]^{-1/2}$
appearing in the general path integral (\ref{eq:Zgen}), deviates
from $[\mathrm{det}'_{0}(-\partial_{\mu}^{2})]^{-1/2}$ pertinent
to a trivial background $\bar{A}\!=\!0$, only in small (disjoint)
neighborhoods of the $\c{N}$ localized instantons. The $\c{N}$-instanton correction to the trivial
determinant can be defined via $[\mathrm{det}'_{0}(-\bar{D}_{\mu}^{2})]^{-1/2}\equiv[\mathrm{det}'_{0}(-\partial_{\mu}^{2})]^{-1/2}K^{\c{N}}$~\cite{coleman1985}. Normalizing the path integral against the trivial background $\bar{A}\!=\!0$, the $\c{N}$-instanton contribution to the partition function is
\begin{equation}
\frac{Z_{\c{N}}}{Z_{0}}\!=\!\mathrm{vol}(G/H)^{\c{N}}\left(\frac{\mathrm{vol}M}{2g^{2}}\right)^{\frac{\c{N}\dim(G/H)}{2}}e^{-\bar{S}_{\mathrm{YM}}}\int_{\ker_{1}\bar{D}_{\mu}}\mathscr{D}\alpha\int_{\ker_{0}\bar{D}_{\mu}^{\perp}}\mathscr{D}(\bar{\eta},\eta)\int\mathscr{D}\Psi\frac{K^{\c{N}}}{\mathcal{K}}e^{\!-\!S_{\alpha}[\bar{A}]\!-\!S_{F}[\bar{A},\alpha]\!-\!S_{\mathrm{gh}}[\bar{A},\alpha]},
\end{equation}
where the action for the FP ghost fields $\eta,\bar{\eta}$ is
\begin{equation}\label{Sgh}
S_{\mathrm{gh}}=\int\dd^{D}{x}\tr\bar{\eta}\bar{D}_{\mu}D_{\mu}\eta,
\end{equation}
and the normalization $\mathcal{K}$ is the transverse mode $(\alpha)$
and ghost path integrals evaluated in the trivial background $\bar{A}\!=\!0$. 

As stated in the introduction to this Appendix, the coset space $G/H\!=\!SO(N)/S[O(2)\!\times\!O(N\!-\!2)]$ has a
collective coordinate interpretation. It is the space of global rotations
that move an instanton between distinct $SO(2)$ subgroups of $SO(N)$.
Furthermore, the fact that the $\alpha$ integral is restricted to
$\ker_{1}\bar{D}_{\mu}$ means that such global rotations that change
$\bar{A}$ are excluded from that path integral. However, there are
still ZMs corresponding to other non-gauge collective coordinates,
such as a translation of an instanton in spacetime. For the monopole-instanton considered here, it is clear that there are no other collective
coordinates besides these. Explicitly separating out the collective
coordinates $\{z_i\}$ corresponding to the locations of the instantons (for
which the Jacobian is a trivial constant that can be absorbed into $\mathcal{K}$),
the final result is
\begin{align}
\frac{Z_{\c{N}}}{Z_{0}}&=\mathrm{vol}(G/H)^{\c{N}}\left(\frac{\mathrm{vol}M}{2g^{2}}\right)^{\frac{\c{N}\dim(G/H)}{2}}\nn\\
&\hspace{20mm}\times e^{-\bar{S}_{\mathrm{YM}}}\int\left(\prod_{i=1}^{\c{N}}\dd{z_{i}}\right)\int_{\ker_{1}\bar{D}_{\mu}}\mathscr{D}'\alpha\int_{\ker_{0}\bar{D}_{\mu}^{\perp}}\mathscr{D}(\bar{\eta},\eta)\int\mathscr{D}\Psi\frac{K^{\c{N}}}{\mathcal{K}}e^{\!-\!S_{\alpha}[\bar{A}]\!-\!S_{F}[\bar{A},\alpha]\!-\!S_{\mathrm{gh}}[\bar{A},\alpha]},\label{eq:Zfin}
\end{align}
where the primed measure $\mathscr{D}'\alpha$ means that ZM
solutions of $\alpha$ are excluded from the domain of integration.
More precisely, if the Gaussian part of the fluctuation action is $S_{\alpha}\!=\!\ev{\alpha,\Omega\alpha}$,
then ZMs of the operator $\Omega$ are to be discarded in a mode expansion
of $\alpha$. Therefore, the only ZMs still present in the path integral are those of fermions bound to the instantons, which have physical consequences for symmetry breaking.
\twocolumngrid

\bibliography{parton.bib}

\begin{thebibliography}{124}%
\makeatletter
\providecommand \@ifxundefined [1]{%
 \@ifx{#1\undefined}
}%
\providecommand \@ifnum [1]{%
 \ifnum #1\expandafter \@firstoftwo
 \else \expandafter \@secondoftwo
 \fi
}%
\providecommand \@ifx [1]{%
 \ifx #1\expandafter \@firstoftwo
 \else \expandafter \@secondoftwo
 \fi
}%
\providecommand \natexlab [1]{#1}%
\providecommand \enquote  [1]{``#1''}%
\providecommand \bibnamefont  [1]{#1}%
\providecommand \bibfnamefont [1]{#1}%
\providecommand \citenamefont [1]{#1}%
\providecommand \href@noop [0]{\@secondoftwo}%
\providecommand \href [0]{\begingroup \@sanitize@url \@href}%
\providecommand \@href[1]{\@@startlink{#1}\@@href}%
\providecommand \@@href[1]{\endgroup#1\@@endlink}%
\providecommand \@sanitize@url [0]{\catcode `\\12\catcode `\$12\catcode
  `\&12\catcode `\#12\catcode `\^12\catcode `\_12\catcode `\%12\relax}%
\providecommand \@@startlink[1]{}%
\providecommand \@@endlink[0]{}%
\providecommand \url  [0]{\begingroup\@sanitize@url \@url }%
\providecommand \@url [1]{\endgroup\@href {#1}{\urlprefix }}%
\providecommand \urlprefix  [0]{URL }%
\providecommand \Eprint [0]{\href }%
\providecommand \doibase [0]{http://dx.doi.org/}%
\providecommand \selectlanguage [0]{\@gobble}%
\providecommand \bibinfo  [0]{\@secondoftwo}%
\providecommand \bibfield  [0]{\@secondoftwo}%
\providecommand \translation [1]{[#1]}%
\providecommand \BibitemOpen [0]{}%
\providecommand \bibitemStop [0]{}%
\providecommand \bibitemNoStop [0]{.\EOS\space}%
\providecommand \EOS [0]{\spacefactor3000\relax}%
\providecommand \BibitemShut  [1]{\csname bibitem#1\endcsname}%
\let\auto@bib@innerbib\@empty
\bibitem [{\citenamefont {Savary}\ and\ \citenamefont
  {Balents}(2017)}]{savary2017}%
  \BibitemOpen
  \bibfield  {author} {\bibinfo {author} {\bibfnamefont {L.}~\bibnamefont
  {Savary}}\ and\ \bibinfo {author} {\bibfnamefont {L.}~\bibnamefont
  {Balents}},\ }\href {\doibase 10.1088/0034-4885/80/1/016502} {\bibfield
  {journal} {\bibinfo  {journal} {Rep. Prog. Phys.}\ }\textbf {\bibinfo
  {volume} {80}},\ \bibinfo {pages} {016502} (\bibinfo {year}
  {2017})}\BibitemShut {NoStop}%
\bibitem [{\citenamefont {Zhou}\ \emph {et~al.}(2017)\citenamefont {Zhou},
  \citenamefont {Kanoda},\ and\ \citenamefont {Ng}}]{zhou2017}%
  \BibitemOpen
  \bibfield  {author} {\bibinfo {author} {\bibfnamefont {Y.}~\bibnamefont
  {Zhou}}, \bibinfo {author} {\bibfnamefont {K.}~\bibnamefont {Kanoda}}, \ and\
  \bibinfo {author} {\bibfnamefont {T.-K.}\ \bibnamefont {Ng}},\ }\href
  {\doibase 10.1103/RevModPhys.89.025003} {\bibfield  {journal} {\bibinfo
  {journal} {Rev. Mod. Phys.}\ }\textbf {\bibinfo {volume} {89}},\ \bibinfo
  {pages} {025003} (\bibinfo {year} {2017})}\BibitemShut {NoStop}%
\bibitem [{\citenamefont {Vojta}(2018)}]{vojta2018}%
  \BibitemOpen
  \bibfield  {author} {\bibinfo {author} {\bibfnamefont {M.}~\bibnamefont
  {Vojta}},\ }\href {\doibase 10.1088/1361-6633/aab6be} {\bibfield  {journal}
  {\bibinfo  {journal} {Rep. Prog. Phys.}\ }\textbf {\bibinfo {volume} {81}},\
  \bibinfo {pages} {064501} (\bibinfo {year} {2018})}\BibitemShut {NoStop}%
\bibitem [{\citenamefont {Wen}(2007)}]{WenBook}%
  \BibitemOpen
  \bibfield  {author} {\bibinfo {author} {\bibfnamefont {X.-G.}\ \bibnamefont
  {Wen}},\ }\href {\doibase 10.1093/acprof:oso/9780199227259.001.0001} {\emph
  {\bibinfo {title} {Quantum {Field} {Theory} of {Many}-{Body} {Systems}:
  {From} the {Origin} of {Sound} to an {Origin} of {Light} and {Electrons}}}},\
  Oxford {Graduate} {Texts}\ (\bibinfo  {publisher} {Oxford University Press},\
  \bibinfo {address} {Oxford},\ \bibinfo {year} {2007})\BibitemShut {NoStop}%
\bibitem [{\citenamefont {Read}\ and\ \citenamefont
  {Sachdev}(1991)}]{read1991}%
  \BibitemOpen
  \bibfield  {author} {\bibinfo {author} {\bibfnamefont {N.}~\bibnamefont
  {Read}}\ and\ \bibinfo {author} {\bibfnamefont {S.}~\bibnamefont {Sachdev}},\
  }\href {\doibase 10.1103/PhysRevLett.66.1773} {\bibfield  {journal} {\bibinfo
   {journal} {Phys. Rev. Lett.}\ }\textbf {\bibinfo {volume} {66}},\ \bibinfo
  {pages} {1773} (\bibinfo {year} {1991})}\BibitemShut {NoStop}%
\bibitem [{\citenamefont {Xu}\ and\ \citenamefont {Sachdev}(2009)}]{xu2009}%
  \BibitemOpen
  \bibfield  {author} {\bibinfo {author} {\bibfnamefont {C.}~\bibnamefont
  {Xu}}\ and\ \bibinfo {author} {\bibfnamefont {S.}~\bibnamefont {Sachdev}},\
  }\href {\doibase 10.1103/PhysRevB.79.064405} {\bibfield  {journal} {\bibinfo
  {journal} {Phys. Rev. B}\ }\textbf {\bibinfo {volume} {79}},\ \bibinfo
  {pages} {064405} (\bibinfo {year} {2009})}\BibitemShut {NoStop}%
\bibitem [{\citenamefont {Hermele}\ \emph {et~al.}(2005)\citenamefont
  {Hermele}, \citenamefont {Senthil},\ and\ \citenamefont
  {Fisher}}]{hermele2005}%
  \BibitemOpen
  \bibfield  {author} {\bibinfo {author} {\bibfnamefont {M.}~\bibnamefont
  {Hermele}}, \bibinfo {author} {\bibfnamefont {T.}~\bibnamefont {Senthil}}, \
  and\ \bibinfo {author} {\bibfnamefont {M.~P.~A.}\ \bibnamefont {Fisher}},\
  }\href {\doibase 10.1103/PhysRevB.72.104404} {\bibfield  {journal} {\bibinfo
  {journal} {Phys. Rev. B}\ }\textbf {\bibinfo {volume} {72}},\ \bibinfo
  {pages} {104404} (\bibinfo {year} {2005})}\BibitemShut {NoStop}%
\bibitem [{\citenamefont {Ghaemi}\ and\ \citenamefont
  {Senthil}(2006)}]{ghaemi2006}%
  \BibitemOpen
  \bibfield  {author} {\bibinfo {author} {\bibfnamefont {P.}~\bibnamefont
  {Ghaemi}}\ and\ \bibinfo {author} {\bibfnamefont {T.}~\bibnamefont
  {Senthil}},\ }\href {\doibase 10.1103/PhysRevB.73.054415} {\bibfield
  {journal} {\bibinfo  {journal} {Phys. Rev. B}\ }\textbf {\bibinfo {volume}
  {73}},\ \bibinfo {pages} {054415} (\bibinfo {year} {2006})}\BibitemShut
  {NoStop}%
\bibitem [{\citenamefont {Lu}\ \emph {et~al.}(2017)\citenamefont {Lu},
  \citenamefont {Cho},\ and\ \citenamefont {Vishwanath}}]{lu2017}%
  \BibitemOpen
  \bibfield  {author} {\bibinfo {author} {\bibfnamefont {Y.-M.}\ \bibnamefont
  {Lu}}, \bibinfo {author} {\bibfnamefont {G.~Y.}\ \bibnamefont {Cho}}, \ and\
  \bibinfo {author} {\bibfnamefont {A.}~\bibnamefont {Vishwanath}},\ }\href
  {\doibase 10.1103/PhysRevB.96.205150} {\bibfield  {journal} {\bibinfo
  {journal} {Phys. Rev. B}\ }\textbf {\bibinfo {volume} {96}},\ \bibinfo
  {pages} {205150} (\bibinfo {year} {2017})}\BibitemShut {NoStop}%
\bibitem [{\citenamefont {Song}\ \emph {et~al.}(2019)\citenamefont {Song},
  \citenamefont {Wang}, \citenamefont {Vishwanath},\ and\ \citenamefont
  {He}}]{song2019}%
  \BibitemOpen
  \bibfield  {author} {\bibinfo {author} {\bibfnamefont {X.-Y.}\ \bibnamefont
  {Song}}, \bibinfo {author} {\bibfnamefont {C.}~\bibnamefont {Wang}}, \bibinfo
  {author} {\bibfnamefont {A.}~\bibnamefont {Vishwanath}}, \ and\ \bibinfo
  {author} {\bibfnamefont {Y.-C.}\ \bibnamefont {He}},\ }\href {\doibase
  10.1038/s41467-019-11727-3} {\bibfield  {journal} {\bibinfo  {journal} {Nat.
  Commun.}\ }\textbf {\bibinfo {volume} {10}},\ \bibinfo {pages} {1} (\bibinfo
  {year} {2019})}\BibitemShut {NoStop}%
\bibitem [{\citenamefont {Dupuis}\ \emph {et~al.}(2019)\citenamefont {Dupuis},
  \citenamefont {Paranjape},\ and\ \citenamefont
  {Witczak-Krempa}}]{dupuis2019}%
  \BibitemOpen
  \bibfield  {author} {\bibinfo {author} {\bibfnamefont {{\'E}.}~\bibnamefont
  {Dupuis}}, \bibinfo {author} {\bibfnamefont {M.~B.}\ \bibnamefont
  {Paranjape}}, \ and\ \bibinfo {author} {\bibfnamefont {W.}~\bibnamefont
  {Witczak-Krempa}},\ }\href {\doibase 10.1103/PhysRevB.100.094443} {\bibfield
  {journal} {\bibinfo  {journal} {Phys. Rev. B}\ }\textbf {\bibinfo {volume}
  {100}},\ \bibinfo {pages} {094443} (\bibinfo {year} {2019})}\BibitemShut
  {NoStop}%
\bibitem [{\citenamefont {Zerf}\ \emph {et~al.}(2019)\citenamefont {Zerf},
  \citenamefont {Boyack}, \citenamefont {Marquard}, \citenamefont {Gracey},\
  and\ \citenamefont {Maciejko}}]{zerf2019}%
  \BibitemOpen
  \bibfield  {author} {\bibinfo {author} {\bibfnamefont {N.}~\bibnamefont
  {Zerf}}, \bibinfo {author} {\bibfnamefont {R.}~\bibnamefont {Boyack}},
  \bibinfo {author} {\bibfnamefont {P.}~\bibnamefont {Marquard}}, \bibinfo
  {author} {\bibfnamefont {J.~A.}\ \bibnamefont {Gracey}}, \ and\ \bibinfo
  {author} {\bibfnamefont {J.}~\bibnamefont {Maciejko}},\ }\href {\doibase
  10.1103/PhysRevB.100.235130} {\bibfield  {journal} {\bibinfo  {journal}
  {Phys. Rev. B}\ }\textbf {\bibinfo {volume} {100}},\ \bibinfo {pages}
  {235130} (\bibinfo {year} {2019})}\BibitemShut {NoStop}%
\bibitem [{\citenamefont {Zerf}\ \emph {et~al.}(2020)\citenamefont {Zerf},
  \citenamefont {Boyack}, \citenamefont {Marquard}, \citenamefont {Gracey},\
  and\ \citenamefont {Maciejko}}]{zerf2020}%
  \BibitemOpen
  \bibfield  {author} {\bibinfo {author} {\bibfnamefont {N.}~\bibnamefont
  {Zerf}}, \bibinfo {author} {\bibfnamefont {R.}~\bibnamefont {Boyack}},
  \bibinfo {author} {\bibfnamefont {P.}~\bibnamefont {Marquard}}, \bibinfo
  {author} {\bibfnamefont {J.~A.}\ \bibnamefont {Gracey}}, \ and\ \bibinfo
  {author} {\bibfnamefont {J.}~\bibnamefont {Maciejko}},\ }\href {\doibase
  10.1103/PhysRevD.101.094505} {\bibfield  {journal} {\bibinfo  {journal}
  {Phys. Rev. D}\ }\textbf {\bibinfo {volume} {101}},\ \bibinfo {pages}
  {094505} (\bibinfo {year} {2020})}\BibitemShut {NoStop}%
\bibitem [{\citenamefont {Janssen}\ \emph {et~al.}(2020)\citenamefont
  {Janssen}, \citenamefont {Wang}, \citenamefont {Scherer}, \citenamefont
  {Meng},\ and\ \citenamefont {Xu}}]{janssen2020}%
  \BibitemOpen
  \bibfield  {author} {\bibinfo {author} {\bibfnamefont {L.}~\bibnamefont
  {Janssen}}, \bibinfo {author} {\bibfnamefont {W.}~\bibnamefont {Wang}},
  \bibinfo {author} {\bibfnamefont {M.~M.}\ \bibnamefont {Scherer}}, \bibinfo
  {author} {\bibfnamefont {Z.~Y.}\ \bibnamefont {Meng}}, \ and\ \bibinfo
  {author} {\bibfnamefont {X.~Y.}\ \bibnamefont {Xu}},\ }\href {\doibase
  10.1103/PhysRevB.101.235118} {\bibfield  {journal} {\bibinfo  {journal}
  {Phys. Rev. B}\ }\textbf {\bibinfo {volume} {101}},\ \bibinfo {pages}
  {235118} (\bibinfo {year} {2020})}\BibitemShut {NoStop}%
\bibitem [{\citenamefont {{Kr\"uger}}\ and\ \citenamefont
  {Janssen}(2021)}]{kruger2021}%
  \BibitemOpen
  \bibfield  {author} {\bibinfo {author} {\bibfnamefont {W.~G.~F.}\
  \bibnamefont {{Kr\"uger}}}\ and\ \bibinfo {author} {\bibfnamefont
  {L.}~\bibnamefont {Janssen}},\ }\href {\doibase 10.1103/PhysRevB.104.165133}
  {\bibfield  {journal} {\bibinfo  {journal} {Phys. Rev. B}\ }\textbf {\bibinfo
  {volume} {104}},\ \bibinfo {pages} {165133} (\bibinfo {year}
  {2021})}\BibitemShut {NoStop}%
\bibitem [{\citenamefont {Polyakov}(1975)}]{polyakov1975}%
  \BibitemOpen
  \bibfield  {author} {\bibinfo {author} {\bibfnamefont {A.~M.}\ \bibnamefont
  {Polyakov}},\ }\href {\doibase 10.1016/0370-2693(75)90162-8} {\bibfield
  {journal} {\bibinfo  {journal} {Phys. Lett. B}\ }\textbf {\bibinfo {volume}
  {59}},\ \bibinfo {pages} {82} (\bibinfo {year} {1975})}\BibitemShut {NoStop}%
\bibitem [{\citenamefont {Polyakov}(1977)}]{polyakov1977}%
  \BibitemOpen
  \bibfield  {author} {\bibinfo {author} {\bibfnamefont {A.~M.}\ \bibnamefont
  {Polyakov}},\ }\href {\doibase 10.1016/0550-3213(77)90086-4} {\bibfield
  {journal} {\bibinfo  {journal} {Nucl. Phys. B}\ }\textbf {\bibinfo {volume}
  {120}},\ \bibinfo {pages} {429} (\bibinfo {year} {1977})}\BibitemShut
  {NoStop}%
\bibitem [{\citenamefont {Polyakov}(1987)}]{polyakov1987}%
  \BibitemOpen
  \bibfield  {author} {\bibinfo {author} {\bibfnamefont {A.~M.}\ \bibnamefont
  {Polyakov}},\ }\href {\doibase 10.1201/9780203755082} {\emph {\bibinfo
  {title} {Gauge {Fields} and {Strings}}}}\ (\bibinfo  {publisher} {Taylor \&
  Francis},\ \bibinfo {address} {London},\ \bibinfo {year} {1987})\BibitemShut
  {NoStop}%
\bibitem [{\citenamefont {Alicea}(2008)}]{alicea2008}%
  \BibitemOpen
  \bibfield  {author} {\bibinfo {author} {\bibfnamefont {J.}~\bibnamefont
  {Alicea}},\ }\href {\doibase 10.1103/PhysRevB.78.035126} {\bibfield
  {journal} {\bibinfo  {journal} {Phys. Rev. B}\ }\textbf {\bibinfo {volume}
  {78}},\ \bibinfo {pages} {035126} (\bibinfo {year} {2008})}\BibitemShut
  {NoStop}%
\bibitem [{\citenamefont {Hermele}\ \emph {et~al.}(2008)\citenamefont
  {Hermele}, \citenamefont {Ran}, \citenamefont {Lee},\ and\ \citenamefont
  {Wen}}]{hermele2008}%
  \BibitemOpen
  \bibfield  {author} {\bibinfo {author} {\bibfnamefont {M.}~\bibnamefont
  {Hermele}}, \bibinfo {author} {\bibfnamefont {Y.}~\bibnamefont {Ran}},
  \bibinfo {author} {\bibfnamefont {P.~A.}\ \bibnamefont {Lee}}, \ and\
  \bibinfo {author} {\bibfnamefont {X.-G.}\ \bibnamefont {Wen}},\ }\href
  {\doibase 10.1103/PhysRevB.77.224413} {\bibfield  {journal} {\bibinfo
  {journal} {Phys. Rev. B}\ }\textbf {\bibinfo {volume} {77}},\ \bibinfo
  {pages} {224413} (\bibinfo {year} {2008})}\BibitemShut {NoStop}%
\bibitem [{\citenamefont {Song}\ \emph {et~al.}(2020)\citenamefont {Song},
  \citenamefont {He}, \citenamefont {Vishwanath},\ and\ \citenamefont
  {Wang}}]{song2020}%
  \BibitemOpen
  \bibfield  {author} {\bibinfo {author} {\bibfnamefont {X.-Y.}\ \bibnamefont
  {Song}}, \bibinfo {author} {\bibfnamefont {Y.-C.}\ \bibnamefont {He}},
  \bibinfo {author} {\bibfnamefont {A.}~\bibnamefont {Vishwanath}}, \ and\
  \bibinfo {author} {\bibfnamefont {C.}~\bibnamefont {Wang}},\ }\href {\doibase
  10.1103/PhysRevX.10.011033} {\bibfield  {journal} {\bibinfo  {journal} {Phys.
  Rev. X}\ }\textbf {\bibinfo {volume} {10}},\ \bibinfo {pages} {011033}
  (\bibinfo {year} {2020})}\BibitemShut {NoStop}%
\bibitem [{\citenamefont {Dupuis}\ and\ \citenamefont
  {Witczak-Krempa}(2021)}]{dupuis2021}%
  \BibitemOpen
  \bibfield  {author} {\bibinfo {author} {\bibfnamefont {{\'E}.}~\bibnamefont
  {Dupuis}}\ and\ \bibinfo {author} {\bibfnamefont {W.}~\bibnamefont
  {Witczak-Krempa}},\ }\href {\doibase 10.1016/j.aop.2021.168496} {\bibfield
  {journal} {\bibinfo  {journal} {Ann. Phys.}\ }\textbf {\bibinfo {volume}
  {435}},\ \bibinfo {pages} {168496} (\bibinfo {year} {2021})}\BibitemShut
  {NoStop}%
\bibitem [{\citenamefont {Witczak-Krempa}\ \emph {et~al.}(2014)\citenamefont
  {Witczak-Krempa}, \citenamefont {Chen}, \citenamefont {Kim},\ and\
  \citenamefont {Balents}}]{witczak-krempa2014}%
  \BibitemOpen
  \bibfield  {author} {\bibinfo {author} {\bibfnamefont {W.}~\bibnamefont
  {Witczak-Krempa}}, \bibinfo {author} {\bibfnamefont {G.}~\bibnamefont
  {Chen}}, \bibinfo {author} {\bibfnamefont {Y.~B.}\ \bibnamefont {Kim}}, \
  and\ \bibinfo {author} {\bibfnamefont {L.}~\bibnamefont {Balents}},\ }\href
  {\doibase 10.1146/annurev-conmatphys-020911-125138} {\bibfield  {journal}
  {\bibinfo  {journal} {Annu. Rev. Condens. Matter Phys.}\ }\textbf {\bibinfo
  {volume} {5}},\ \bibinfo {pages} {57} (\bibinfo {year} {2014})}\BibitemShut
  {NoStop}%
\bibitem [{\citenamefont {Rau}\ \emph {et~al.}(2016)\citenamefont {Rau},
  \citenamefont {Lee},\ and\ \citenamefont {Kee}}]{rau2016}%
  \BibitemOpen
  \bibfield  {author} {\bibinfo {author} {\bibfnamefont {J.~G.}\ \bibnamefont
  {Rau}}, \bibinfo {author} {\bibfnamefont {E.~K.-H.}\ \bibnamefont {Lee}}, \
  and\ \bibinfo {author} {\bibfnamefont {H.-Y.}\ \bibnamefont {Kee}},\ }\href
  {https://doi.org/10.1146/annurev-conmatphys-031115-011319} {\bibfield
  {journal} {\bibinfo  {journal} {Annu. Rev. Condens. Matter Phys.}\ }\textbf
  {\bibinfo {volume} {7}},\ \bibinfo {pages} {195} (\bibinfo {year}
  {2016})}\BibitemShut {NoStop}%
\bibitem [{\citenamefont {Jackeli}\ and\ \citenamefont
  {Khaliullin}(2009)}]{jackeli2009}%
  \BibitemOpen
  \bibfield  {author} {\bibinfo {author} {\bibfnamefont {G.}~\bibnamefont
  {Jackeli}}\ and\ \bibinfo {author} {\bibfnamefont {G.}~\bibnamefont
  {Khaliullin}},\ }\href {\doibase 10.1103/PhysRevLett.102.017205} {\bibfield
  {journal} {\bibinfo  {journal} {Phys. Rev. Lett.}\ }\textbf {\bibinfo
  {volume} {102}},\ \bibinfo {pages} {017205} (\bibinfo {year}
  {2009})}\BibitemShut {NoStop}%
\bibitem [{\citenamefont {Takagi}\ \emph {et~al.}(2019)\citenamefont {Takagi},
  \citenamefont {Takayama}, \citenamefont {Jackeli}, \citenamefont
  {Khaliullin},\ and\ \citenamefont {Nagler}}]{takagi2019}%
  \BibitemOpen
  \bibfield  {author} {\bibinfo {author} {\bibfnamefont {H.}~\bibnamefont
  {Takagi}}, \bibinfo {author} {\bibfnamefont {T.}~\bibnamefont {Takayama}},
  \bibinfo {author} {\bibfnamefont {G.}~\bibnamefont {Jackeli}}, \bibinfo
  {author} {\bibfnamefont {G.}~\bibnamefont {Khaliullin}}, \ and\ \bibinfo
  {author} {\bibfnamefont {S.~E.}\ \bibnamefont {Nagler}},\ }\href {\doibase
  10.1038/s42254-019-0038-2} {\bibfield  {journal} {\bibinfo  {journal} {Nat.
  Rev. Phys.}\ }\textbf {\bibinfo {volume} {1}},\ \bibinfo {pages} {264}
  (\bibinfo {year} {2019})}\BibitemShut {NoStop}%
\bibitem [{\citenamefont {Kitaev}(2006)}]{kitaev2006}%
  \BibitemOpen
  \bibfield  {author} {\bibinfo {author} {\bibfnamefont {A.}~\bibnamefont
  {Kitaev}},\ }\href {\doibase https://doi.org/10.1016/j.aop.2005.10.005}
  {\bibfield  {journal} {\bibinfo  {journal} {Ann. Phys.}\ }\textbf {\bibinfo
  {volume} {321}},\ \bibinfo {pages} {2} (\bibinfo {year} {2006})}\BibitemShut
  {NoStop}%
\bibitem [{\citenamefont {Banerjee}\ \emph {et~al.}(2016)\citenamefont
  {Banerjee}, \citenamefont {Bridges}, \citenamefont {Yan}, \citenamefont
  {Aczel}, \citenamefont {Li}, \citenamefont {Stone}, \citenamefont {Granroth},
  \citenamefont {Lumsden}, \citenamefont {Yiu}, \citenamefont {Knolle},
  \citenamefont {Bhattacharjee}, \citenamefont {Kovrizhin}, \citenamefont
  {Moessner}, \citenamefont {Tennant}, \citenamefont {Mandrus},\ and\
  \citenamefont {Nagler}}]{banerjee2016}%
  \BibitemOpen
  \bibfield  {author} {\bibinfo {author} {\bibfnamefont {A.}~\bibnamefont
  {Banerjee}}, \bibinfo {author} {\bibfnamefont {C.~A.}\ \bibnamefont
  {Bridges}}, \bibinfo {author} {\bibfnamefont {J.-Q.}\ \bibnamefont {Yan}},
  \bibinfo {author} {\bibfnamefont {A.~A.}\ \bibnamefont {Aczel}}, \bibinfo
  {author} {\bibfnamefont {L.}~\bibnamefont {Li}}, \bibinfo {author}
  {\bibfnamefont {M.~B.}\ \bibnamefont {Stone}}, \bibinfo {author}
  {\bibfnamefont {G.~E.}\ \bibnamefont {Granroth}}, \bibinfo {author}
  {\bibfnamefont {M.~D.}\ \bibnamefont {Lumsden}}, \bibinfo {author}
  {\bibfnamefont {Y.}~\bibnamefont {Yiu}}, \bibinfo {author} {\bibfnamefont
  {J.}~\bibnamefont {Knolle}}, \bibinfo {author} {\bibfnamefont
  {S.}~\bibnamefont {Bhattacharjee}}, \bibinfo {author} {\bibfnamefont {D.~L.}\
  \bibnamefont {Kovrizhin}}, \bibinfo {author} {\bibfnamefont {R.}~\bibnamefont
  {Moessner}}, \bibinfo {author} {\bibfnamefont {D.~A.}\ \bibnamefont
  {Tennant}}, \bibinfo {author} {\bibfnamefont {D.~G.}\ \bibnamefont
  {Mandrus}}, \ and\ \bibinfo {author} {\bibfnamefont {S.~E.}\ \bibnamefont
  {Nagler}},\ }\href {\doibase 10.1038/nmat4604} {\bibfield  {journal}
  {\bibinfo  {journal} {Nat. Mater.}\ }\textbf {\bibinfo {volume} {15}},\
  \bibinfo {pages} {733} (\bibinfo {year} {2016})}\BibitemShut {NoStop}%
\bibitem [{\citenamefont {Banerjee}\ \emph {et~al.}(2017)\citenamefont
  {Banerjee}, \citenamefont {Yan}, \citenamefont {Knolle}, \citenamefont
  {Bridges}, \citenamefont {Stone}, \citenamefont {Lumsden}, \citenamefont
  {Mandrus}, \citenamefont {Tennant}, \citenamefont {Moessner},\ and\
  \citenamefont {Nagler}}]{banerjee2017}%
  \BibitemOpen
  \bibfield  {author} {\bibinfo {author} {\bibfnamefont {A.}~\bibnamefont
  {Banerjee}}, \bibinfo {author} {\bibfnamefont {J.}~\bibnamefont {Yan}},
  \bibinfo {author} {\bibfnamefont {J.}~\bibnamefont {Knolle}}, \bibinfo
  {author} {\bibfnamefont {C.~A.}\ \bibnamefont {Bridges}}, \bibinfo {author}
  {\bibfnamefont {M.~B.}\ \bibnamefont {Stone}}, \bibinfo {author}
  {\bibfnamefont {M.~D.}\ \bibnamefont {Lumsden}}, \bibinfo {author}
  {\bibfnamefont {D.~G.}\ \bibnamefont {Mandrus}}, \bibinfo {author}
  {\bibfnamefont {D.~A.}\ \bibnamefont {Tennant}}, \bibinfo {author}
  {\bibfnamefont {R.}~\bibnamefont {Moessner}}, \ and\ \bibinfo {author}
  {\bibfnamefont {S.~E.}\ \bibnamefont {Nagler}},\ }\href {\doibase
  10.1126/science.aah6015} {\bibfield  {journal} {\bibinfo  {journal}
  {Science}\ }\textbf {\bibinfo {volume} {356}},\ \bibinfo {pages} {1055}
  (\bibinfo {year} {2017})}\BibitemShut {NoStop}%
\bibitem [{\citenamefont {Jan\v{s}a}\ \emph {et~al.}(2018)\citenamefont
  {Jan\v{s}a}, \citenamefont {Zorko}, \citenamefont {Gomil\v{s}ek},
  \citenamefont {Pregelj}, \citenamefont {Kr\"amer}, \citenamefont {Biner},
  \citenamefont {Biffin}, \citenamefont {R\"uegg},\ and\ \citenamefont
  {Klanj\v{s}ek}}]{jansa2018}%
  \BibitemOpen
  \bibfield  {author} {\bibinfo {author} {\bibfnamefont {N.}~\bibnamefont
  {Jan\v{s}a}}, \bibinfo {author} {\bibfnamefont {A.}~\bibnamefont {Zorko}},
  \bibinfo {author} {\bibfnamefont {M.}~\bibnamefont {Gomil\v{s}ek}}, \bibinfo
  {author} {\bibfnamefont {M.}~\bibnamefont {Pregelj}}, \bibinfo {author}
  {\bibfnamefont {K.~W.}\ \bibnamefont {Kr\"amer}}, \bibinfo {author}
  {\bibfnamefont {D.}~\bibnamefont {Biner}}, \bibinfo {author} {\bibfnamefont
  {A.}~\bibnamefont {Biffin}}, \bibinfo {author} {\bibfnamefont
  {C.}~\bibnamefont {R\"uegg}}, \ and\ \bibinfo {author} {\bibfnamefont
  {M.}~\bibnamefont {Klanj\v{s}ek}},\ }\href {\doibase
  10.1038/s41567-018-0129-5} {\bibfield  {journal} {\bibinfo  {journal} {Nat.
  Phys.}\ }\textbf {\bibinfo {volume} {14}},\ \bibinfo {pages} {786} (\bibinfo
  {year} {2018})}\BibitemShut {NoStop}%
\bibitem [{\citenamefont {Banerjee}\ \emph {et~al.}(2018)\citenamefont
  {Banerjee}, \citenamefont {Lampen-Kelley}, \citenamefont {Knolle},
  \citenamefont {Balz}, \citenamefont {Aczel}, \citenamefont {Winn},
  \citenamefont {Liu}, \citenamefont {Pajerowski}, \citenamefont {Yan},
  \citenamefont {Bridges}, \citenamefont {Savici}, \citenamefont {Chakoumakos},
  \citenamefont {Lumsden}, \citenamefont {Tennant}, \citenamefont {Moessner},
  \citenamefont {Mandrus},\ and\ \citenamefont {Nagler}}]{banerjee2018}%
  \BibitemOpen
  \bibfield  {author} {\bibinfo {author} {\bibfnamefont {A.}~\bibnamefont
  {Banerjee}}, \bibinfo {author} {\bibfnamefont {P.}~\bibnamefont
  {Lampen-Kelley}}, \bibinfo {author} {\bibfnamefont {J.}~\bibnamefont
  {Knolle}}, \bibinfo {author} {\bibfnamefont {C.}~\bibnamefont {Balz}},
  \bibinfo {author} {\bibfnamefont {A.~A.}\ \bibnamefont {Aczel}}, \bibinfo
  {author} {\bibfnamefont {B.}~\bibnamefont {Winn}}, \bibinfo {author}
  {\bibfnamefont {Y.}~\bibnamefont {Liu}}, \bibinfo {author} {\bibfnamefont
  {D.}~\bibnamefont {Pajerowski}}, \bibinfo {author} {\bibfnamefont
  {J.}~\bibnamefont {Yan}}, \bibinfo {author} {\bibfnamefont {C.~A.}\
  \bibnamefont {Bridges}}, \bibinfo {author} {\bibfnamefont {A.~T.}\
  \bibnamefont {Savici}}, \bibinfo {author} {\bibfnamefont {B.~C.}\
  \bibnamefont {Chakoumakos}}, \bibinfo {author} {\bibfnamefont {M.~D.}\
  \bibnamefont {Lumsden}}, \bibinfo {author} {\bibfnamefont {D.~A.}\
  \bibnamefont {Tennant}}, \bibinfo {author} {\bibfnamefont {R.}~\bibnamefont
  {Moessner}}, \bibinfo {author} {\bibfnamefont {D.~G.}\ \bibnamefont
  {Mandrus}}, \ and\ \bibinfo {author} {\bibfnamefont {S.~E.}\ \bibnamefont
  {Nagler}},\ }\href {\doibase 10.1038/s41535-018-0079-2} {\bibfield  {journal}
  {\bibinfo  {journal} {npj Quantum Mater.}\ }\textbf {\bibinfo {volume} {3}},\
  \bibinfo {pages} {1} (\bibinfo {year} {2018})}\BibitemShut {NoStop}%
\bibitem [{\citenamefont {Balz}\ \emph {et~al.}(2021)\citenamefont {Balz},
  \citenamefont {Janssen}, \citenamefont {Lampen-Kelley}, \citenamefont
  {Banerjee}, \citenamefont {Liu}, \citenamefont {Yan}, \citenamefont
  {Mandrus}, \citenamefont {Vojta},\ and\ \citenamefont {Nagler}}]{balz2021}%
  \BibitemOpen
  \bibfield  {author} {\bibinfo {author} {\bibfnamefont {C.}~\bibnamefont
  {Balz}}, \bibinfo {author} {\bibfnamefont {L.}~\bibnamefont {Janssen}},
  \bibinfo {author} {\bibfnamefont {P.}~\bibnamefont {Lampen-Kelley}}, \bibinfo
  {author} {\bibfnamefont {A.}~\bibnamefont {Banerjee}}, \bibinfo {author}
  {\bibfnamefont {Y.~H.}\ \bibnamefont {Liu}}, \bibinfo {author} {\bibfnamefont
  {J.-Q.}\ \bibnamefont {Yan}}, \bibinfo {author} {\bibfnamefont {D.~G.}\
  \bibnamefont {Mandrus}}, \bibinfo {author} {\bibfnamefont {M.}~\bibnamefont
  {Vojta}}, \ and\ \bibinfo {author} {\bibfnamefont {S.~E.}\ \bibnamefont
  {Nagler}},\ }\href {\doibase 10.1103/PhysRevB.103.174417} {\bibfield
  {journal} {\bibinfo  {journal} {Phys. Rev. B}\ }\textbf {\bibinfo {volume}
  {103}},\ \bibinfo {pages} {174417} (\bibinfo {year} {2021})}\BibitemShut
  {NoStop}%
\bibitem [{\citenamefont {Kasahara}\ \emph {et~al.}(2018)\citenamefont
  {Kasahara}, \citenamefont {Ohnishi}, \citenamefont {Mizukami}, \citenamefont
  {Tanaka}, \citenamefont {Ma}, \citenamefont {Sugii}, \citenamefont {Kurita},
  \citenamefont {Tanaka}, \citenamefont {Nasu}, \citenamefont {Motome},
  \citenamefont {Shibauchi},\ and\ \citenamefont {Matsuda}}]{kasahara2018}%
  \BibitemOpen
  \bibfield  {author} {\bibinfo {author} {\bibfnamefont {Y.}~\bibnamefont
  {Kasahara}}, \bibinfo {author} {\bibfnamefont {T.}~\bibnamefont {Ohnishi}},
  \bibinfo {author} {\bibfnamefont {Y.}~\bibnamefont {Mizukami}}, \bibinfo
  {author} {\bibfnamefont {O.}~\bibnamefont {Tanaka}}, \bibinfo {author}
  {\bibfnamefont {S.}~\bibnamefont {Ma}}, \bibinfo {author} {\bibfnamefont
  {K.}~\bibnamefont {Sugii}}, \bibinfo {author} {\bibfnamefont
  {N.}~\bibnamefont {Kurita}}, \bibinfo {author} {\bibfnamefont
  {H.}~\bibnamefont {Tanaka}}, \bibinfo {author} {\bibfnamefont
  {J.}~\bibnamefont {Nasu}}, \bibinfo {author} {\bibfnamefont {Y.}~\bibnamefont
  {Motome}}, \bibinfo {author} {\bibfnamefont {T.}~\bibnamefont {Shibauchi}}, \
  and\ \bibinfo {author} {\bibfnamefont {Y.}~\bibnamefont {Matsuda}},\ }\href
  {\doibase 10.1038/s41586-018-0274-0} {\bibfield  {journal} {\bibinfo
  {journal} {Nature}\ }\textbf {\bibinfo {volume} {559}},\ \bibinfo {pages}
  {227} (\bibinfo {year} {2018})}\BibitemShut {NoStop}%
\bibitem [{\citenamefont {Yokoi}\ \emph {et~al.}(2021)\citenamefont {Yokoi},
  \citenamefont {Ma}, \citenamefont {Kasahara}, \citenamefont {Kasahara},
  \citenamefont {Shibauchi}, \citenamefont {Kurita}, \citenamefont {Tanaka},
  \citenamefont {Nasu}, \citenamefont {Motome}, \citenamefont {Hickey},
  \citenamefont {Trebst},\ and\ \citenamefont {Matsuda}}]{yokoi2021}%
  \BibitemOpen
  \bibfield  {author} {\bibinfo {author} {\bibfnamefont {T.}~\bibnamefont
  {Yokoi}}, \bibinfo {author} {\bibfnamefont {S.}~\bibnamefont {Ma}}, \bibinfo
  {author} {\bibfnamefont {Y.}~\bibnamefont {Kasahara}}, \bibinfo {author}
  {\bibfnamefont {S.}~\bibnamefont {Kasahara}}, \bibinfo {author}
  {\bibfnamefont {T.}~\bibnamefont {Shibauchi}}, \bibinfo {author}
  {\bibfnamefont {N.}~\bibnamefont {Kurita}}, \bibinfo {author} {\bibfnamefont
  {H.}~\bibnamefont {Tanaka}}, \bibinfo {author} {\bibfnamefont
  {J.}~\bibnamefont {Nasu}}, \bibinfo {author} {\bibfnamefont {Y.}~\bibnamefont
  {Motome}}, \bibinfo {author} {\bibfnamefont {C.}~\bibnamefont {Hickey}},
  \bibinfo {author} {\bibfnamefont {S.}~\bibnamefont {Trebst}}, \ and\ \bibinfo
  {author} {\bibfnamefont {Y.}~\bibnamefont {Matsuda}},\ }\href {\doibase
  10.1126/science.aay5551} {\bibfield  {journal} {\bibinfo  {journal}
  {Science}\ }\textbf {\bibinfo {volume} {373}},\ \bibinfo {pages} {568}
  (\bibinfo {year} {2021})}\BibitemShut {NoStop}%
\bibitem [{\citenamefont {Zou}\ and\ \citenamefont {He}(2020)}]{zou2020}%
  \BibitemOpen
  \bibfield  {author} {\bibinfo {author} {\bibfnamefont {L.}~\bibnamefont
  {Zou}}\ and\ \bibinfo {author} {\bibfnamefont {Y.-C.}\ \bibnamefont {He}},\
  }\href {\doibase 10.1103/PhysRevResearch.2.013072} {\bibfield  {journal}
  {\bibinfo  {journal} {Phys. Rev. Research}\ }\textbf {\bibinfo {volume}
  {2}},\ \bibinfo {pages} {013072} (\bibinfo {year} {2020})}\BibitemShut
  {NoStop}%
\bibitem [{\citenamefont {Barkeshli}\ and\ \citenamefont
  {McGreevy}(2014)}]{Barkeshli1}%
  \BibitemOpen
  \bibfield  {author} {\bibinfo {author} {\bibfnamefont {M.}~\bibnamefont
  {Barkeshli}}\ and\ \bibinfo {author} {\bibfnamefont {J.}~\bibnamefont
  {McGreevy}},\ }\href {\doibase 10.1103/PhysRevB.89.235116} {\bibfield
  {journal} {\bibinfo  {journal} {Phys. Rev. B}\ }\textbf {\bibinfo {volume}
  {89}},\ \bibinfo {pages} {235116} (\bibinfo {year} {2014})}\BibitemShut
  {NoStop}%
\bibitem [{\citenamefont {Shankar}\ and\ \citenamefont
  {Maciejko}(2021)}]{shankar2021}%
  \BibitemOpen
  \bibfield  {author} {\bibinfo {author} {\bibfnamefont {G.}~\bibnamefont
  {Shankar}}\ and\ \bibinfo {author} {\bibfnamefont {J.}~\bibnamefont
  {Maciejko}},\ }\href {\doibase 10.1103/PhysRevB.104.035134} {\bibfield
  {journal} {\bibinfo  {journal} {Phys. Rev. B}\ }\textbf {\bibinfo {volume}
  {104}},\ \bibinfo {pages} {035134} (\bibinfo {year} {2021})}\BibitemShut
  {NoStop}%
\bibitem [{\citenamefont {{'t Hooft}}(1976{\natexlab{a}})}]{thooft1976}%
  \BibitemOpen
  \bibfield  {author} {\bibinfo {author} {\bibfnamefont {G.}~\bibnamefont {{'t
  Hooft}}},\ }\href {\doibase 10.1103/PhysRevLett.37.8} {\bibfield  {journal}
  {\bibinfo  {journal} {Phys. Rev. Lett.}\ }\textbf {\bibinfo {volume} {37}},\
  \bibinfo {pages} {8} (\bibinfo {year} {1976}{\natexlab{a}})}\BibitemShut
  {NoStop}%
\bibitem [{\citenamefont {{'t Hooft}}(1976{\natexlab{b}})}]{thooft1976a}%
  \BibitemOpen
  \bibfield  {author} {\bibinfo {author} {\bibfnamefont {G.}~\bibnamefont {{'t
  Hooft}}},\ }\href {\doibase 10.1103/PhysRevD.14.3432} {\bibfield  {journal}
  {\bibinfo  {journal} {Phys. Rev. D}\ }\textbf {\bibinfo {volume} {14}},\
  \bibinfo {pages} {3432} (\bibinfo {year} {1976}{\natexlab{b}})}\BibitemShut
  {NoStop}%
\bibitem [{\citenamefont {{'t Hooft}}(1986)}]{thooft1986}%
  \BibitemOpen
  \bibfield  {author} {\bibinfo {author} {\bibfnamefont {G.}~\bibnamefont {{'t
  Hooft}}},\ }\href {\doibase 10.1016/0370-1573(86)90117-1} {\bibfield
  {journal} {\bibinfo  {journal} {Phys. Rep.}\ }\textbf {\bibinfo {volume}
  {142}},\ \bibinfo {pages} {357} (\bibinfo {year} {1986})}\BibitemShut
  {NoStop}%
\bibitem [{\citenamefont {Komargodski}\ and\ \citenamefont
  {Seiberg}(2018)}]{komargodski2018}%
  \BibitemOpen
  \bibfield  {author} {\bibinfo {author} {\bibfnamefont {Z.}~\bibnamefont
  {Komargodski}}\ and\ \bibinfo {author} {\bibfnamefont {N.}~\bibnamefont
  {Seiberg}},\ }\href {\doibase 10.1007/JHEP01(2018)109} {\bibfield  {journal}
  {\bibinfo  {journal} {JHEP}\ }\textbf {\bibinfo {volume} {01}},\ \bibinfo
  {pages} {109} (\bibinfo {year} {2018})}\BibitemShut {NoStop}%
\bibitem [{\citenamefont {Benini}(2018)}]{benini2018}%
  \BibitemOpen
  \bibfield  {author} {\bibinfo {author} {\bibfnamefont {F.}~\bibnamefont
  {Benini}},\ }\href {\doibase 10.1007/JHEP02(2018)068} {\bibfield  {journal}
  {\bibinfo  {journal} {JHEP}\ }\textbf {\bibinfo {volume} {02}},\ \bibinfo
  {pages} {068} (\bibinfo {year} {2018})}\BibitemShut {NoStop}%
\bibitem [{\citenamefont {Chen}\ \emph {et~al.}(1993)\citenamefont {Chen},
  \citenamefont {Fisher},\ and\ \citenamefont {Wu}}]{ChenFisherWu}%
  \BibitemOpen
  \bibfield  {author} {\bibinfo {author} {\bibfnamefont {W.}~\bibnamefont
  {Chen}}, \bibinfo {author} {\bibfnamefont {M.~P.~A.}\ \bibnamefont {Fisher}},
  \ and\ \bibinfo {author} {\bibfnamefont {Y.-S.}\ \bibnamefont {Wu}},\ }\href
  {\doibase 10.1103/PhysRevB.48.13749} {\bibfield  {journal} {\bibinfo
  {journal} {Phys. Rev. B}\ }\textbf {\bibinfo {volume} {48}},\ \bibinfo
  {pages} {13749} (\bibinfo {year} {1993})}\BibitemShut {NoStop}%
\bibitem [{\citenamefont {Son}(2015)}]{son2015}%
  \BibitemOpen
  \bibfield  {author} {\bibinfo {author} {\bibfnamefont {D.~T.}\ \bibnamefont
  {Son}},\ }\href {\doibase 10.1103/PhysRevX.5.031027} {\bibfield  {journal}
  {\bibinfo  {journal} {Phys. Rev. X}\ }\textbf {\bibinfo {volume} {5}},\
  \bibinfo {pages} {031027} (\bibinfo {year} {2015})}\BibitemShut {NoStop}%
\bibitem [{\citenamefont {Metlitski}\ and\ \citenamefont
  {Vishwanath}(2016)}]{Metlitskiduality}%
  \BibitemOpen
  \bibfield  {author} {\bibinfo {author} {\bibfnamefont {M.~A.}\ \bibnamefont
  {Metlitski}}\ and\ \bibinfo {author} {\bibfnamefont {A.}~\bibnamefont
  {Vishwanath}},\ }\href {\doibase 10.1103/PhysRevB.93.245151} {\bibfield
  {journal} {\bibinfo  {journal} {Phys. Rev. B}\ }\textbf {\bibinfo {volume}
  {93}},\ \bibinfo {pages} {245151} (\bibinfo {year} {2016})}\BibitemShut
  {NoStop}%
\bibitem [{\citenamefont {Wang}\ and\ \citenamefont
  {Senthil}(2015)}]{Wangduality}%
  \BibitemOpen
  \bibfield  {author} {\bibinfo {author} {\bibfnamefont {C.}~\bibnamefont
  {Wang}}\ and\ \bibinfo {author} {\bibfnamefont {T.}~\bibnamefont {Senthil}},\
  }\href {\doibase 10.1103/PhysRevX.5.041031} {\bibfield  {journal} {\bibinfo
  {journal} {Phys. Rev. X}\ }\textbf {\bibinfo {volume} {5}},\ \bibinfo {pages}
  {041031} (\bibinfo {year} {2015})}\BibitemShut {NoStop}%
\bibitem [{\citenamefont {Seiberg}\ \emph {et~al.}(2016)\citenamefont
  {Seiberg}, \citenamefont {Senthil}, \citenamefont {Wang},\ and\ \citenamefont
  {Witten}}]{SEIBERGduality}%
  \BibitemOpen
  \bibfield  {author} {\bibinfo {author} {\bibfnamefont {N.}~\bibnamefont
  {Seiberg}}, \bibinfo {author} {\bibfnamefont {T.}~\bibnamefont {Senthil}},
  \bibinfo {author} {\bibfnamefont {C.}~\bibnamefont {Wang}}, \ and\ \bibinfo
  {author} {\bibfnamefont {E.}~\bibnamefont {Witten}},\ }\href {\doibase
  http://dx.doi.org/10.1016/j.aop.2016.08.007} {\bibfield  {journal} {\bibinfo
  {journal} {Ann. Phys.}\ }\textbf {\bibinfo {volume} {374}},\ \bibinfo {pages}
  {395 } (\bibinfo {year} {2016})}\BibitemShut {NoStop}%
\bibitem [{\citenamefont {Karch}\ and\ \citenamefont
  {Tong}(2016)}]{DualityKarchTong}%
  \BibitemOpen
  \bibfield  {author} {\bibinfo {author} {\bibfnamefont {A.}~\bibnamefont
  {Karch}}\ and\ \bibinfo {author} {\bibfnamefont {D.}~\bibnamefont {Tong}},\
  }\href {\doibase 10.1103/PhysRevX.6.031043} {\bibfield  {journal} {\bibinfo
  {journal} {Phys. Rev. X}\ }\textbf {\bibinfo {volume} {6}},\ \bibinfo {pages}
  {031043} (\bibinfo {year} {2016})}\BibitemShut {NoStop}%
\bibitem [{\citenamefont {Murugan}\ and\ \citenamefont
  {Nastase}(2017)}]{murugan2016}%
  \BibitemOpen
  \bibfield  {author} {\bibinfo {author} {\bibfnamefont {J.}~\bibnamefont
  {Murugan}}\ and\ \bibinfo {author} {\bibfnamefont {H.}~\bibnamefont
  {Nastase}},\ }\href {\doibase 10.1007/JHEP05(2017)159} {\bibfield  {journal}
  {\bibinfo  {journal} {JHEP}\ }\textbf {\bibinfo {volume} {05}},\ \bibinfo
  {pages} {159} (\bibinfo {year} {2017})}\BibitemShut {NoStop}%
\bibitem [{\citenamefont {Mross}\ \emph {et~al.}(2016)\citenamefont {Mross},
  \citenamefont {Alicea},\ and\ \citenamefont {Motrunich}}]{mross2016}%
  \BibitemOpen
  \bibfield  {author} {\bibinfo {author} {\bibfnamefont {D.~F.}\ \bibnamefont
  {Mross}}, \bibinfo {author} {\bibfnamefont {J.}~\bibnamefont {Alicea}}, \
  and\ \bibinfo {author} {\bibfnamefont {O.~I.}\ \bibnamefont {Motrunich}},\
  }\href {\doibase 10.1103/PhysRevLett.117.016802} {\bibfield  {journal}
  {\bibinfo  {journal} {Phys. Rev. Lett.}\ }\textbf {\bibinfo {volume} {117}},\
  \bibinfo {pages} {016802} (\bibinfo {year} {2016})}\BibitemShut {NoStop}%
\bibitem [{\citenamefont {Kachru}\ \emph {et~al.}(2017)\citenamefont {Kachru},
  \citenamefont {Mulligan}, \citenamefont {Torroba},\ and\ \citenamefont
  {Wang}}]{kachru2017}%
  \BibitemOpen
  \bibfield  {author} {\bibinfo {author} {\bibfnamefont {S.}~\bibnamefont
  {Kachru}}, \bibinfo {author} {\bibfnamefont {M.}~\bibnamefont {Mulligan}},
  \bibinfo {author} {\bibfnamefont {G.}~\bibnamefont {Torroba}}, \ and\
  \bibinfo {author} {\bibfnamefont {H.}~\bibnamefont {Wang}},\ }\href {\doibase
  10.1103/PhysRevLett.118.011602} {\bibfield  {journal} {\bibinfo  {journal}
  {Phys. Rev. Lett.}\ }\textbf {\bibinfo {volume} {118}},\ \bibinfo {pages}
  {011602} (\bibinfo {year} {2017})}\BibitemShut {NoStop}%
\bibitem [{\citenamefont {Karch}\ \emph {et~al.}(2017)\citenamefont {Karch},
  \citenamefont {Robinson},\ and\ \citenamefont {Tong}}]{karch2017}%
  \BibitemOpen
  \bibfield  {author} {\bibinfo {author} {\bibfnamefont {A.}~\bibnamefont
  {Karch}}, \bibinfo {author} {\bibfnamefont {B.}~\bibnamefont {Robinson}}, \
  and\ \bibinfo {author} {\bibfnamefont {D.}~\bibnamefont {Tong}},\ }\href
  {\doibase 10.1007/JHEP01(2017)017} {\bibfield  {journal} {\bibinfo  {journal}
  {JHEP}\ }\textbf {\bibinfo {volume} {01}},\ \bibinfo {pages} {017} (\bibinfo
  {year} {2017})}\BibitemShut {NoStop}%
\bibitem [{\citenamefont {Wen}(1991)}]{Wenparton1}%
  \BibitemOpen
  \bibfield  {author} {\bibinfo {author} {\bibfnamefont {X.~G.}\ \bibnamefont
  {Wen}},\ }\href {\doibase 10.1103/PhysRevLett.66.802} {\bibfield  {journal}
  {\bibinfo  {journal} {Phys. Rev. Lett.}\ }\textbf {\bibinfo {volume} {66}},\
  \bibinfo {pages} {802} (\bibinfo {year} {1991})}\BibitemShut {NoStop}%
\bibitem [{\citenamefont {Wen}(1999)}]{Wenparton2}%
  \BibitemOpen
  \bibfield  {author} {\bibinfo {author} {\bibfnamefont {X.-G.}\ \bibnamefont
  {Wen}},\ }\href {\doibase 10.1103/PhysRevB.60.8827} {\bibfield  {journal}
  {\bibinfo  {journal} {Phys. Rev. B}\ }\textbf {\bibinfo {volume} {60}},\
  \bibinfo {pages} {8827} (\bibinfo {year} {1999})}\BibitemShut {NoStop}%
\bibitem [{\citenamefont {Lu}\ and\ \citenamefont {Vishwanath}(2012)}]{lu12}%
  \BibitemOpen
  \bibfield  {author} {\bibinfo {author} {\bibfnamefont {Y.-M.}\ \bibnamefont
  {Lu}}\ and\ \bibinfo {author} {\bibfnamefont {A.}~\bibnamefont
  {Vishwanath}},\ }\href {\doibase 10.1103/PhysRevB.86.125119} {\bibfield
  {journal} {\bibinfo  {journal} {Phys. Rev. B}\ }\textbf {\bibinfo {volume}
  {86}},\ \bibinfo {pages} {125119} (\bibinfo {year} {2012})}\BibitemShut
  {NoStop}%
\bibitem [{\citenamefont {Senthil}\ and\ \citenamefont
  {Levin}(2013)}]{senthil2013}%
  \BibitemOpen
  \bibfield  {author} {\bibinfo {author} {\bibfnamefont {T.}~\bibnamefont
  {Senthil}}\ and\ \bibinfo {author} {\bibfnamefont {M.}~\bibnamefont
  {Levin}},\ }\href {\doibase 10.1103/PhysRevLett.110.046801} {\bibfield
  {journal} {\bibinfo  {journal} {Phys. Rev. Lett.}\ }\textbf {\bibinfo
  {volume} {110}},\ \bibinfo {pages} {046801} (\bibinfo {year}
  {2013})}\BibitemShut {NoStop}%
\bibitem [{\citenamefont {Chen}\ \emph {et~al.}(2013)\citenamefont {Chen},
  \citenamefont {Gu}, \citenamefont {Liu},\ and\ \citenamefont {Wen}}]{chen13}%
  \BibitemOpen
  \bibfield  {author} {\bibinfo {author} {\bibfnamefont {X.}~\bibnamefont
  {Chen}}, \bibinfo {author} {\bibfnamefont {Z.-C.}\ \bibnamefont {Gu}},
  \bibinfo {author} {\bibfnamefont {Z.-X.}\ \bibnamefont {Liu}}, \ and\
  \bibinfo {author} {\bibfnamefont {X.-G.}\ \bibnamefont {Wen}},\ }\href
  {\doibase 10.1103/PhysRevB.87.155114} {\bibfield  {journal} {\bibinfo
  {journal} {Phys. Rev. B}\ }\textbf {\bibinfo {volume} {87}},\ \bibinfo
  {pages} {155114} (\bibinfo {year} {2013})}\BibitemShut {NoStop}%
\bibitem [{\citenamefont {Barkeshli}(2013)}]{barkeshli2013}%
  \BibitemOpen
  \bibfield  {author} {\bibinfo {author} {\bibfnamefont {M.}~\bibnamefont
  {Barkeshli}},\ }\href {http://arxiv.org/abs/1307.8194} {\bibfield  {journal}
  {\bibinfo  {journal} {arXiv:1307.8194}\ } (\bibinfo {year}
  {2013})}\BibitemShut {NoStop}%
\bibitem [{\citenamefont {Kalmeyer}\ and\ \citenamefont
  {Laughlin}(1987)}]{kalmeyer1987}%
  \BibitemOpen
  \bibfield  {author} {\bibinfo {author} {\bibfnamefont {V.}~\bibnamefont
  {Kalmeyer}}\ and\ \bibinfo {author} {\bibfnamefont {R.~B.}\ \bibnamefont
  {Laughlin}},\ }\href {\doibase 10.1103/PhysRevLett.59.2095} {\bibfield
  {journal} {\bibinfo  {journal} {Phys. Rev. Lett.}\ }\textbf {\bibinfo
  {volume} {59}},\ \bibinfo {pages} {2095} (\bibinfo {year}
  {1987})}\BibitemShut {NoStop}%
\bibitem [{\citenamefont {Niemi}\ and\ \citenamefont
  {Semenoff}(1983)}]{niemi1983}%
  \BibitemOpen
  \bibfield  {author} {\bibinfo {author} {\bibfnamefont {A.~J.}\ \bibnamefont
  {Niemi}}\ and\ \bibinfo {author} {\bibfnamefont {G.~W.}\ \bibnamefont
  {Semenoff}},\ }\href {\doibase 10.1103/PhysRevLett.51.2077} {\bibfield
  {journal} {\bibinfo  {journal} {Phys. Rev. Lett.}\ }\textbf {\bibinfo
  {volume} {51}},\ \bibinfo {pages} {2077} (\bibinfo {year}
  {1983})}\BibitemShut {NoStop}%
\bibitem [{\citenamefont {Redlich}(1984{\natexlab{a}})}]{redlich1984}%
  \BibitemOpen
  \bibfield  {author} {\bibinfo {author} {\bibfnamefont {A.~N.}\ \bibnamefont
  {Redlich}},\ }\href {\doibase 10.1103/PhysRevLett.52.18} {\bibfield
  {journal} {\bibinfo  {journal} {Phys. Rev. Lett.}\ }\textbf {\bibinfo
  {volume} {52}},\ \bibinfo {pages} {18} (\bibinfo {year}
  {1984}{\natexlab{a}})}\BibitemShut {NoStop}%
\bibitem [{\citenamefont {Redlich}(1984{\natexlab{b}})}]{redlich1984b}%
  \BibitemOpen
  \bibfield  {author} {\bibinfo {author} {\bibfnamefont {A.~N.}\ \bibnamefont
  {Redlich}},\ }\href {\doibase 10.1103/PhysRevD.29.2366} {\bibfield  {journal}
  {\bibinfo  {journal} {Phys. Rev. D}\ }\textbf {\bibinfo {volume} {29}},\
  \bibinfo {pages} {2366} (\bibinfo {year} {1984}{\natexlab{b}})}\BibitemShut
  {NoStop}%
\bibitem [{\citenamefont {Ye}\ and\ \citenamefont {Sachdev}(1998)}]{ye1998}%
  \BibitemOpen
  \bibfield  {author} {\bibinfo {author} {\bibfnamefont {J.}~\bibnamefont
  {Ye}}\ and\ \bibinfo {author} {\bibfnamefont {S.}~\bibnamefont {Sachdev}},\
  }\href {\doibase 10.1103/PhysRevLett.80.5409} {\bibfield  {journal} {\bibinfo
   {journal} {Phys. Rev. Lett.}\ }\textbf {\bibinfo {volume} {80}},\ \bibinfo
  {pages} {5409} (\bibinfo {year} {1998})}\BibitemShut {NoStop}%
\bibitem [{\citenamefont {Wen}\ and\ \citenamefont {Wu}(1993)}]{WenWu}%
  \BibitemOpen
  \bibfield  {author} {\bibinfo {author} {\bibfnamefont {X.-G.}\ \bibnamefont
  {Wen}}\ and\ \bibinfo {author} {\bibfnamefont {Y.-S.}\ \bibnamefont {Wu}},\
  }\href {\doibase 10.1103/PhysRevLett.70.1501} {\bibfield  {journal} {\bibinfo
   {journal} {Phys. Rev. Lett.}\ }\textbf {\bibinfo {volume} {70}},\ \bibinfo
  {pages} {1501} (\bibinfo {year} {1993})}\BibitemShut {NoStop}%
\bibitem [{\citenamefont {Haldane}(1988)}]{haldane1988}%
  \BibitemOpen
  \bibfield  {author} {\bibinfo {author} {\bibfnamefont {F.~D.~M.}\
  \bibnamefont {Haldane}},\ }\href {\doibase 10.1103/PhysRevLett.61.2015}
  {\bibfield  {journal} {\bibinfo  {journal} {Phys. Rev. Lett.}\ }\textbf
  {\bibinfo {volume} {61}},\ \bibinfo {pages} {2015} (\bibinfo {year}
  {1988})}\BibitemShut {NoStop}%
\bibitem [{\citenamefont {Affleck}\ \emph {et~al.}(1982)\citenamefont
  {Affleck}, \citenamefont {Harvey},\ and\ \citenamefont
  {Witten}}]{affleck1982}%
  \BibitemOpen
  \bibfield  {author} {\bibinfo {author} {\bibfnamefont {I.}~\bibnamefont
  {Affleck}}, \bibinfo {author} {\bibfnamefont {J.}~\bibnamefont {Harvey}}, \
  and\ \bibinfo {author} {\bibfnamefont {E.}~\bibnamefont {Witten}},\ }\href
  {\doibase 10.1016/0550-3213(82)90277-2} {\bibfield  {journal} {\bibinfo
  {journal} {Nucl. Phys. B}\ }\textbf {\bibinfo {volume} {206}},\ \bibinfo
  {pages} {413} (\bibinfo {year} {1982})}\BibitemShut {NoStop}%
\bibitem [{\citenamefont {Wu}\ and\ \citenamefont {Yang}(1976)}]{wu1976}%
  \BibitemOpen
  \bibfield  {author} {\bibinfo {author} {\bibfnamefont {T.~T.}\ \bibnamefont
  {Wu}}\ and\ \bibinfo {author} {\bibfnamefont {C.~N.}\ \bibnamefont {Yang}},\
  }\href {\doibase 10.1016/0550-3213(76)90143-7} {\bibfield  {journal}
  {\bibinfo  {journal} {Nucl. Phys. B}\ }\textbf {\bibinfo {volume} {107}},\
  \bibinfo {pages} {365} (\bibinfo {year} {1976})}\BibitemShut {NoStop}%
\bibitem [{\citenamefont {Kazama}\ \emph {et~al.}(1977)\citenamefont {Kazama},
  \citenamefont {Yang},\ and\ \citenamefont {Goldhaber}}]{kazama1977}%
  \BibitemOpen
  \bibfield  {author} {\bibinfo {author} {\bibfnamefont {Y.}~\bibnamefont
  {Kazama}}, \bibinfo {author} {\bibfnamefont {C.~N.}\ \bibnamefont {Yang}}, \
  and\ \bibinfo {author} {\bibfnamefont {A.~S.}\ \bibnamefont {Goldhaber}},\
  }\href {\doibase 10.1103/PhysRevD.15.2287} {\bibfield  {journal} {\bibinfo
  {journal} {Phys. Rev. D}\ }\textbf {\bibinfo {volume} {15}},\ \bibinfo
  {pages} {2287} (\bibinfo {year} {1977})}\BibitemShut {NoStop}%
\bibitem [{\citenamefont {Fu}\ \emph {et~al.}(2018)\citenamefont {Fu},
  \citenamefont {Knolle},\ and\ \citenamefont {Perkins}}]{fu2018}%
  \BibitemOpen
  \bibfield  {author} {\bibinfo {author} {\bibfnamefont {J.}~\bibnamefont
  {Fu}}, \bibinfo {author} {\bibfnamefont {J.}~\bibnamefont {Knolle}}, \ and\
  \bibinfo {author} {\bibfnamefont {N.~B.}\ \bibnamefont {Perkins}},\ }\href
  {\doibase 10.1103/PhysRevB.97.115142} {\bibfield  {journal} {\bibinfo
  {journal} {Phys. Rev. B}\ }\textbf {\bibinfo {volume} {97}},\ \bibinfo
  {pages} {115142} (\bibinfo {year} {2018})}\BibitemShut {NoStop}%
\bibitem [{\citenamefont {Read}\ and\ \citenamefont {Green}(2000)}]{read2000}%
  \BibitemOpen
  \bibfield  {author} {\bibinfo {author} {\bibfnamefont {N.}~\bibnamefont
  {Read}}\ and\ \bibinfo {author} {\bibfnamefont {D.}~\bibnamefont {Green}},\
  }\href {\doibase 10.1103/PhysRevB.61.10267} {\bibfield  {journal} {\bibinfo
  {journal} {Phys. Rev. B}\ }\textbf {\bibinfo {volume} {61}},\ \bibinfo
  {pages} {10267} (\bibinfo {year} {2000})}\BibitemShut {NoStop}%
\bibitem [{\citenamefont {Dijkgraaf}\ and\ \citenamefont
  {Witten}(1990)}]{dijkgraaf1990}%
  \BibitemOpen
  \bibfield  {author} {\bibinfo {author} {\bibfnamefont {R.}~\bibnamefont
  {Dijkgraaf}}\ and\ \bibinfo {author} {\bibfnamefont {E.}~\bibnamefont
  {Witten}},\ }\href {\doibase 10.1007/BF02096988} {\bibfield  {journal}
  {\bibinfo  {journal} {Commun. Math. Phys.}\ }\textbf {\bibinfo {volume}
  {129}},\ \bibinfo {pages} {393} (\bibinfo {year} {1990})}\BibitemShut
  {NoStop}%
\bibitem [{\citenamefont {Metlitski}\ \emph {et~al.}(2017)\citenamefont
  {Metlitski}, \citenamefont {Vishwanath},\ and\ \citenamefont
  {Xu}}]{Maxduality}%
  \BibitemOpen
  \bibfield  {author} {\bibinfo {author} {\bibfnamefont {M.~A.}\ \bibnamefont
  {Metlitski}}, \bibinfo {author} {\bibfnamefont {A.}~\bibnamefont
  {Vishwanath}}, \ and\ \bibinfo {author} {\bibfnamefont {C.}~\bibnamefont
  {Xu}},\ }\href {\doibase 10.1103/PhysRevB.95.205137} {\bibfield  {journal}
  {\bibinfo  {journal} {Phys. Rev. B}\ }\textbf {\bibinfo {volume} {95}},\
  \bibinfo {pages} {205137} (\bibinfo {year} {2017})}\BibitemShut {NoStop}%
\bibitem [{\citenamefont {Aharony}\ \emph {et~al.}(2017)\citenamefont
  {Aharony}, \citenamefont {Benini}, \citenamefont {Hsin},\ and\ \citenamefont
  {Seiberg}}]{Aharony2017}%
  \BibitemOpen
  \bibfield  {author} {\bibinfo {author} {\bibfnamefont {O.}~\bibnamefont
  {Aharony}}, \bibinfo {author} {\bibfnamefont {F.}~\bibnamefont {Benini}},
  \bibinfo {author} {\bibfnamefont {P.-S.}\ \bibnamefont {Hsin}}, \ and\
  \bibinfo {author} {\bibfnamefont {N.}~\bibnamefont {Seiberg}},\ }\href
  {\doibase 10.1007/JHEP02(2017)072} {\bibfield  {journal} {\bibinfo  {journal}
  {JHEP}\ }\textbf {\bibinfo {volume} {02}},\ \bibinfo {pages} {072} (\bibinfo
  {year} {2017})}\BibitemShut {NoStop}%
\bibitem [{\citenamefont {Bursa}\ \emph {et~al.}(2013)\citenamefont {Bursa},
  \citenamefont {Lau},\ and\ \citenamefont {Teper}}]{bursa2013}%
  \BibitemOpen
  \bibfield  {author} {\bibinfo {author} {\bibfnamefont {F.}~\bibnamefont
  {Bursa}}, \bibinfo {author} {\bibfnamefont {R.}~\bibnamefont {Lau}}, \ and\
  \bibinfo {author} {\bibfnamefont {M.}~\bibnamefont {Teper}},\ }\href
  {\doibase 10.1007/JHEP05(2013)025} {\bibfield  {journal} {\bibinfo  {journal}
  {JHEP}\ }\textbf {\bibinfo {volume} {05}},\ \bibinfo {pages} {025} (\bibinfo
  {year} {2013})}\BibitemShut {NoStop}%
\bibitem [{\citenamefont {Lau}\ and\ \citenamefont {Teper}(2017)}]{lau2017}%
  \BibitemOpen
  \bibfield  {author} {\bibinfo {author} {\bibfnamefont {R.}~\bibnamefont
  {Lau}}\ and\ \bibinfo {author} {\bibfnamefont {M.}~\bibnamefont {Teper}},\
  }\href {\doibase 10.1007/JHEP10(2017)022} {\bibfield  {journal} {\bibinfo
  {journal} {JHEP}\ }\textbf {\bibinfo {volume} {10}},\ \bibinfo {pages} {022}
  (\bibinfo {year} {2017})}\BibitemShut {NoStop}%
\bibitem [{\citenamefont {Seiberg}\ and\ \citenamefont
  {Witten}(2016)}]{seiberg2016}%
  \BibitemOpen
  \bibfield  {author} {\bibinfo {author} {\bibfnamefont {N.}~\bibnamefont
  {Seiberg}}\ and\ \bibinfo {author} {\bibfnamefont {E.}~\bibnamefont
  {Witten}},\ }\href {\doibase 10.1093/ptep/ptw083} {\bibfield  {journal}
  {\bibinfo  {journal} {Prog. Theor. Exp. Phys.}\ }\textbf {\bibinfo {volume}
  {2016}},\ \bibinfo {pages} {12C101} (\bibinfo {year} {2016})}\BibitemShut
  {NoStop}%
\bibitem [{\citenamefont {Witten}(1984)}]{witten1984}%
  \BibitemOpen
  \bibfield  {author} {\bibinfo {author} {\bibfnamefont {E.}~\bibnamefont
  {Witten}},\ }\href {https://projecteuclid.org/euclid.cmp/1103940923}
  {\bibfield  {journal} {\bibinfo  {journal} {Commun. Math. Phys.}\ }\textbf
  {\bibinfo {volume} {92}},\ \bibinfo {pages} {455} (\bibinfo {year}
  {1984})}\BibitemShut {NoStop}%
\bibitem [{\citenamefont {Antoniadis}\ and\ \citenamefont
  {Bachas}(1986)}]{antoniadis1986}%
  \BibitemOpen
  \bibfield  {author} {\bibinfo {author} {\bibfnamefont {I.}~\bibnamefont
  {Antoniadis}}\ and\ \bibinfo {author} {\bibfnamefont {C.}~\bibnamefont
  {Bachas}},\ }\href {\doibase 10.1016/0550-3213(86)90217-8} {\bibfield
  {journal} {\bibinfo  {journal} {Nucl. Phys. B}\ }\textbf {\bibinfo {volume}
  {278}},\ \bibinfo {pages} {343} (\bibinfo {year} {1986})}\BibitemShut
  {NoStop}%
\bibitem [{\citenamefont {Naculich}\ \emph {et~al.}(1990)\citenamefont
  {Naculich}, \citenamefont {Riggs},\ and\ \citenamefont
  {Schnitzer}}]{naculich1990}%
  \BibitemOpen
  \bibfield  {author} {\bibinfo {author} {\bibfnamefont {S.~G.}\ \bibnamefont
  {Naculich}}, \bibinfo {author} {\bibfnamefont {H.~A.}\ \bibnamefont {Riggs}},
  \ and\ \bibinfo {author} {\bibfnamefont {H.~J.}\ \bibnamefont {Schnitzer}},\
  }\href {\doibase 10.1016/0370-2693(90)90623-E} {\bibfield  {journal}
  {\bibinfo  {journal} {Phys. Lett. B}\ }\textbf {\bibinfo {volume} {246}},\
  \bibinfo {pages} {417} (\bibinfo {year} {1990})}\BibitemShut {NoStop}%
\bibitem [{\citenamefont {Wen}(1989)}]{wen1989}%
  \BibitemOpen
  \bibfield  {author} {\bibinfo {author} {\bibfnamefont {X.~G.}\ \bibnamefont
  {Wen}},\ }\href {\doibase 10.1103/PhysRevB.40.7387} {\bibfield  {journal}
  {\bibinfo  {journal} {Phys. Rev. B}\ }\textbf {\bibinfo {volume} {40}},\
  \bibinfo {pages} {7387} (\bibinfo {year} {1989})}\BibitemShut {NoStop}%
\bibitem [{\citenamefont {Wen}\ \emph {et~al.}(1989)\citenamefont {Wen},
  \citenamefont {Wilczek},\ and\ \citenamefont {Zee}}]{wen1989b}%
  \BibitemOpen
  \bibfield  {author} {\bibinfo {author} {\bibfnamefont {X.~G.}\ \bibnamefont
  {Wen}}, \bibinfo {author} {\bibfnamefont {F.}~\bibnamefont {Wilczek}}, \ and\
  \bibinfo {author} {\bibfnamefont {A.}~\bibnamefont {Zee}},\ }\href {\doibase
  10.1103/PhysRevB.39.11413} {\bibfield  {journal} {\bibinfo  {journal} {Phys.
  Rev. B}\ }\textbf {\bibinfo {volume} {39}},\ \bibinfo {pages} {11413}
  (\bibinfo {year} {1989})}\BibitemShut {NoStop}%
\bibitem [{\citenamefont {Yang}\ \emph {et~al.}(2019)\citenamefont {Yang},
  \citenamefont {Iadecola}, \citenamefont {Chamon},\ and\ \citenamefont
  {Mudry}}]{yang2019}%
  \BibitemOpen
  \bibfield  {author} {\bibinfo {author} {\bibfnamefont {Z.-C.}\ \bibnamefont
  {Yang}}, \bibinfo {author} {\bibfnamefont {T.}~\bibnamefont {Iadecola}},
  \bibinfo {author} {\bibfnamefont {C.}~\bibnamefont {Chamon}}, \ and\ \bibinfo
  {author} {\bibfnamefont {C.}~\bibnamefont {Mudry}},\ }\href {\doibase
  10.1103/PhysRevB.99.155138} {\bibfield  {journal} {\bibinfo  {journal} {Phys.
  Rev. B}\ }\textbf {\bibinfo {volume} {99}},\ \bibinfo {pages} {155138}
  (\bibinfo {year} {2019})}\BibitemShut {NoStop}%
\bibitem [{\citenamefont {Mirmojarabian}\ \emph {et~al.}(2020)\citenamefont
  {Mirmojarabian}, \citenamefont {Kargarian},\ and\ \citenamefont
  {Langari}}]{mirmojarabian2020}%
  \BibitemOpen
  \bibfield  {author} {\bibinfo {author} {\bibfnamefont {F.}~\bibnamefont
  {Mirmojarabian}}, \bibinfo {author} {\bibfnamefont {M.}~\bibnamefont
  {Kargarian}}, \ and\ \bibinfo {author} {\bibfnamefont {A.}~\bibnamefont
  {Langari}},\ }\href {\doibase 10.1103/PhysRevB.101.115116} {\bibfield
  {journal} {\bibinfo  {journal} {Phys. Rev. B}\ }\textbf {\bibinfo {volume}
  {101}},\ \bibinfo {pages} {115116} (\bibinfo {year} {2020})}\BibitemShut
  {NoStop}%
\bibitem [{\citenamefont {Farjami}\ \emph {et~al.}(2020)\citenamefont
  {Farjami}, \citenamefont {Horner}, \citenamefont {Self}, \citenamefont
  {Papi\ifmmode~\acute{c}\else \'{c}\fi{}},\ and\ \citenamefont
  {Pachos}}]{farjami2020}%
  \BibitemOpen
  \bibfield  {author} {\bibinfo {author} {\bibfnamefont {A.}~\bibnamefont
  {Farjami}}, \bibinfo {author} {\bibfnamefont {M.~D.}\ \bibnamefont {Horner}},
  \bibinfo {author} {\bibfnamefont {C.~N.}\ \bibnamefont {Self}}, \bibinfo
  {author} {\bibfnamefont {Z.}~\bibnamefont {Papi\ifmmode~\acute{c}\else
  \'{c}\fi{}}}, \ and\ \bibinfo {author} {\bibfnamefont {J.~K.}\ \bibnamefont
  {Pachos}},\ }\href {\doibase 10.1103/PhysRevB.101.245116} {\bibfield
  {journal} {\bibinfo  {journal} {Phys. Rev. B}\ }\textbf {\bibinfo {volume}
  {101}},\ \bibinfo {pages} {245116} (\bibinfo {year} {2020})}\BibitemShut
  {NoStop}%
\bibitem [{\citenamefont {Aharony}\ \emph {et~al.}(2013)\citenamefont
  {Aharony}, \citenamefont {Razamat}, \citenamefont {Seiberg},\ and\
  \citenamefont {Willett}}]{aharony2013}%
  \BibitemOpen
  \bibfield  {author} {\bibinfo {author} {\bibfnamefont {O.}~\bibnamefont
  {Aharony}}, \bibinfo {author} {\bibfnamefont {S.~S.}\ \bibnamefont
  {Razamat}}, \bibinfo {author} {\bibfnamefont {N.}~\bibnamefont {Seiberg}}, \
  and\ \bibinfo {author} {\bibfnamefont {B.}~\bibnamefont {Willett}},\ }\href
  {\doibase 10.1007/JHEP08(2013)099} {\bibfield  {journal} {\bibinfo  {journal}
  {JHEP}\ }\textbf {\bibinfo {volume} {08}},\ \bibinfo {pages} {099} (\bibinfo
  {year} {2013})}\BibitemShut {NoStop}%
\bibitem [{\citenamefont {Benini}\ \emph {et~al.}(2017)\citenamefont {Benini},
  \citenamefont {Hsin},\ and\ \citenamefont {Seiberg}}]{benini2017}%
  \BibitemOpen
  \bibfield  {author} {\bibinfo {author} {\bibfnamefont {F.}~\bibnamefont
  {Benini}}, \bibinfo {author} {\bibfnamefont {P.-S.}\ \bibnamefont {Hsin}}, \
  and\ \bibinfo {author} {\bibfnamefont {N.}~\bibnamefont {Seiberg}},\ }\href
  {\doibase 10.1007/JHEP04(2017)135} {\bibfield  {journal} {\bibinfo  {journal}
  {JHEP}\ }\textbf {\bibinfo {volume} {04}},\ \bibinfo {pages} {135} (\bibinfo
  {year} {2017})}\BibitemShut {NoStop}%
\bibitem [{\citenamefont {C\'ordova}\ \emph {et~al.}(2018)\citenamefont
  {C\'ordova}, \citenamefont {Hsin},\ and\ \citenamefont
  {Seiberg}}]{cordova2018}%
  \BibitemOpen
  \bibfield  {author} {\bibinfo {author} {\bibfnamefont {C.}~\bibnamefont
  {C\'ordova}}, \bibinfo {author} {\bibfnamefont {P.-S.}\ \bibnamefont {Hsin}},
  \ and\ \bibinfo {author} {\bibfnamefont {N.}~\bibnamefont {Seiberg}},\ }\href
  {\doibase 10.21468/SciPostPhys.4.4.021} {\bibfield  {journal} {\bibinfo
  {journal} {SciPost Phys.}\ }\textbf {\bibinfo {volume} {4}},\ \bibinfo
  {pages} {021} (\bibinfo {year} {2018})}\BibitemShut {NoStop}%
\bibitem [{\citenamefont {Senthil}\ \emph {et~al.}(2019)\citenamefont
  {Senthil}, \citenamefont {Son}, \citenamefont {Wang},\ and\ \citenamefont
  {Xu}}]{senthil2019}%
  \BibitemOpen
  \bibfield  {author} {\bibinfo {author} {\bibfnamefont {T.}~\bibnamefont
  {Senthil}}, \bibinfo {author} {\bibfnamefont {D.~T.}\ \bibnamefont {Son}},
  \bibinfo {author} {\bibfnamefont {C.}~\bibnamefont {Wang}}, \ and\ \bibinfo
  {author} {\bibfnamefont {C.}~\bibnamefont {Xu}},\ }\href {\doibase
  10.1016/j.physrep.2019.09.001} {\bibfield  {journal} {\bibinfo  {journal}
  {Phys. Rep.}\ }\textbf {\bibinfo {volume} {827}},\ \bibinfo {pages} {1}
  (\bibinfo {year} {2019})}\BibitemShut {NoStop}%
\bibitem [{\citenamefont {Ahn}\ \emph {et~al.}(2018)\citenamefont {Ahn},
  \citenamefont {Kim}, \citenamefont {Kim},\ and\ \citenamefont
  {Yang}}]{ahn2018}%
  \BibitemOpen
  \bibfield  {author} {\bibinfo {author} {\bibfnamefont {J.}~\bibnamefont
  {Ahn}}, \bibinfo {author} {\bibfnamefont {D.}~\bibnamefont {Kim}}, \bibinfo
  {author} {\bibfnamefont {Y.}~\bibnamefont {Kim}}, \ and\ \bibinfo {author}
  {\bibfnamefont {B.-J.}\ \bibnamefont {Yang}},\ }\href {\doibase
  10.1103/PhysRevLett.121.106403} {\bibfield  {journal} {\bibinfo  {journal}
  {Phys. Rev. Lett.}\ }\textbf {\bibinfo {volume} {121}},\ \bibinfo {pages}
  {106403} (\bibinfo {year} {2018})}\BibitemShut {NoStop}%
\bibitem [{\citenamefont {Ahn}\ \emph {et~al.}(2019)\citenamefont {Ahn},
  \citenamefont {Park}, \citenamefont {Kim}, \citenamefont {Kim},\ and\
  \citenamefont {Yang}}]{ahn2019}%
  \BibitemOpen
  \bibfield  {author} {\bibinfo {author} {\bibfnamefont {J.}~\bibnamefont
  {Ahn}}, \bibinfo {author} {\bibfnamefont {S.}~\bibnamefont {Park}}, \bibinfo
  {author} {\bibfnamefont {D.}~\bibnamefont {Kim}}, \bibinfo {author}
  {\bibfnamefont {Y.}~\bibnamefont {Kim}}, \ and\ \bibinfo {author}
  {\bibfnamefont {B.-J.}\ \bibnamefont {Yang}},\ }\href {\doibase
  10.1088/1674-1056/ab4d3b} {\bibfield  {journal} {\bibinfo  {journal} {Chin.
  Phys. B}\ }\textbf {\bibinfo {volume} {28}},\ \bibinfo {pages} {117101}
  (\bibinfo {year} {2019})}\BibitemShut {NoStop}%
\bibitem [{\citenamefont {Goddard}\ \emph {et~al.}(1977)\citenamefont
  {Goddard}, \citenamefont {Nuyts},\ and\ \citenamefont {Olive}}]{goddard1977}%
  \BibitemOpen
  \bibfield  {author} {\bibinfo {author} {\bibfnamefont {P.}~\bibnamefont
  {Goddard}}, \bibinfo {author} {\bibfnamefont {J.}~\bibnamefont {Nuyts}}, \
  and\ \bibinfo {author} {\bibfnamefont {D.}~\bibnamefont {Olive}},\ }\href
  {\doibase 10.1016/0550-3213(77)90221-8} {\bibfield  {journal} {\bibinfo
  {journal} {Nucl. Phys. B}\ }\textbf {\bibinfo {volume} {125}},\ \bibinfo
  {pages} {1} (\bibinfo {year} {1977})}\BibitemShut {NoStop}%
\bibitem [{\citenamefont {Tong}(2005)}]{tong2005}%
  \BibitemOpen
  \bibfield  {author} {\bibinfo {author} {\bibfnamefont {D.}~\bibnamefont
  {Tong}},\ }\href {https://arxiv.org/abs/hep-th/0509216} {\bibfield  {journal}
  {\bibinfo  {journal} {arXiv:hep-th/0509216}\ } (\bibinfo {year}
  {2005})}\BibitemShut {NoStop}%
\bibitem [{\citenamefont {Borokhov}\ \emph {et~al.}(2002)\citenamefont
  {Borokhov}, \citenamefont {Kapustin},\ and\ \citenamefont
  {Wu}}]{borokhov2002}%
  \BibitemOpen
  \bibfield  {author} {\bibinfo {author} {\bibfnamefont {V.}~\bibnamefont
  {Borokhov}}, \bibinfo {author} {\bibfnamefont {A.}~\bibnamefont {Kapustin}},
  \ and\ \bibinfo {author} {\bibfnamefont {X.}~\bibnamefont {Wu}},\ }\href
  {\doibase 10.1088/1126-6708/2002/11/049} {\bibfield  {journal} {\bibinfo
  {journal} {JHEP}\ }\textbf {\bibinfo {volume} {11}},\ \bibinfo {pages} {049}
  (\bibinfo {year} {2002})}\BibitemShut {NoStop}%
\bibitem [{\citenamefont {Brandt}\ and\ \citenamefont
  {Neri}(1979)}]{brandt1979}%
  \BibitemOpen
  \bibfield  {author} {\bibinfo {author} {\bibfnamefont {R.~A.}\ \bibnamefont
  {Brandt}}\ and\ \bibinfo {author} {\bibfnamefont {F.}~\bibnamefont {Neri}},\
  }\href {\doibase https://doi.org/10.1016/0550-3213(79)90211-6} {\bibfield
  {journal} {\bibinfo  {journal} {Nucl. Phys. B}\ }\textbf {\bibinfo {volume}
  {161}},\ \bibinfo {pages} {253} (\bibinfo {year} {1979})}\BibitemShut
  {NoStop}%
\bibitem [{\citenamefont {Coleman}(1983)}]{coleman1983}%
  \BibitemOpen
  \bibfield  {author} {\bibinfo {author} {\bibfnamefont {S.}~\bibnamefont
  {Coleman}},\ }in\ \href {\doibase 10.1007/978-1-4613-3655-6_2} {\emph
  {\bibinfo {booktitle} {The {{Unity}} of the {{Fundamental Interactions}}}}},\
  \bibinfo {editor} {edited by\ \bibinfo {editor} {\bibfnamefont
  {A.}~\bibnamefont {Zichichi}}}\ (\bibinfo  {publisher} {{Springer US}},\
  \bibinfo {address} {{Boston, MA}},\ \bibinfo {year} {1983})\BibitemShut
  {NoStop}%
\bibitem [{\citenamefont {Marston}(1990)}]{marston1990}%
  \BibitemOpen
  \bibfield  {author} {\bibinfo {author} {\bibfnamefont {J.~B.}\ \bibnamefont
  {Marston}},\ }\href {\doibase 10.1103/PhysRevLett.64.1166} {\bibfield
  {journal} {\bibinfo  {journal} {Phys. Rev. Lett.}\ }\textbf {\bibinfo
  {volume} {64}},\ \bibinfo {pages} {1166} (\bibinfo {year}
  {1990})}\BibitemShut {NoStop}%
\bibitem [{\citenamefont {{\"Unsal}}(2008)}]{unsal2008}%
  \BibitemOpen
  \bibfield  {author} {\bibinfo {author} {\bibfnamefont {M.}~\bibnamefont
  {{\"Unsal}}},\ }\href {http://arxiv.org/abs/0804.4664} {\bibfield  {journal}
  {\bibinfo  {journal} {arXiv:0804.4664}\ } (\bibinfo {year}
  {2008})}\BibitemShut {NoStop}%
\bibitem [{\citenamefont {Stone}(2020)}]{stone2020}%
  \BibitemOpen
  \bibfield  {author} {\bibinfo {author} {\bibfnamefont {M.}~\bibnamefont
  {Stone}},\ }\href {https://arxiv.org/abs/2009.00518} {\bibfield  {journal}
  {\bibinfo  {journal} {arXiv:2009.00518}\ } (\bibinfo {year}
  {2020})}\BibitemShut {NoStop}%
\bibitem [{\citenamefont {Hui}\ \emph {et~al.}(2018)\citenamefont {Hui},
  \citenamefont {Mulligan},\ and\ \citenamefont {Kim}}]{hui2018}%
  \BibitemOpen
  \bibfield  {author} {\bibinfo {author} {\bibfnamefont {A.}~\bibnamefont
  {Hui}}, \bibinfo {author} {\bibfnamefont {M.}~\bibnamefont {Mulligan}}, \
  and\ \bibinfo {author} {\bibfnamefont {E.-A.}\ \bibnamefont {Kim}},\ }\href
  {\doibase 10.1103/PhysRevB.97.085112} {\bibfield  {journal} {\bibinfo
  {journal} {Phys. Rev. B}\ }\textbf {\bibinfo {volume} {97}},\ \bibinfo
  {pages} {085112} (\bibinfo {year} {2018})}\BibitemShut {NoStop}%
\bibitem [{\citenamefont {Hui}\ \emph {et~al.}(2019)\citenamefont {Hui},
  \citenamefont {Kim},\ and\ \citenamefont {Mulligan}}]{hui2019}%
  \BibitemOpen
  \bibfield  {author} {\bibinfo {author} {\bibfnamefont {A.}~\bibnamefont
  {Hui}}, \bibinfo {author} {\bibfnamefont {E.-A.}\ \bibnamefont {Kim}}, \ and\
  \bibinfo {author} {\bibfnamefont {M.}~\bibnamefont {Mulligan}},\ }\href
  {\doibase 10.1103/PhysRevB.99.125135} {\bibfield  {journal} {\bibinfo
  {journal} {Phys. Rev. B}\ }\textbf {\bibinfo {volume} {99}},\ \bibinfo
  {pages} {125135} (\bibinfo {year} {2019})}\BibitemShut {NoStop}%
\bibitem [{\citenamefont {{'t Hooft}}(1974)}]{hooft1974}%
  \BibitemOpen
  \bibfield  {author} {\bibinfo {author} {\bibfnamefont {G.}~\bibnamefont {{'t
  Hooft}}},\ }\href {\doibase 10.1016/0550-3213(74)90154-0} {\bibfield
  {journal} {\bibinfo  {journal} {Nucl. Phys. B}\ }\textbf {\bibinfo {volume}
  {72}},\ \bibinfo {pages} {461} (\bibinfo {year} {1974})}\BibitemShut
  {NoStop}%
\bibitem [{\citenamefont {Pufu}(2014)}]{pufu2014}%
  \BibitemOpen
  \bibfield  {author} {\bibinfo {author} {\bibfnamefont {S.~S.}\ \bibnamefont
  {Pufu}},\ }\href {\doibase 10.1103/PhysRevD.89.065016} {\bibfield  {journal}
  {\bibinfo  {journal} {Phys. Rev. D}\ }\textbf {\bibinfo {volume} {89}},\
  \bibinfo {pages} {065016} (\bibinfo {year} {2014})}\BibitemShut {NoStop}%
\bibitem [{\citenamefont {Dupuis}\ \emph {et~al.}(2022)\citenamefont {Dupuis},
  \citenamefont {Boyack},\ and\ \citenamefont {Witczak-Krempa}}]{dupuis2022}%
  \BibitemOpen
  \bibfield  {author} {\bibinfo {author} {\bibfnamefont {{\'E}.}~\bibnamefont
  {Dupuis}}, \bibinfo {author} {\bibfnamefont {R.}~\bibnamefont {Boyack}}, \
  and\ \bibinfo {author} {\bibfnamefont {W.}~\bibnamefont {Witczak-Krempa}},\
  }\href {\doibase 10.1103/PhysRevX.12.031012} {\bibfield  {journal} {\bibinfo
  {journal} {Phys. Rev. X}\ }\textbf {\bibinfo {volume} {12}},\ \bibinfo
  {pages} {031012} (\bibinfo {year} {2022})}\BibitemShut {NoStop}%
\bibitem [{\citenamefont {Polyakov}(1988)}]{polyakov1988}%
  \BibitemOpen
  \bibfield  {author} {\bibinfo {author} {\bibfnamefont {A.~M.}\ \bibnamefont
  {Polyakov}},\ }\href {\doibase 10.1142/S0217732388000398} {\bibfield
  {journal} {\bibinfo  {journal} {Mod. Phys. Lett. A}\ }\textbf {\bibinfo
  {volume} {03}},\ \bibinfo {pages} {325} (\bibinfo {year} {1988})}\BibitemShut
  {NoStop}%
\bibitem [{\citenamefont {Fradkin}\ and\ \citenamefont
  {Schaposnik}(1994)}]{fradkin1994}%
  \BibitemOpen
  \bibfield  {author} {\bibinfo {author} {\bibfnamefont {E.}~\bibnamefont
  {Fradkin}}\ and\ \bibinfo {author} {\bibfnamefont {F.~A.}\ \bibnamefont
  {Schaposnik}},\ }\href {\doibase 10.1016/0370-2693(94)91374-9} {\bibfield
  {journal} {\bibinfo  {journal} {Phys. Lett. B}\ }\textbf {\bibinfo {volume}
  {338}},\ \bibinfo {pages} {253} (\bibinfo {year} {1994})}\BibitemShut
  {NoStop}%
\bibitem [{wit()}]{witten2003}%
  \BibitemOpen
  \href@noop {} {}\bibinfo {note} {E. Witten, in {\it From Fields to Strings:
  Circumnavigating Theoretical Physics (Ian Kogan Memorial Collection)}, edited
  by M. Shifman, A. Vainshtein, and J. Wheater (World Scientific, Singapore,
  2005), Vol. 2, pp. 1173-1200 [arXiv:hep-th/0307041].}\BibitemShut {Stop}%
\bibitem [{\citenamefont {Kogut}\ and\ \citenamefont
  {Susskind}(1975)}]{kogut1975}%
  \BibitemOpen
  \bibfield  {author} {\bibinfo {author} {\bibfnamefont {J.}~\bibnamefont
  {Kogut}}\ and\ \bibinfo {author} {\bibfnamefont {L.}~\bibnamefont
  {Susskind}},\ }\href {\doibase 10.1103/PhysRevD.11.395} {\bibfield  {journal}
  {\bibinfo  {journal} {Phys. Rev. D}\ }\textbf {\bibinfo {volume} {11}},\
  \bibinfo {pages} {395} (\bibinfo {year} {1975})}\BibitemShut {NoStop}%
\bibitem [{\citenamefont {Kogut}(1979)}]{kogut1979}%
  \BibitemOpen
  \bibfield  {author} {\bibinfo {author} {\bibfnamefont {J.~B.}\ \bibnamefont
  {Kogut}},\ }\href {\doibase 10.1103/RevModPhys.51.659} {\bibfield  {journal}
  {\bibinfo  {journal} {Rev. Mod. Phys.}\ }\textbf {\bibinfo {volume} {51}},\
  \bibinfo {pages} {659} (\bibinfo {year} {1979})}\BibitemShut {NoStop}%
\bibitem [{\citenamefont {Creutz}(1983)}]{creutz}%
  \BibitemOpen
  \bibfield  {author} {\bibinfo {author} {\bibfnamefont {M.}~\bibnamefont
  {Creutz}},\ }\href@noop {} {\emph {\bibinfo {title} {Quarks, Gluons and
  Lattices}}}\ (\bibinfo  {publisher} {Cambridge University Press},\ \bibinfo
  {address} {Cambridge},\ \bibinfo {year} {1983})\BibitemShut {NoStop}%
\bibitem [{\citenamefont {Fradkin}\ and\ \citenamefont
  {Susskind}(1978)}]{fradkin1978}%
  \BibitemOpen
  \bibfield  {author} {\bibinfo {author} {\bibfnamefont {E.}~\bibnamefont
  {Fradkin}}\ and\ \bibinfo {author} {\bibfnamefont {L.}~\bibnamefont
  {Susskind}},\ }\href {\doibase 10.1103/PhysRevD.17.2637} {\bibfield
  {journal} {\bibinfo  {journal} {Phys. Rev. D}\ }\textbf {\bibinfo {volume}
  {17}},\ \bibinfo {pages} {2637} (\bibinfo {year} {1978})}\BibitemShut
  {NoStop}%
\bibitem [{\citenamefont {Rossi}\ and\ \citenamefont
  {Wolff}(1984)}]{rossi1984}%
  \BibitemOpen
  \bibfield  {author} {\bibinfo {author} {\bibfnamefont {P.}~\bibnamefont
  {Rossi}}\ and\ \bibinfo {author} {\bibfnamefont {U.}~\bibnamefont {Wolff}},\
  }\href {\doibase 10.1016/0550-3213(84)90589-3} {\bibfield  {journal}
  {\bibinfo  {journal} {Nucl. Phys. B}\ }\textbf {\bibinfo {volume} {248}},\
  \bibinfo {pages} {105} (\bibinfo {year} {1984})}\BibitemShut {NoStop}%
\bibitem [{\citenamefont {Wolff}(1985)}]{wolff1985}%
  \BibitemOpen
  \bibfield  {author} {\bibinfo {author} {\bibfnamefont {U.}~\bibnamefont
  {Wolff}},\ }\href {\doibase 10.1016/0370-2693(85)91448-0} {\bibfield
  {journal} {\bibinfo  {journal} {Phys. Lett. B}\ }\textbf {\bibinfo {volume}
  {153}},\ \bibinfo {pages} {92} (\bibinfo {year} {1985})}\BibitemShut
  {NoStop}%
\bibitem [{\citenamefont {Collins}\ and\ \citenamefont
  {Matsumoto}(2009)}]{collins2009}%
  \BibitemOpen
  \bibfield  {author} {\bibinfo {author} {\bibfnamefont {B.}~\bibnamefont
  {Collins}}\ and\ \bibinfo {author} {\bibfnamefont {S.}~\bibnamefont
  {Matsumoto}},\ }\href {\doibase 10.1063/1.3251304} {\bibfield  {journal}
  {\bibinfo  {journal} {J. Math. Phys.}\ }\textbf {\bibinfo {volume} {50}},\
  \bibinfo {pages} {113516} (\bibinfo {year} {2009})}\BibitemShut {NoStop}%
\bibitem [{\citenamefont {Chen}\ and\ \citenamefont {Zimet}(2018)}]{chen2018}%
  \BibitemOpen
  \bibfield  {author} {\bibinfo {author} {\bibfnamefont {J.-Y.}\ \bibnamefont
  {Chen}}\ and\ \bibinfo {author} {\bibfnamefont {M.}~\bibnamefont {Zimet}},\
  }\href {\doibase 10.1007/JHEP08(2018)015} {\bibfield  {journal} {\bibinfo
  {journal} {JHEP}\ }\textbf {\bibinfo {volume} {08}},\ \bibinfo {pages} {015}
  (\bibinfo {year} {2018})}\BibitemShut {NoStop}%
\bibitem [{\citenamefont {Sahoo}\ \emph {et~al.}(2016)\citenamefont {Sahoo},
  \citenamefont {Zhang},\ and\ \citenamefont {Teo}}]{sahoo2016}%
  \BibitemOpen
  \bibfield  {author} {\bibinfo {author} {\bibfnamefont {S.}~\bibnamefont
  {Sahoo}}, \bibinfo {author} {\bibfnamefont {Z.}~\bibnamefont {Zhang}}, \ and\
  \bibinfo {author} {\bibfnamefont {J.~C.~Y.}\ \bibnamefont {Teo}},\ }\href
  {\doibase 10.1103/PhysRevB.94.165142} {\bibfield  {journal} {\bibinfo
  {journal} {Phys. Rev. B}\ }\textbf {\bibinfo {volume} {94}},\ \bibinfo
  {pages} {165142} (\bibinfo {year} {2016})}\BibitemShut {NoStop}%
\bibitem [{\citenamefont {Cheng}(2018)}]{cheng2018}%
  \BibitemOpen
  \bibfield  {author} {\bibinfo {author} {\bibfnamefont {M.}~\bibnamefont
  {Cheng}},\ }\href {\doibase 10.1103/PhysRevLett.120.036801} {\bibfield
  {journal} {\bibinfo  {journal} {Phys. Rev. Lett.}\ }\textbf {\bibinfo
  {volume} {120}},\ \bibinfo {pages} {036801} (\bibinfo {year}
  {2018})}\BibitemShut {NoStop}%
\bibitem [{\citenamefont {Di~Francesco}\ \emph {et~al.}(1997)\citenamefont
  {Di~Francesco}, \citenamefont {Mathieu},\ and\ \citenamefont
  {S\'en\'echal}}]{CFT}%
  \BibitemOpen
  \bibfield  {author} {\bibinfo {author} {\bibfnamefont {P.}~\bibnamefont
  {Di~Francesco}}, \bibinfo {author} {\bibfnamefont {P.}~\bibnamefont
  {Mathieu}}, \ and\ \bibinfo {author} {\bibfnamefont {D.}~\bibnamefont
  {S\'en\'echal}},\ }\href@noop {} {\emph {\bibinfo {title} {Conformal Field
  Theory}}}\ (\bibinfo  {publisher} {Springer},\ \bibinfo {address} {New
  York},\ \bibinfo {year} {1997})\BibitemShut {NoStop}%
\bibitem [{\citenamefont {Affleck}(1986)}]{affleck1986}%
  \BibitemOpen
  \bibfield  {author} {\bibinfo {author} {\bibfnamefont {I.}~\bibnamefont
  {Affleck}},\ }\href {\doibase 10.1016/0550-3213(86)90167-7} {\bibfield
  {journal} {\bibinfo  {journal} {Nucl. Phys. B}\ }\textbf {\bibinfo {volume}
  {265}},\ \bibinfo {pages} {409} (\bibinfo {year} {1986})}\BibitemShut
  {NoStop}%
\bibitem [{\citenamefont {Chamon}(2000)}]{chamon2000}%
  \BibitemOpen
  \bibfield  {author} {\bibinfo {author} {\bibfnamefont {C.}~\bibnamefont
  {Chamon}},\ }\href {\doibase 10.1103/PhysRevB.62.2806} {\bibfield  {journal}
  {\bibinfo  {journal} {Phys. Rev. B}\ }\textbf {\bibinfo {volume} {62}},\
  \bibinfo {pages} {2806} (\bibinfo {year} {2000})}\BibitemShut {NoStop}%
\bibitem [{\citenamefont {Hou}\ \emph {et~al.}(2007)\citenamefont {Hou},
  \citenamefont {Chamon},\ and\ \citenamefont {Mudry}}]{hou2007}%
  \BibitemOpen
  \bibfield  {author} {\bibinfo {author} {\bibfnamefont {C.-Y.}\ \bibnamefont
  {Hou}}, \bibinfo {author} {\bibfnamefont {C.}~\bibnamefont {Chamon}}, \ and\
  \bibinfo {author} {\bibfnamefont {C.}~\bibnamefont {Mudry}},\ }\href
  {\doibase 10.1103/PhysRevLett.98.186809} {\bibfield  {journal} {\bibinfo
  {journal} {Phys. Rev. Lett.}\ }\textbf {\bibinfo {volume} {98}},\ \bibinfo
  {pages} {186809} (\bibinfo {year} {2007})}\BibitemShut {NoStop}%
\bibitem [{\citenamefont {Bernard}(1979)}]{bernard1979}%
  \BibitemOpen
  \bibfield  {author} {\bibinfo {author} {\bibfnamefont {C.}~\bibnamefont
  {Bernard}},\ }\href {\doibase 10.1103/PhysRevD.19.3013} {\bibfield  {journal}
  {\bibinfo  {journal} {Phys. Rev. D}\ }\textbf {\bibinfo {volume} {19}},\
  \bibinfo {pages} {3013} (\bibinfo {year} {1979})}\BibitemShut {NoStop}%
\bibitem [{\citenamefont {Osborn}(1981)}]{osborn1981}%
  \BibitemOpen
  \bibfield  {author} {\bibinfo {author} {\bibfnamefont {H.}~\bibnamefont
  {Osborn}},\ }\href {\doibase https://doi.org/10.1016/0003-4916(81)90159-7}
  {\bibfield  {journal} {\bibinfo  {journal} {Ann. Phys.}\ }\textbf {\bibinfo
  {volume} {135}},\ \bibinfo {pages} {373} (\bibinfo {year}
  {1981})}\BibitemShut {NoStop}%
\bibitem [{\citenamefont {Mari{\~n}o}(2015)}]{marino2015}%
  \BibitemOpen
  \bibfield  {author} {\bibinfo {author} {\bibfnamefont {M.}~\bibnamefont
  {Mari{\~n}o}},\ }\href@noop {} {\emph {\bibinfo {title} {Instantons and Large
  $N$}}}\ (\bibinfo  {publisher} {Cambridge University Press},\ \bibinfo
  {address} {Cambridge},\ \bibinfo {year} {2015})\BibitemShut {NoStop}%
\bibitem [{\citenamefont {Coleman}(1985)}]{coleman1985}%
  \BibitemOpen
  \bibfield  {author} {\bibinfo {author} {\bibfnamefont {S.}~\bibnamefont
  {Coleman}},\ }\href@noop {} {\emph {\bibinfo {title} {Aspects of Symmetry}}}\
  (\bibinfo  {publisher} {Cambridge University Press},\ \bibinfo {address}
  {Cambridge},\ \bibinfo {year} {1985})\BibitemShut {NoStop}%
\end{thebibliography}%
\end{document}